\documentclass[journal,onecolumn, 12pt]{IEEEtran}

\ifCLASSINFOpdf
\else
\fi

\usepackage{amsmath,amssymb,amsfonts}
\usepackage{cite}
\usepackage{url}
\usepackage{xcolor}

\usepackage{nicefrac}
\usepackage{mathtools}
\usepackage{hyperref}
\usepackage{enumitem}

\newtheorem{theorem}{Theorem}
\newtheorem{proposition}{Proposition}
\newtheorem{lemma}{Lemma}

\newtheorem{remark}{Remark}

\newcommand{\ttS}{{\mathtt{S}}}
\newcommand{\ttX}{{\mathtt{X}}}

\newcommand{\bi}{{\mathbf{i}}}
\newcommand{\bs}{{\mathbf{s}}}
\newcommand{\bS}{{\mathbf{S}}}

\newcommand{\bU}{{\mathbf{U}}}

\newcommand{\bV}{{\mathbf{V}}}
\newcommand{\bW}{{\mathbf{W}}}
\newcommand{\bx}{{\mathbf{x}}}
\newcommand{\bX}{{\mathbf{X}}}
\newcommand{\by}{{\mathbf{y}}}
\newcommand{\bY}{{\mathbf{Y}}}

\newcommand{\bZ}{{\mathbf{Z}}}
\newcommand{\sT}{{\mathrm{T}}}
\newcommand\smallO{
	\mathchoice
	{{\scriptstyle\mathcal{O}}}
	{{\scriptstyle\mathcal{O}}}
	{{\scriptscriptstyle\mathcal{O}}}
	{\scalebox{.7}{$\scriptscriptstyle\mathcal{O}$}}
}
\newcommand{\R}{\mathbb{R}}
\newcommand{\cI}{\mathcal{I}}
\newcommand{\cL}{\mathcal{L}}
\newcommand{\cN}{\mathcal{N}}
\newcommand{\cH}{\mathcal{H}}
\newcommand{\cO}{\mathcal{O}}
\newcommand{\cZ}{\mathcal{Z}}
\newcommand{\E}{\mathbb{E}}
\newcommand{\Var}{{\mathbb{V}\mathrm{ar}}}
\newcommand{\MMSE}{\mathrm{MMSE}}
\newcommand{\simiid}[1][0pt]{\mathrel{\raisebox{#1}{$\sim$}}}
\newcommand{\iid}{\overset{\text{\tiny i.i.d.}}{\simiid[-2pt]}}

\begin{document}
%
\title{Tensor estimation with structured priors}
%
%
%
\author{
	Cl\'{e}ment Luneau and Nicolas Macris\\
	\'{E}cole Polytechnique F\'{e}d\'{e}rale de Lausanne, Switzerland.\\
	Emails: clement.luneau@epfl.ch; nicolas.macris@epfl.ch
}

%


\maketitle

\begin{abstract}
We consider rank-one symmetric tensor estimation when the tensor is corrupted by gaussian noise and the spike forming the tensor is a structured signal coming from a generalized linear model. The latter is a mathematically tractable model of a non-trivial hidden lower-dimensional latent structure in a signal. We work in a large dimensional regime with fixed ratio of signal-to-latent space dimensions.
Remarkably, in this asymptotic regime, the mutual information between the spike and the observations can be expressed as a finite-dimensional variational problem, and it is possible to deduce the minimum-mean-square-error from its solution. We discuss, on examples, properties of the phase transitions as a function of the signal-to-noise ratio. Typically, the critical signal-to-noise ratio decreases with increasing signal-to-latent space dimensions.  We discuss the limit of vanishing ratio of signal-to-latent space dimensions and determine the limiting
tensor estimation problem. We also point out similarities and differences with the case of matrices.
\end{abstract}


%
\IEEEpeerreviewmaketitle

\section{Introduction}\label{section:introduction}
\IEEEPARstart{N}{a}tural signals have an underlying structure, an insight that has triggered a paradigm shift in the last fifteen years, and spurred fundamental  progress in estimation and inference. Compressive sensing \cite{CandesRombergTao_2006,Donoho_CompressedSensing2006} takes sparsity as the model of structure when a signal $\bX\in\mathbb{R}^n$ has a sparse representation in an appropriate basis, that is, $\bX = \Psi \bZ$ with $\Psi$ an $n\times n$ change of basis matrix and $\bZ\in \mathbb{R}^n$ a sparse vector with $p \ll n$ non-zero components.
For example, $\bX$ can represent a natural image and $\Psi$ a wavelet basis \cite{Mallat_book_1999}.
Despite its success, this model of structure is often too constrained because the appropriate basis  may be unknown and, more generally, the linearity of the transformation may be a severe limitation.
Deep networks have been proposed as an alternative \cite{MousaviPatel_2015} and,
with the advent of generative adversarial networks (GAN) \cite{GoodfellowGAN_2014} and variational auto-encoders (VAE) \cite{Hinton504},
such flexible and non-linear ``generative models'' of structure have been the object of intense interest. Roughly speaking, a generative model can be viewed as a mapping $G: \bS\in \mathbb{R}^p \mapsto \bX = G(\bS)\in \mathbb{R}^n$ with $p\ll n$ and satisfying certain general regularity assumptions \cite{Bora_2017}.
In other words, the signal $\bX$ lies on a low $p$-dimensional ``manifold'' parametrized by $\bS$.
Such models have been studied in the framework of classical denoising problems from observations $\bY=A \bX + \bZ$ where $A$ is a sensing matrix and $\bZ$ some Gaussian noise.
In particular, \cite{Bora_2017} studies fundamental limits under minimal Lipshitz conditions on $G$ and empirically investigates the problem with learned mappings coming from GAN and VAE
Another kind of generative model takes $G$ equal to a one-layer or multi-layer neural network with fixed weights (i.e., frozen and not learned) drawn from a random matrix ensemble \cite{ManoelKrzakala_2017, DBLP:journals/tit/HandV20, heckel2018rateoptimal, HandLeongVoroninski_2018, DBLP:journals/corr/abs-1803-09319}.
Such mappings $G$ are often referred to as \textit{generalized linear models} and this is the terminology that we adopt here.
The simplification of fixed random weights has the virtue of being much more amenable to mathematical (or at least analytical) analysis.
Especially, the mutual information as well as the message passing algorithmic behaviour for classical denoising have been discussed in depth in a Bayesian setting at various levels of rigor \cite{ManoelKrzakala_2017, Gabrie_TwoLayerGLM_JSTAT_2019}. 

In this work we investigate generalized models of structure in the context of non-linear estimation (or factorization) of noisy tensors. Tensors representing data have found many modern applications in signal processing, graph analysis, data mining and machine learning \cite{sidiropoulos2016, cichoki2015, kolda2009}, with a large part of the literature focusing on tensor decompositions, either in deterministic settings, or in random settings with independent structureless components.
Here we focus on a simple statistical model of noisy symmetric rank-one tensors. A {\it structured} signal
$\bX=(X_1, \cdots, X_n) \in \mathbb{R}^n$ is generated by a one-layer GLM $X_i = \varphi( (\bW \bS)_i/\sqrt p)$ where the latent vector $\bS\in \mathbb{R}^p$ has independent and identically distributed (i.i.d.) entries and $\bW$ is a \textit{known} random matrix with independent standard Gaussian entries. We only observe a noisy version of the rank-one tensor $\bX^{\otimes r}$ ($r\geq 2$) through an additive white Gaussian noise channel, i.e., $\bY = \frac{\sqrt{\lambda}}{n^{(r-1)/2}} \bX^{\otimes r} + \bZ$ where the noise $\bZ$ is a symmetric tensor with independent standard Gaussians entries and $\lambda > 0$ is the signal-to-noise ratio. 
We study the high dimensional limit $n, p\to \infty$ such that $n/p\to \alpha = \Theta(1)$ and show that, quite remarkably,
the asymptotic mutual information $\lim_{n\to +\infty}I(\bX;\bY\vert \bW)/n$ is given by a finite-dimensional variational problem (see Theorem~\ref{theorem:limit_mutual_information} in Section~\ref{subsec:results}).
We also rigorously deduce the corresponding asymptotic \textit{minimum mean square error} (MMSE), which is given by a simple function of the solution to the variational problem (see Theorem ~\ref{theorem:tensor_MMSE} in Section~\ref{subsec:results}).
For concreteness, and to keep the analysis as simple as possible, we focus on the case $r=3$ and one-layer GLM.
However, extensions to any order $r > 3$, multi-layer GLM and asymmetric tensors are possible with the techniques used here.
An extensive recent study of the matrix case $r=2$ can be found in \cite{aubin2019spiked}.

The analysis and results presented here go beyond many recent works dealing with i.i.d.\ components for $\bX$, for matrices $r=2$ \cite{XXT, Lelarge_fundamental_2019, miolane2017fundamental}, and tensors $r\geq 3$ \cite{LesieurMiolane_2017, barbier2017layered}.
There is a rich phenomenology of phase transitions already for the i.i.d.\ case which stems from the (simpler) variational formula for the mutual information.
In Section~\ref{subsec:examples} we discuss the (numerical) solutions to the new variational problem obtained for structured signals for various examples of priors and activation functions, and we illustrate properties of phase transitions. Furthermore we discuss the similarities and differences between the genuine tensor and matrix cases. 

Let us say a few words about the techniques used in this work.
There is a long history in the literature connecting Bayesian inference problems with spin-glass models of statistical mechanics \cite{nishimori01, mezard2009information} and it has been conjectured for some time
that the true variational expressions for the mutual information should coincide with the so-called ``replica-symmetric''
formula for the free energy derived by analytical non-rigorous methods.
The veracity of these conjectures has now been established by a variety of methods for various problems, e.g., 
coding theory \cite{Giurgiu_SCproof}, random linear estimation \cite{8606971, 9079920}, matrix and tensor estimation \cite{koradamacris, XXT, Lelarge_fundamental_2019, miolane2017fundamental, barbier2017layered}.
In all these cases the signal has i.i.d.\ components.
For structured signals, rigorous proofs of the low-dimensional variational expression for the asymptotic mutual information are virtually non-existent.
To the best of our knowledge, besides the case where $\bX$ is uniformly distributed on the sphere \cite{luneau2020highdimensional} (which turns out to be equivalent to an i.i.d.\ Gaussian prior), there are two recent exceptions:
\cite{Gabrie_TwoLayerGLM_JSTAT_2019} which includes the rigorous calculation of a mutual information for a GLM with input generated by another GLM, and \cite{aubin2019spiked} which treats the rank-one matrix case with input coming from a GLM.
The later work uses two different flavors of the interpolation method \cite{Guerra-Toninelli-2002, Alaoui2018} which do not extend to odd-order tensors nor asymmetric ones.
Moreover, certain (reasonable) assumptions are required.
In this work we leverage on recent progress on the proofs of replica-symmetric formulas by the \textit{adaptive interpolation method} \cite{barbier_adaptive_2019, Barbier_Macris_jphysA_2019} which is a powerful evolution of the celebrated Guerra-Toninelli interpolation scheme \cite{Guerra-Toninelli-2002}.
Our treatment is completely self-contained, leverages on only one method, and can also deal with asymmetric matrices and tensors.
 
In Section~\ref{section:main_results} we formulate the model, present the main theorems for the asymptotic mutual information and MMSE along with examples and illustrations of phase transitions, and explain key ideas behind the proofs.
In Sections~\ref{sec:proofthm1} and \ref{sec:proofthm2} we go through the proofs and
in Section~\ref{section:alphatendstozero} we give an analysis of the limit $\alpha \to 0$.
The appendices contain technical derivations. 

\section{Asymptotic mutual information and MMSE for tensor decomposition with a generative prior}\label{section:main_results}
We formulate a statistical model of rank-one tensor decomposition given noisy observations, when the spike is itself generated from another latent vector.
We observe the entries of a symmetric tensor $\bY \in (\R^n)^{\otimes 3}$ given by:
\begin{equation}\label{eq:entries_Y}
	Y_{ijk} = \frac{\sqrt{\lambda}}{n} X_{i} X_{j} X_{k} + Z_{ijk} \;, 1 \leq i \leq j \leq k \leq n\: ;
\end{equation}
where the positive real number $\lambda$ plays the role of a SNR, $Z_{ijk} \iid \cN(0,1)$, $1 \leq i \leq j \leq k$, is an additive white Gaussian noise and $X_1, \dots, X_n$ are the entries of the spike $\bX \in \R^n$. This spike is generated by a latent vector $\bS \in \R^p$ -- whose entries are i.i.d. with respect to (w.r.t.) some probability distribution $P_S$ on the real numbers -- via a \textit{generalized linear model} (GLM):
\begin{equation}\label{eq:definition_X}
	X_i \triangleq \varphi\bigg(\frac{(\bW \bS)_i}{\sqrt{p}}\bigg),\quad i=1,\cdots, n \;.
\end{equation}
The $n \times p$ random matrix $\bW$ has entries i.i.d.\ with respect to $\cN(0,1)$. 
It is often customary to summarize \eqref{eq:definition_X} by $\bX = \varphi\big(\bW \bS / \sqrt{p}\big)$ where it is understood that the function $\varphi: \R \to \R$ is applied componentwise.

\subsection{Main results}\label{subsec:results}
Our main results are stated in the next two theorems. They provide a complete information-theoretic characterization of the problem.
Theorem 1 expresses the normalized mutual information $n^{-1}I(\bX ; \bY \vert \bW)$, in the high-dimensional regime where $n\to +\infty$ while $n/p =\alpha$ is kept fixed, as a \textit{low-dimensional} explicit  variational problem. This variational problem involves an optimization over three parameters and can be solved numerically given the activation function $\varphi$ and the prior distribution $P_S$.
%

\begin{theorem}[Mutual information between $\bX$ and $\bY$ given $\bW$ in the high-dimensional regime]\label{theorem:limit_mutual_information}
Suppose that the following hypotheses hold:
\begin{enumerate}[label=(H\arabic*), nosep]
	\item \label{hyp:S_bounded_support}
	There exists $M_S > 0$ such that the support of $P_S$ is included in $[-M_S,M_S]$.
	\item \label{hyp:varphi} $\varphi$ is bounded and twice differentiable with its first and second derivatives being bounded and continuous. They are denoted $\varphi^\prime$, $\varphi^{\prime\prime}$.
\end{enumerate}
Let $S \sim P_S$ and $U,V,Z,\widetilde{Z} \sim \cN(0,1)$ independent scalar random variables.
Define the second moments $\rho_s = \E[S^2]$ and $\rho_x = \E[\varphi(T)^2]$ with $T \sim \cN(0,\rho_s)$. 
Define the potential function $\psi_{\lambda, \alpha}: [0,+\infty)^2 \times [0,\rho_s]$:
\begin{multline}\label{def_potential_psi}
\psi_{\lambda,\alpha}(q_x, q_s, r_s) \triangleq
\frac{1}{\alpha}I(S;\sqrt{r_s}\,S + Z) + I\big(U; \sqrt{\lambda q_x^2/2}\,\varphi(\sqrt{\rho_s - q_s} \, U + \sqrt{q_s} \, V) + \widetilde{Z} \,\big\vert\, V \big)\\
- \frac{r_s(\rho_s - q_s)}{2\alpha}
+\frac{\lambda}{12} (\rho_x - q_x)^2(\rho_x + 2 q_x)\;.
\end{multline}
If $n$, $p$ go to infinity such that $\nicefrac{n}{p} \to \alpha >0$ then:
\begin{equation}\label{eq:main_I}
\lim_{n \to +\infty} \frac{I(\bX ; \bY \vert \bW)}{n}
= \mathop{\vphantom{p}\inf}_{q_x \in [0,\rho_x]}\adjustlimits{\inf}_{q_s \in [0,\rho_s]} {\sup}_{r_s \geq 0}\; \psi_{\lambda,\alpha}(q_x , q_s, r_s) \;.
\end{equation}
\end{theorem}

%

One important quantity to assess the performance of an algorithm designed to recover $\bX^{\otimes 3}$ from the knowledge of $\bY$ and $\bW$ is the \textit{minimum mean square error} (MMSE).
The later serves as a lower bar on the error of any estimator, and as a limit to approach as closely as possible for any algorithm striving to estimate $\bX^{\otimes 3}$.
It is well-known that the mean square error of an estimator of $\bX^{\otimes 3}$ that is a function of $\bY, \bW$ \textit{only} is minimized by the posterior mean $\E[\bX^{\otimes 3} \vert \bY, \bW]$.
We denote the tensor-MMSE by $\MMSE_n(\bX^{\otimes 3}\vert \bY, \bW)$, i.e.,
\begin{equation}\label{def:tensor_MMSE}
\MMSE_n(\bX^{\otimes 3}\vert \bY, \bW) 
\triangleq \frac{\E\,\big\Vert \bX^{\otimes 3} - \E[\bX^{\otimes 3} \vert \bY, \bW] \big\Vert^2 }{n^3}\;.
\end{equation}
It depends on $\lambda$ through the observations $\bY$.
Combining Theorem~\ref{theorem:limit_mutual_information} with the \textit{I-MMSE relation} (see \cite{GuoShamaiVerdu_IMMSE_2005})
\begin{equation}\label{eq:I-MMSE}
\frac{\partial}{\partial \lambda}\bigg(\frac{I(\bX, \bY \vert \bW)}{n}\bigg)
= \frac{1}{12}\MMSE_n(\bX^{\otimes 3}\vert \bY, \bW) + \cO(n^{-1})
\end{equation}
yields Theorem~\ref{theorem:tensor_MMSE}. 
It gives a formula for the tensor-MMSE in the high-dimensional regime that can be calculated from the solution to the variational problem \eqref{eq:main_I}.
Its proof is given in Section~\ref{sec:proofthm2}.

\begin{theorem}[Tensor-MMSE]\label{theorem:tensor_MMSE}
Suppose that \ref{hyp:S_bounded_support} and \ref{hyp:varphi} hold.
Define for all $\lambda \in (0, +\infty)$:
\begin{multline*}
\mathcal{Q}_x^*(\lambda)
\triangleq \bigg\{q_x^* \in [0,\rho_x] :
\adjustlimits{\inf}_{q_s \in [0,\rho_s]} {\sup}_{r_s \geq 0}\; \psi_{\lambda,\alpha}(q_x^* , q_s, r_s) 
= \mathop{\vphantom{p}\inf}_{q_x \in [0,\rho_x]}\adjustlimits{\inf}_{q_s \in [0,\rho_s]} {\sup}_{r_s \geq 0}\; \psi_{\lambda,\alpha}(q_x , q_s, r_s) 
\bigg\}\;.
\end{multline*}
For every $\lambda > 0$, $\mathcal{Q}_x^*(\lambda)$ is nonempty and the set
$\mathcal{D} \triangleq \big\{\lambda \in (0,+\infty): \mathcal{Q}_x^*(\lambda) \text{ is a singleton} \big\}$
is equal to $(0,+\infty)$ minus a countable set.
For every $\lambda \in \mathcal{D}$, letting $\mathcal{Q}_x^*(\lambda) = \{q_x^*(\lambda)\}$, we have:
\begin{equation}
\lim_{\substack{n \to +\infty\\ \nicefrac{n}{p}\to \alpha}} \MMSE_n(\bX^{\otimes 3}\vert \bY, \bW) 
= \rho_x^3 - \big(q_x^*(\lambda)\big)^3 \;.
\end{equation}
\end{theorem}

Extensions in various directions of Theorems~\ref{theorem:limit_mutual_information} and \ref{theorem:tensor_MMSE} are possible by the methods of the present paper, but at the expense of more technical work.
First, the analysis for rank-one tensors of any rank $r \geq 3$ is identical. The potential is given by
\begin{multline*}
\psi_{\lambda,\alpha}(q_x, q_s, r_s) \triangleq
\frac{1}{\alpha}I(S;\sqrt{r_s}\,S + Z) + I\big(U; \sqrt{\nicefrac{\lambda q_x^{r-1}}{(r-1)!}}\,\varphi(\sqrt{\rho_s - q_s} \, U + \sqrt{q_s} \, V) + \widetilde{Z} \,\big\vert\, V \big)\\
- \frac{r_s(\rho_s - q_s)}{2\alpha}
+\frac{\lambda}{2(r!)}\big(\rho_x^r
+ r q_x^r
- rq_x^{r-1}\rho_x\big)\;,
\end{multline*}
while the asymptotic tensor-MMSE is $\rho_x^r - (q_x^*(\lambda))^r$.
Second, the results can be extended to unbounded activation functions and priors with unbounded support but finite third moments.
This involves a technical limiting process on both sides of equation \eqref{eq:main_I} using the methods in \cite{barbierGLM}.
Another direction that should be amenable to analysis with our methods is the case of asymmetric tensors, e.g., $\bX^{\otimes 3}$ is replaced by $\bU \otimes \bV \otimes \bW$ where each of the three different vectors is given by a GLM.
The structureless case where all three vectors $\bU$, $\bV$, $\bW$ have i.i.d.\ entries is treated in \cite{barbier2017layered}, and the variational problem already displays a rich phenomenology in the highly asymmetric case \cite{Kadmon_2019}.

A high level summary on how we prove the theorems is given in Section~\ref{subsec:key_ideas_proofs} while the proofs themselves are carried out in Sections~\ref{sec:proofthm1} and \ref{sec:proofthm2}.  
\subsection{Examples of phase transitions and their properties}\label{subsec:examples}
This section illustrates features of the phase transitions found when numerically solving the variational problem \eqref{eq:main_I} for $r=3$.
We also discuss similarities and differences with the matrix case $r=2$.
To find solutions to the variational problem \eqref{eq:main_I}, we write down the stationary point equations of the potential function \eqref{def_potential_psi}.
It yields a fixed point equation for $(q_x,q_s,r_s)$ that we solve with a fixed-point iteration starting from several different initializations.
When multiple fixed points exist, we keep the one corresponding to the smallest potential value as it should be clear from the form of the optimization problem \eqref{eq:main_I}.

\begin{figure}[hbt]
	\centering
	\includegraphics[width=0.85\textwidth]{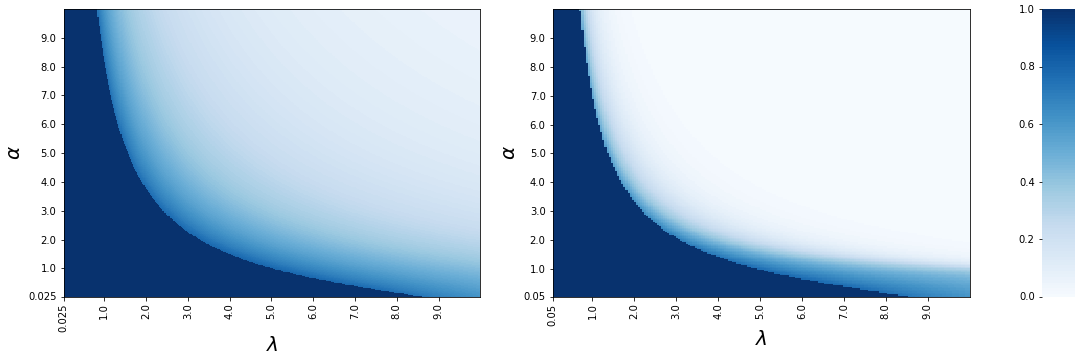}
	\caption{\label{MMSE_alpha_lambda}
		Asymptotic tensor-MMSE for $r=3$ as a function of $(\lambda,\alpha)$ for a linear activation $\varphi(x) = x$.
		{\it Left:} Gaussian prior $P_S \sim \cN(0,1)$.
		{\it Right:} Rademacher prior $P_S(1) = P_S(-1) = \frac{1}{2}$.
		We observe a unique discontinuity line $\lambda_c(\alpha)$ below which the MMSE equals its maximum $\rho_x^3=1$.
		Above the line, the MMSE is strictly less than $1$ and decreases to zero.
		For $\alpha$ close to $0$, the threshold $\lambda_c(\alpha) \approx 8.73$ is the same threshold than in the i.i.d.\ case with a Gaussian prior $X_1,\dots,X_n \iid \cN(0,1)$.}
\end{figure}

\begin{figure}[hbt]
	\centering
	\includegraphics[width=0.7\textwidth]{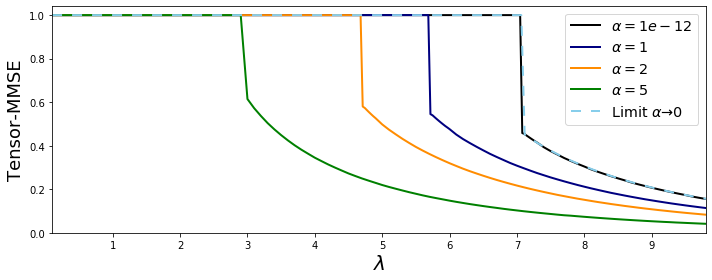}
	\caption{\label{MMSE_sign_activation}
		Asymptotic tensor-MMSE for $r=3$, $P_S = \cN(0,1)$ and $\varphi(z) =\mathrm{sign}(z)$ as a function of $\lambda$.
		The location $\lambda_c(\alpha)$ of the discontinuity decreases with increasing $\alpha$. For $\alpha=10^{-12}$ the threshold $\lambda_c(\alpha) \approx 7.07$ is the same than for the i.i.d. case with Rademacher prior $X_1, \dots, X_n \iid P_X(\pm 1)=\frac{1}{2}$ (whose asymptotic MMSE is given by the curve ``Limit $\alpha \to 0$'').
	}
\end{figure}

We first focus on the case of odd activation functions $\varphi(-z) = -\varphi(z)$ and centered priors $\E_{S \sim P_S}[S] = 0$.
This implies $\E\,X_i = 0$ and, if $\varphi$ is not identically zero, this is a necessary and sufficient condition for the existence of a fixed point $(q_x, q_s, r_s)$ such that $q_x=0$ (in which case we also have $q_s = r_s = 0$).
The same condition arises in the matrix case \cite{aubin2019spiked} but, contrary to what happens there, we find that all eigenvalues of the Jacobian matrix at the all-zero fixed point are zero indicating that it is asymptotically stable for order-$3$ tensors.
Numerically, we observe that for all $\lambda < \lambda_c(\alpha)$ this \textit{uninformative} fixed point yields the smallest potential.
This means that in this phase the asymptotic tensor-MMSE is equal to its maximum $\rho_x^3$: one cannot estimate the signal better than random guessing.
When $\lambda > \lambda_c(\alpha)$ a fixed point with a lower potential value appears.
The asymptotic MMSE has a jump discontinuity at $\lambda = \lambda_c(\alpha)$ and decreases for $\lambda > \lambda_c(\alpha)$.
These features are already observed for the structureless i.i.d.\ case.
In the structured case, we observe that  $\lambda_c(\alpha)$ has a monotone decrease with increasing $\alpha$. This is illustrated in Figure~\ref{MMSE_alpha_lambda} for a linear activation function and in Figure~\ref{MMSE_sign_activation} for a $\mathrm{sign}$ activation function\footnote{Our theorems are proven here for bounded and smooth activation functions but, as explained, the proofs can be extended to unbounded and piecewise differentiable ones. Numerical solutions involve non-trivial integrals that are much easier to handle for piecewise linear functions.}
%

In Section~\ref{section:alphatendstozero} we present a \textit{non-rigorous} calculation which shows that, in the limit $\alpha \to 0$, the asymptotic tensor-MMSE -- and in particular the threshold $\lambda_c(\alpha)$ -- is the same than for a tensor denoising problem
$\widetilde{Y}_{ijk} = \frac{\sqrt{\lambda}}{n} \widetilde{X}_i\widetilde{X}_j \widetilde{X}_k + \widetilde{Z}_{ijk}$
with $\widetilde{X}_i = \varphi(\sqrt{\rho_s - \E[S]^2} \, U_i + \vert \E S \vert\, V_i)$,
where $U_1,\dots, U_n\iid \cN(0,1)$ are latent variables and $V_1,\dots, V_n \iid \cN(0,1)$ are \textit{known}.
The latter take into account the bias that is present when $\E S \neq 0$.
We stress that when $\E S \neq 0$ the asymptotic mutual information of this problem (given by \eqref{limit_mutual_info_alpha=0} in Section~\ref{section:alphatendstozero}) is not quite the same as the one known in the literature for rank-one tensor problems with i.i.d.\ $X_i$'s.
However, it is not difficult to adapt the proof to account for the side information $\bV$ and obtain \eqref{limit_mutual_info_alpha=0}.
When the prior is centered $(\E S = 0)$, the limiting problem is just the usual rank-one tensor denoising problem with spike signal $\widetilde X_i \iid \varphi(\cN(0, \rho_s))$.
Numerically, we indeed observe in Figure~\ref{MMSE_alpha_lambda} that for both kinds of priors and for $\alpha$ close to $0$ the threshold $\lambda_c(\alpha) \approx 8.73$ is the same than for a signal $X_1,\dots,X_n \iid \cN(0,1)$.
Similarly, in Figure~\ref{MMSE_sign_activation}, the curve for $\alpha=10^{-12}$ agrees with the one labelled ``Limit $\alpha \to 0$'' corresponding to the asymptotic tensor-MMSE of the limiting tensor problem and that is computed using the formulas known in the literature.

\begin{figure}[ht]
	\centering
	\includegraphics[width=0.85\textwidth]{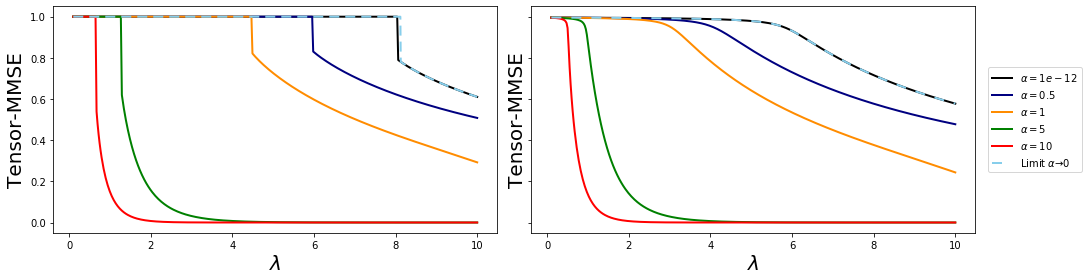}
	\caption{\label{MMSE_asym}
		Asymptotic tensor-MMSE for $\varphi(z)=z$ and an asymmetric Rademacher prior $P_S(1) = 1-P_S(-1) =p$.
		\textit{Left:} $p=0.6$.
		\textit{Right:} $p=0.7$.
	}
\end{figure}

We next discuss an example of non-centered latent prior $P_S$.
In Figure~\ref{MMSE_asym} we draw the asymptotic tensor-MMSE for a linear activation function and a Rademacher prior $P_S(1) = p$, $P_S(-1) = 1-p$ with $p \in \{0.6, 0.7\}$.
We observe that for a small asymmetry the asymptotic MMSE has a jump discontinuity just as in the centered case, while it becomes continuous once the asymmetry is large enough.
Here $\E S = 2p-1$ and the asymptotic MMSE of the predicted limiting problem \eqref{limit_mutual_info_alpha=0} is again in agreement with the one for $\alpha=10^{-12}$ close to $0$.

\begin{figure}[hbt]
	\centering
	\includegraphics[width=0.85\textwidth]{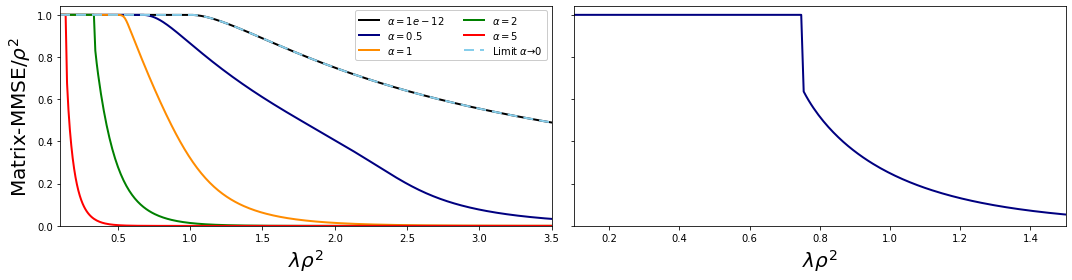}
	\caption{\label{MMSE_bernouilli_rademacher}
		Asymptotic matrix-MMSE when estimating $\bX^{\otimes 2}$ from $\bY = \sqrt{\nicefrac{\lambda}{n}} \, \bX^{\otimes 2} + \bZ$.
		We use a Bernouilli-Rademacher prior $P_S(0) = 1-\rho$, $P_S(\pm 1) = \rho/2$ with $\rho = 0.05$.
		\textit{Left:} generative prior $\bX = \bW \bS / \sqrt{p}$ with $\bS \iid P_S$.
		\textit{Right:} $\bX \iid P_S$.
	}
\end{figure}

To conclude this section we wish to briefly discuss the matrix case $r=2$, and point out similarities and differences with genuine tensors $r\geq 3$.
In the matrix case, \cite{aubin2019spiked} observe for \textit{a set} of centred priors and odd activations that the asymptotic matrix-MMSE is equal to its maximum $\rho_x^2$ for $\lambda < \lambda_c(\alpha)$ and decreases for $\lambda > \lambda_c(\alpha)$  \textit{while remaining continuous at $\lambda_c(\alpha)$}.
Again $\lambda_c(\alpha)$ decreases with increasing $\alpha$.
We give an example on the left panel of Figure~\ref{MMSE_bernouilli_rademacher}.
The continuity of the phase transition is an important qualitative difference with what we observe here for order-$3$ tensors.
Such continuity for Bayesian inference problems is known to go hand in hand with the optimality of the AMP algorithm and, as shown in 
\cite{aubin2019spiked}, matrix factorization with generative prior is no exception.
Because the continuity of the phase transition is observed for all the priors and activations used in \cite{aubin2019spiked}, it supports the claim that such model of structure makes estimation algorithmically easier.
In contrast, the persisting discontinuity of the transition for tensors of order $r\geq 3$ suggests that structure does not make the problem algorithmically easier here.
The observations of \cite{aubin2019spiked} should also be nuanced as it is not difficult to come up with a situation where the phase transition is discontinuous.
E.g., consider the spiked matrix model with generative prior $\bX = \varphi(\nicefrac{\bW \bS}{\sqrt{p}})$ for the \textit{odd activation} function $\varphi(x) = 0$ if $\vert x \vert \leq \epsilon$ and $\varphi(x)=\mathrm{sign}(x)$ otherwise, and the \textit{centered latent prior} $P_S = \cN(0,1)$.
Similarly to what is done in Section~\ref{section:alphatendstozero}, we can show that when $\alpha$ vanishes the asymptotic matrix-MMSE approaches the one of the spiked matrix model $\widetilde{Y}_{ij} = \sqrt{\frac{\lambda}{n}} \widetilde{X}_i\widetilde{X}_j + \widetilde{Z}_{ij}$ where $\widetilde{X}_1,\dots,\widetilde{X}_n \iid \varphi(\cN(0,1))$ are i.i.d. Bernouilli-Rademacher random variables.
We can make $\mathbb{P}(\widetilde{X}_i = 0)  = 1 - 2 \mathbb{P}(\cN(0,1) < -\epsilon) = 1-\rho$ as large as needed by increasing $\epsilon$ (then  $\mathbb{P}(\widetilde{X}_i = 1) = \mathbb{P}(\widetilde{X}_i = -1) =\rho/2$).
It is known that the asymptotic matrix-MMSE has a jump discontinuity for such prior when the probability of being $0$ is large enough, e.g., see the right panel in Figure~\ref{MMSE_bernouilli_rademacher}.
Therefore, when $\epsilon$ is large enough, the asymptotic matrix-MMSE of the original spiked matrix model with generative prior also has a jump discontinuity, at least for small $\alpha$.
An interesting question for future research is whether or not the discontinuity disappears when $\alpha$ is made large enough.
If so, it would further support the claim that such generative prior makes estimation algorithmically easier when the ratio $\alpha$ of signal-to-latent space dimensions is large enough.
If not, the existence of a jump discontinuity would then merely depend on the choice of activation function and not on the ratio of signal-to-latent space dimensions.
\subsection{Key ideas in the proofs of Theorems \ref{theorem:limit_mutual_information} and \ref{theorem:tensor_MMSE}}\label{subsec:key_ideas_proofs}
The proof of Theorem~\ref{theorem:limit_mutual_information} is based on the adaptive interpolation method \cite{barbier_adaptive_2019, Barbier_Macris_jphysA_2019} whose main difference with the canonical interpolation method \cite{guerra2002thermodynamic,Guerra-2003} is the increased flexibility given to the path followed by the interpolation between its two extremes.
The method has been developed separately for symmetric rank-one tensor problems where the spike has i.i.d. components \cite{barbier_adaptive_2019,Barbier_Macris_jphysA_2019}, and for one-layer GLMs whose input signal has again i.i.d. components \cite{barbierGLM}.
The problem studied in this contribution combines the two aforementioned models and our proof shows that the two interpolations combine well in a modular way.
This modular feature of the adaptive interpolation method has also been used for \textit{non-symmetric} order-three tensors \cite{barbier2017layered} and two-layer GLMs\cite{Gabrie_TwoLayerGLM_JSTAT_2019}.

An essential ingredient is an interpolating inference problem. 
Let $t\in [0,1]$ an interpolation parameter and $R(t)$ a smooth interpolation function that will be suitably \textit{adapted}.
We consider the pair of observations 
$(\bY^{(t)}, \widetilde{\bY}^{(t)} ) = \big(\frac{\sqrt{\lambda (1-t)}}{n} \bX^{\otimes 3} + \bZ, \sqrt{\frac{\lambda R(t)}{2}}\, \bX \;\, + \widetilde{\bZ}\big)$
where $\bX \triangleq \varphi(\nicefrac{\bW \bS}{\sqrt{p}})$
and the noise vector $\widetilde{\bZ}$ and the symmetric noise tensor $\bZ$ have entries $Z_{ijk}, \widetilde{Z}_{\ell} \iid \cN(0,1)$ for $1 \leq i \leq j \leq k \leq n$, $1 \leq \ell \leq n$.
At $t=0$ we recover the original problem while at $t=1$ we have a pure GLM with signal-to-noise ratio $\frac{\lambda R(1)}{2}$.
From the fundamental theorem of calculus, we have
$I(\bX;\bY \vert \bW)/n = I(\bX; \widetilde{\bY}^{(1)}\vert \bW)/n - \int_0^1 n^{-1}\big(\nicefrac{\partial I(\bX; \bY^{(t)}, \widetilde\bY^{(t)}\vert W)}{\partial t}\big)dt$.
The first term on the right-hand side is the normalized mutual information of a GLM given in the high-dimensional regime by the variational formula (proved in \cite{barbierGLM} with the adapative interpolation method):
\begin{equation*}
\adjustlimits{\inf}_{q_s \in [0,\rho_s]} {\sup}_{r_s \geq 0}\bigg\{\frac{I(S;\sqrt{r_s}\,S \! + \! Z) }{\alpha}\! + \! I\big(U; \sqrt{\frac{\lambda R(1)}{2}}\,\varphi(\sqrt{\rho_s - q_s} \, U \! + \! \sqrt{q_s} \, V) \! + \! \widetilde{Z} \,\big\vert\, V \big)
\! - \!  \frac{r_s(\rho_s - q_s)}{2\alpha}\bigg\}\;.
\end{equation*}
Comparing with \eqref{def_potential_psi} and \eqref{eq:main_I} we see that, if we set for the end point $R(1) = q_x^2$, we are missing the term $\frac{\lambda}{12}(\rho_x - q_x)^2 (\rho_x + 2 q_x)$.
In other words, \textit{and roughly speaking}, Theorem~\ref{theorem:limit_mutual_information} follows if we can show that $-n^{-1}\frac{\partial I(\bX; \bY^{(t)}, \widetilde\bY^{(t)}\vert W) }{\partial t} \approx \frac{\lambda}{12}(\rho_x+ q_x^2)(\rho_x + 2 q_x)$ for a \textit{suitable choice} of the interpolating function $R(t)$.
Remarkably, this condition essentially reduces to an \textit{ordinary differential equation} (ODE) for $R(t)$. The existence of a solution to this ODE is guaranteed by the standard Cauchy-Lipshitz theorem.
Obtaining the ODE is non-trivial and involves:
(i) remarkable identities stemming from Bayes' law; 
(ii) concentration theorems for the \textit{overlap} $Q = \frac{1}{n}\sum_{i=1}^n x_i X_i$ akin to a correlation between the ground truth $\bX$ and a vector $\bx$ distributed with respect to the posterior of the interpolating inference problem.

In order to prove Theorem~\ref{theorem:tensor_MMSE} we use the I-MMSE relation \eqref{eq:I-MMSE}.
This involves the computation of the \textit{derivative with respect to $\lambda$ of the variational formula} \eqref{eq:main_I} for the asymptotic mutual information.
The computation requires a careful application of an envelope theorem \cite{Milgrom_Envelope_Theorems} which eventually allows to show that, except for a countable set of $\lambda$'s, it is enough to evaluate the \textit{partial derivative with respect to $\lambda$ of the potential} \eqref{def_potential_psi} at the solution to the variational problem.
\section{Proof of the variational formula for the mutual information}\label{sec:proofthm1}
In this section we present the main steps of the proof of Theorem \ref{theorem:limit_mutual_information}. Intermediate results are found in the appendices.
\subsection{Adaptive path interpolation}
We introduce a ``time'' parameter $t \in [0,1]$.
The adaptive interpolation interpolates from the original model \eqref{eq:entries_Y} at $t=0$ to a GLM whose asymptotic mutual information is known \cite{barbierGLM}.
In between, we follow an interpolation path $R(\cdot,\epsilon): [0,1] \to (0,+\infty)$ which is a continuously differentiable function of $t$ parametrized by a ``small'' perturbation $\epsilon \in (0,+\infty)$ and is such that $R(0,\epsilon)=\epsilon$.
More precisely, for $t \in [0,1]$, the observations are:
\begin{align}\label{interpolation_model}
\begin{cases}
\bY^{(t)} \;\; = \frac{\sqrt{\lambda (1-t)}}{n} \bX^{\otimes 3} + \bZ\\
\widetilde{\bY}^{(t,\epsilon)} = \:\sqrt{\frac{\lambda R(t,\epsilon)}{2}}\, \bX \;\, + \widetilde{\bZ}
\end{cases}
\end{align}
where $\bX \triangleq \varphi(\nicefrac{\bW \bS}{\sqrt{p}})$.
The noise vector $\widetilde{\bZ} \in \R^n$ has entries $\widetilde{Z}_1,\dots,\widetilde{Z}_n \iid \cN(0,1)$, while the symmetric noise tensor $\bZ \in (\R^n)^{\otimes 3}$ has entries $\bZ_{\bi} \iid \cN(0,1)$ for $\bi \in \cI \triangleq \{(i_1, i_2, i_3) \in [n]: i_1 \leq i_2 \leq i_3\}$.

Before diving further, we introduce some important quantities and notations. We denote $i_n(t,\epsilon)$ the normalized mutual information between $\bX$ and $(\bY^{(t)},\widetilde{\bY}^{(t,\epsilon)})$ given $\bW$, that is:
\begin{equation}\label{definition_i_n}
	i_n(t,\epsilon) \triangleq \frac{1}{n}I(\bX ; \bY^{(t)},\widetilde{\bY}^{(t,\epsilon)} \vert \bW)
	=\frac{1}{n}I(\bS ; \bY^{(t)},\widetilde{\bY}^{(t,\epsilon)} \vert \bW) \;.
\end{equation}
The last equality holds because $\bX$ is a deterministic function of $\bS$ when $\bW$ is known. Set $dP_s(\bs) = \prod_{i=1}^{p} dP_s(s_i)$ for the prior
distribution of $\bS$.
The {\it usual} Bayesian posterior distribution of $\bS$ given $(\bY^{(t)},\widetilde{\bY}^{(t,\epsilon)}, \bW)$ reads:
\begin{equation}\label{posterior_interpolation}
dP(\bs ; \bY^{(t)},\widetilde{\bY}^{(t,\epsilon)}, \bW) =
\frac{1}{\cZ_{t,\epsilon}(\bY^{(t)},\widetilde{\bY}^{(t,\epsilon)}, \bW)}
dP_s(\bs) \, e^{-\cH_{t,\epsilon}(\bs \,;\, \bY^{(t)},\widetilde{\bY}^{(t,\epsilon)}, \bW)}\;,
\end{equation}
where the normalization factor $\cZ_{t,\epsilon}(\bY^{(t)},\widetilde{\bY}^{(t,\epsilon)}, \bW)$ is simply:
\begin{equation}\label{partition_function}
\cZ_{t,\epsilon}(\bY^{(t)},\widetilde{\bY}^{(t,\epsilon)}, \bW) \triangleq \int dP_s(\bs) \, e^{-\cH_{t,\epsilon}(\bs \,;\, \bY^{(t)},\widetilde{\bY}^{(t,\epsilon)}, \bW)} \;.
\end{equation}
and 
\begin{multline}\label{interpolating_hamiltonian}
\cH_{t,\epsilon}(\bs ; \bY^{(t)}, \widetilde{\bY}^{(t,\epsilon)}, \bW)
\triangleq
\sum_{\bi \in \cI}\biggl(
\frac{\lambda(1-t)}{2n^2} x_{i_1}^2 x_{i_2}^2 x_{i_3}^2 - \frac{\sqrt{\lambda(1-t)}}{n}\, Y_{\bi}^{(t)} x_{i_1} x_{i_2} x_{i_3}\bigg)\\
+
\sum_{j=1}^{n} \biggl(\frac{\lambda R(t,\epsilon)}{4} x_j^2 - \sqrt{\frac{\lambda R(t,\epsilon)}{2}}\,\widetilde{Y}_j^{(t,\epsilon)} x_j\biggr) \,,
\end{multline}
with $x_1,\dots,x_n$ the entries of $\bx \triangleq \varphi(\nicefrac{\bW \bs}{\sqrt{p}})$.
This dependence on $\bs$ must be kept in mind each time we use the notation $\bx$. It is common to adopt the statistical mechanics interpretation and call 
\eqref{interpolating_hamiltonian} a Hamiltonian, \eqref{partition_function} the partition function and \eqref{posterior_interpolation} the Gibbs distribution.

To deal with future computations, it is useful to introduce the angular brackets $\langle - \rangle_{t,\epsilon}$ (also called Gibbs brackets) which denote an expectation with respect to the posterior distribution \eqref{posterior_interpolation}.
That is, for a generic function $g: \R^p \to \R$, we have: 
\begin{equation}
\langle g(\bs) \rangle_{t,\epsilon} \triangleq \int \!g(\bs)\,dP(\bs ; \bY^{(t)},\widetilde{\bY}^{(t,\epsilon)}, \bW)\:.
\end{equation}
Finally, we define the so-called average free entropy:
\begin{equation}\label{interpolating_free_entropy}
f_n(t,\epsilon) \triangleq \frac1n \E \ln \cZ_{t,\epsilon}(\bY^{(t)},\widetilde{\bY}^{(t,\epsilon)}, \bW)\;.
\end{equation}
This is equal to the mutual information $i_n(t,\epsilon)$ up to some additive term (see formula \eqref{link_i_n_f_n} in Lemma~\ref{lemma:link_i_n_f_n} in Appendix~\ref{app:establishing_sum_rule}). It is often easier to work directly with $f_n(t,\epsilon)$ instead of $i_n(t,\epsilon)$. 

We now focus on the mutual information \eqref{definition_i_n} at both extremes of the interpolation path.
Letting $t=0$ in \eqref{interpolation_model}, we see that the observation $\bY^{(0)}$ is exactly \eqref{eq:entries_Y},
while $\widetilde{\bY}^{(0,\epsilon)} = \sqrt{\frac{\lambda \epsilon}{2}} \bX + \widetilde{\bZ}$.
This latter channel induces a perturbation to the normalized mutual information of the former channel of the order of $\epsilon$ (see Lemma~\eqref{lemma:link_i_n_f_n} in Appendix~\ref{app:establishing_sum_rule} for the proof), that is:
\begin{equation}\label{i_n_t=0}
i_n(0,\epsilon)
\triangleq \frac{1}{n}I(\bX ; \bY^{(0)},\widetilde{\bY}^{(0,\epsilon)} \vert \bW)
=\frac{I(\bX ; \bY \vert \bW)}{n} + \cO(\epsilon) \;,
\end{equation}
where $\vert \cO(\epsilon) \vert \leq C \epsilon$.
At $t=1$ the observation $\bY^{(1)}$ is pure noise, while the normalized mutual information between $\bS$ and $\widetilde{\bY}^{(1,\epsilon)} = \sqrt{\nicefrac{\lambda R(1,\epsilon)}{2}} \, \varphi(\nicefrac{\bW \bS}{\sqrt{p}}) + \widetilde{\bZ}$ is given by a variational formula in the high-dimensional regime $\nicefrac{n}{p} \to \alpha$ \cite{barbierGLM}.
Let $S \sim P_S$ and $U,V,Z,\widetilde{Z} \sim \cN(0,1)$ independent scalar random variables. Define the potential function $\widetilde{\psi}_\alpha:  [0,+\infty)^2 \times [0,\rho_s]$:
\begin{multline}
\widetilde{\psi}_\alpha(r, r_s, q_s) \triangleq
I(S;\sqrt{r_s}\,S + Z) + \alpha I\big(U; \sqrt{r}\,\varphi(\sqrt{\rho_s - q_s} \, U + \sqrt{q_s} \, V) + \widetilde{Z} \,\big\vert\, V \big) - \frac{r_s(\rho_s - q_s)}{2} \;.
\end{multline}
By \cite[Corollary 1]{barbierGLM}, we have:
\begin{equation}\label{i_n_t=1}
i_n(1,\epsilon)
= \frac{1}{n}I(\bX ; \widetilde{\bY}^{(1,\epsilon)} \vert \bW)
= \smallO_n(1) + \frac{1}{\alpha} \,\adjustlimits{\inf}_{q_s \in [0,\rho_s]} {\sup}_{r_s \geq 0}\;
\widetilde{\psi}_\alpha\bigg(\frac{\lambda R(1,\epsilon)}{2} , r_s, q_s\bigg) \;.
\end{equation}
Combining \eqref{i_n_t=0}, \eqref{i_n_t=1} and the fundamental theorem of calculus $i_n(0,\epsilon)=i_n(1,\epsilon)-\int_{0}^{1}i_n^{\prime}(t,\epsilon) dt$, where $i_n^{\prime}(\cdot,\epsilon)$ is the derivative of $i_n(\cdot,\epsilon)$ w.r.t.\ its first argument, we obtain the sum-rule of the adaptive interpolation.
\begin{proposition}[Sum-rule]\label{prop:sum_rule}
Suppose that \ref{hyp:S_bounded_support} and \ref{hyp:varphi} hold, and that $R'(t,\epsilon)$ is uniformly bounded in $(t,\epsilon) \in [0,1] \times [0,+\infty)$ where $R'(\cdot,\epsilon)$ denotes the derivative of $R(\cdot,\epsilon)$ with respect to its first argument.
Define the scalar overlap
$$
Q \triangleq \frac{1}{n} \sum\limits_{i=1}^{n} \varphi\big(\big[\nicefrac{\bW \bs}{\sqrt{p}}\big]_i \,\big) \varphi\big(\big[\nicefrac{\bW \bS}{\sqrt{p}}\big]_i \,\big)
= \frac{1}{n} \sum_{i=1}^{n} x_i X_i \;.
$$
Then:
\begin{multline}\label{sum_rule}
\frac{I(\bX ; \bY \vert \bW)}{n}
= \cO(\epsilon) + \smallO_n(1) 
 + \frac{1}{\alpha}\,\adjustlimits{\inf}_{q_s \in [0,\rho_s]} {\sup}_{r_s \geq 0}\;
\widetilde{\psi}_\alpha\bigg(\frac{\lambda R(1,\epsilon)}{2} , r_s, q_s\bigg)\\
-\frac{\lambda}{12} \int_0^1 \big(\E\,\langle Q^3 \rangle_{t,\epsilon} - \rho_x^3\big)dt
- \frac{\lambda}{4} \int_0^1 R'(t,\epsilon)\big(\rho_x - \E\,\langle Q \rangle_{t,\epsilon}\big)dt\;,
\end{multline}
where $\smallO_n(1)$ and $\mathcal{O}(\epsilon)$ are independent of $\epsilon$ and $n$, respectively.
\begin{IEEEproof}
See Lemma~\ref{lemma:formula_derivative_free_entropy} in Appendix~\ref{app:establishing_sum_rule} for the computation 
of the derivative $i_n^{\prime}(t,\epsilon)$.
\end{IEEEproof}
\end{proposition}

The sum rule of Proposition~\ref{prop:sum_rule} is valid for the general class of differentiable interpolating paths. By choosing two appropriate interpolation paths we can prove matching upper and lower bounds on the asymptotic normalized mutual information. This is discussed in the next two paragraphs.

\subsection{Upper bound on the asymptotic normalized mutual information}\label{subsec:upper_bound}
\begin{proposition}\label{prop:upperbound_mutual_info}
Suppose that \ref{hyp:S_bounded_support} and \ref{hyp:varphi} hold. Then:
\begin{equation}
\limsup_{n \to +\infty} \frac{I(\bX ; \bY \vert \bW)}{n} \leq \inf_{q_x \in [0,\rho_x]} \adjustlimits{\inf}_{q_s \in [0,\rho_s]} {\sup}_{r_s \geq 0}\;\psi_{\lambda,\alpha}\big(q_x, q_s, r_s\big)\;.
\end{equation}
\end{proposition}
\begin{IEEEproof}
Fix $\epsilon > 0$ and pick the linear interpolation path $R(t,\epsilon) = \epsilon + t q^2$ where $q \in [0,\rho_x]$.
Then the sum-rule \eqref{sum_rule} in Proposition~\ref{prop:sum_rule} reads:
\begin{align}
& \frac{I(\bX ; \bY \vert \bW)}{n}
= \cO(\epsilon) + \smallO_n(1) 
+ \frac{1}{\alpha} \adjustlimits{\inf}_{q_s \in [0,\rho_s]} {\sup}_{r_s \geq 0}\;
\widetilde{\psi}_\alpha\bigg(\frac{\lambda \epsilon}{2} + \frac{\lambda q^2}{2}, r_s, q_s\bigg)
+\frac{\lambda}{12} \rho_x^3
- \frac{\lambda}{4} q^2 \rho_x
\nonumber\\&
\qquad\qquad\qquad\qquad\qquad 
-\frac{\lambda}{12} \int_0^1 \bigg(\E\,\langle Q^3 \rangle_{t,\epsilon} - \E\bigg[\bigg\Vert \frac{\langle \bx \rangle_{t,\epsilon}}{\sqrt{n}} \bigg\Vert^4\langle Q \rangle_{t,\epsilon}\bigg]\bigg)dt
\nonumber\\&
\qquad\qquad\qquad\qquad\qquad 
-\frac{\lambda}{12}\int_0^1 \bigg(\E\bigg[\bigg\Vert \frac{\langle \bx \rangle_{t,\epsilon}}{\sqrt{n}} \bigg\Vert^4\langle Q \rangle_{t,\epsilon}\bigg]- 3 q^2 \, \E\langle Q \rangle_{t,\epsilon} \bigg) dt\;. \label{sum_rule_linear_path}
\end{align}
In this last identity, we "artificially" added and subtracted the term $\E\big[\big\Vert \frac{\langle \bx \rangle_{t,\epsilon}}{\sqrt{n}} \big\Vert^4\langle Q \rangle_{t,\epsilon}\big]$ for reasons that will appear immediately.
By the Nishimori identity\footnote{
	In our setting, the  Nishimori identity states that ${\E\langle g(\bs,\bS)\rangle_{t,\epsilon} = \E\langle g(\bs,\bs') \rangle_{t,\epsilon} = \E\langle g(\bS,\bs) \rangle_{t,\epsilon}}$ where $\bs,\bs'$ are two samples drawn independently from the posterior distribution of $\bS$ given $(\bY^{(t)}, \widetilde{\bY}^{(t,\epsilon)}, \bW)$.
	It is a direct consequence of Bayes' theorem.
	Here $g$ can also explicitly depend on $\bY^{(t)}, \widetilde{\bY}^{(t,\epsilon)}, \bW$ so the identity holds for $\bX = \varphi(\frac{\bW \bS}{\sqrt{p}}), \bx = \varphi(\frac{\bW \bs}{\sqrt{p}}), \bx' = \varphi(\frac{\bW \bs'}{\sqrt{p}})$ too.},
we have
\begin{equation}
\E\bigg[\bigg\Vert \frac{\langle \bx \rangle_{t,\epsilon}}{\sqrt{n}} \bigg\Vert^4\langle Q \rangle_{t,\epsilon}\bigg]
= \E\,\bigg\Vert \frac{\langle \bx \rangle_{t,\epsilon}}{\sqrt{n}} \bigg\Vert^6, 
\quad \E\langle Q \rangle_{t,\epsilon} = \E\,\bigg\Vert \frac{\langle \bx \rangle_{t,\epsilon}}{\sqrt{n}} \bigg\Vert^2 \;,
\end{equation}
and, by convexity of $x \mapsto x^3$ on $[0,+\infty)$, we have $\forall a,b \geq 0: a^3 - 3b^2 a \geq -2 b^3$.
Hence the integrand of the last integral on the right-hand side of \eqref{sum_rule_linear_path} satisfies:
\begin{equation}\label{lowerbound_convexity_x^3}
\E\bigg[\bigg\Vert \frac{\langle \bx \rangle_{t,\epsilon}}{\sqrt{n}} \bigg\Vert^4\langle Q \rangle_{t,\epsilon}\bigg]- 3 q^2 \, \E\langle Q \rangle_{t,\epsilon}
= \E\bigg[\bigg\Vert \frac{\langle \bx \rangle_{t,\epsilon}}{\sqrt{n}} \bigg\Vert^6 - 3 q^2 \bigg\Vert \frac{\langle \bx \rangle_{t,\epsilon}}{\sqrt{n}} \bigg\Vert^2 \bigg]
\geq - 2 q^3 \;.
\end{equation}
Besides, by Lemma~\ref{lemma:properties_functions} in Appendix \ref{app:useful_lemmas}, the function $r \mapsto \adjustlimits{\inf}_{q_s \in [0,\rho_s]} {\sup}_{r_s \geq 0}\; \widetilde{\psi}_\alpha(q, r_s, q_s)$ is nondecreasing and $(\alpha/2) \Vert \varphi \Vert_{\infty}^2$-Lipschitz. Thus:
\begin{equation}\label{upperbound_psi_epsilon}
\adjustlimits{\inf}_{q_s \in [0,\rho_s]} {\sup}_{r_s \geq 0}\; \widetilde{\psi}_\alpha\bigg(\frac{\lambda \epsilon}{2} + \frac{\lambda q^2}{2}, r_s, q_s\bigg) \leq \frac{\lambda \alpha \Vert \varphi \Vert_{\infty}^2}{4} \epsilon + \adjustlimits{\inf}_{q_s \in [0,\rho_s]} {\sup}_{r_s \geq 0}\; \widetilde{\psi}_\alpha\bigg(\frac{\lambda q^2}{2}, r_s, q_s\bigg) \;.
\end{equation}
Therefore, making use of \eqref{lowerbound_convexity_x^3} and \eqref{upperbound_psi_epsilon} to upper bound \eqref{sum_rule_linear_path} yields:
\begin{align}
\frac{I(\bX ; \bY \vert \bW)}{n}
&\leq \cO(\epsilon) + \smallO_n(1) 
+ \adjustlimits{\inf}_{q_s \in [0,\rho_s]} {\sup}_{r_s \geq 0}\;
\frac{1}{\alpha}\widetilde{\psi}_\alpha\bigg(\frac{\lambda q^2}{2}, r_s, q_s\bigg)
+\frac{\lambda}{12} \rho_x^3
- \frac{\lambda}{4} q^2 \rho_x
+ \frac{\lambda}{6} q^3\nonumber\\
&\qquad\qquad\qquad\qquad\qquad\quad
-\frac{\lambda}{12} \int_0^1 \bigg(\E\,\langle Q^3 \rangle_{t,\epsilon} - \E\bigg[\bigg\Vert \frac{\langle \bx \rangle_{t,\epsilon}}{\sqrt{n}} \bigg\Vert^4\langle Q \rangle_{t,\epsilon}\bigg]\bigg)dt\nonumber\\
&= \cO(\epsilon) + \smallO_n(1) 
+ \adjustlimits{\inf}_{q_s \in [0,\rho_s]} {\sup}_{r_s \geq 0}\;
\psi_{\lambda,\alpha}\big(q, q_s, r_s\big)\nonumber\\
&\qquad\qquad\qquad\qquad\qquad\quad
-\frac{\lambda}{12} \int_0^1 \bigg(\E\,\langle Q^3 \rangle_{t,\epsilon} - \E\bigg[\bigg\Vert \frac{\langle \bx \rangle_{t,\epsilon}}{\sqrt{n}} \bigg\Vert^4\langle Q \rangle_{t,\epsilon}\bigg]\bigg)dt\;. \label{sum_rule_upper_bound}
\end{align}
where the last equality follows from the trivial identity:
\begin{equation}\label{link_psi_tilde_and_psi}
\psi_{\lambda,\alpha}\big(q, q_s, r_s\big)
= \frac{1}{\alpha}\widetilde{\psi}_\alpha\bigg(\frac{\lambda q^2}{2}, r_s, q_s\bigg)
+\frac{\lambda}{12} \rho_x^3
- \frac{\lambda}{4} q^2 \rho_x
+ \frac{\lambda}{6} q^3 \;.
\end{equation}
It now remains to get rid of the integral on the right-hand side of \eqref{sum_rule_upper_bound}. The integrand satisfies:
\begin{multline}\label{inequality_remainder}
\bigg\vert \E\,\langle Q^3 \rangle_{t,\epsilon} - \E\bigg[\bigg\Vert \frac{\langle \bx \rangle_{t,\epsilon}}{\sqrt{n}} \bigg\Vert^4\langle Q \rangle_{t,\epsilon}\bigg]\bigg\vert
= \bigg\vert \E\,\bigg\langle Q \bigg(Q + \bigg\Vert \frac{\langle \bx \rangle_{t,\epsilon}}{\sqrt{n}} \bigg\Vert^2\bigg) 
 \bigg(Q - \bigg\Vert \frac{\langle \bx \rangle_{t,\epsilon}}{\sqrt{n}} \bigg\Vert^2\bigg)\bigg\rangle_{\!\! t,\epsilon} \bigg\vert\\
\leq 2 \Vert \varphi \Vert_{\infty}^4 \E\,\bigg\langle \bigg\vert Q - \bigg\Vert \frac{\langle \bx \rangle_{t,\epsilon}}{\sqrt{n}} \bigg\Vert^2\bigg\vert \bigg\rangle_{\!\! t,\epsilon}
\leq 2 \Vert \varphi \Vert_{\infty}^4 \sqrt{\E\,\bigg\langle \!\bigg( Q - \bigg\Vert \frac{\langle \bx \rangle_{t,\epsilon}}{\sqrt{n}}\bigg\Vert^2 \bigg)^{\!\! 2} \bigg\rangle_{\!\! t,\epsilon}} \;.
\end{multline}
We see that if the overlap $Q \triangleq \nicefrac{\bx^{\sT} \bX}{n}$ would concentrate on $\nicefrac{\langle \bx \rangle_{t, \epsilon}^{\sT} \langle \bx \rangle_{t, \epsilon}}{n}$ then the remaining integral in \eqref{sum_rule_upper_bound} would be negligible.
However, proving such a concentration property is only holds when we average on a well-chosen set of ``perturbations'' $\epsilon$.
In essence, the average over $\epsilon$ smoothens the phase transitions that might appear for particular choices of $\epsilon$ when $n$ goes to infinity.\\
We now take $\epsilon \in [s_n, 2s_n]$ where $s_n \triangleq n^{-\eta}$, $\eta > 0$, and 
integrate w.r.t.\ $\epsilon$ on both sides of \eqref{sum_rule_upper_bound}:
\begin{align}\label{sumrule_lowerbound_integration_epsilon}
&\frac{I(\bX ; \bY \vert \bW)}{n} = \int_{s_n}^{2s_n}\frac{I(\bX ; \bY \vert \bW)}{n} \frac{d\epsilon}{s_n}\nonumber\\
&\leq \smallO_n(1) + \adjustlimits{\inf}_{q_s \in [0,\rho_s]} {\sup}_{r_s \geq 0}\;\psi_{\lambda,\alpha}\big(q, q_s, r_s\big)
-\frac{\lambda}{12} \int_0^1 dt \int_{s_n}^{2s_n} \bigg(\E\,\langle Q^3 \rangle_{t,\epsilon} - \E\bigg[\bigg\Vert \frac{\langle \bx \rangle_{t,\epsilon}}{\sqrt{n}} \bigg\Vert^4\langle Q \rangle_{t,\epsilon}\bigg]\bigg)\frac{d\epsilon}{s_n}\nonumber\\
&\leq \smallO_n(1) + \adjustlimits{\inf}_{q_s \in [0,\rho_s]} {\sup}_{r_s \geq 0}\;\psi_{\lambda,\alpha}(q, q_s, r_s)
+ \frac{\lambda \Vert \varphi \Vert_{\infty}^4}{6} \! \int_0^1 \! dt \int_{s_n}^{2s_n} \!\sqrt{\E\,\bigg\langle \!\bigg( Q - \bigg\Vert \frac{\langle \bx \rangle_{t,\epsilon}}{\sqrt{n}}\bigg\Vert^2 \bigg)^{\!\! 2} \bigg\rangle_{\!\! t,\epsilon}}\,\frac{d\epsilon}{s_n}\;.
\end{align}
Since $R(t,\cdot)$ is a $\mathcal{C}^1$-diffeomorphism from $[s_n,2s_n]$ to its image $R(t, [s_n,2s_n]) \subseteq [s_n, 2s_n + \rho_x^2]$, we  make the change of variables $\epsilon \to R \equiv R(t,\epsilon)$ and obtain (using Cauchy-Schwarz for the first inequality)
for all $t \in [0,1]$:
\begin{align}
\int_{s_n}^{2s_n} \sqrt{\E\,\bigg\langle \!\bigg( Q - \bigg\Vert \frac{\langle \bx \rangle_{t,\epsilon}}{\sqrt{n}}\bigg\Vert^2 \bigg)^{\!\! 2} \bigg\rangle_{\!\! t,\epsilon}}\,\frac{d\epsilon}{s_n}
&\leq \sqrt{\int_{s_n}^{2s_n} \E\,\bigg\langle \!\bigg( Q - \bigg\Vert \frac{\langle \bx \rangle_{t,\epsilon}}{\sqrt{n}}\bigg\Vert^2 \bigg)^{\!\! 2} \bigg\rangle_{\!\! t,\epsilon}\,\frac{d\epsilon}{s_n}}\nonumber\\
&=\sqrt{\int_{R(t, [s_n,2s_n])}
\E\,\bigg\langle \!\bigg( Q - \bigg\Vert \frac{\langle \bx \rangle_{t,R}}{\sqrt{n}}\bigg\Vert^2 \bigg)^{\!\! 2} \bigg\rangle_{\!\! t,R}\, \frac{dR}{s_n}}\nonumber\\
&\leq \sqrt{\int_{s_n}^{2s_n + \rho_x^2}
\E\,\bigg\langle \bigg( Q - \bigg\Vert \frac{\langle \bx \rangle_{t,R}}{\sqrt{n}} \bigg\Vert^2\bigg)^{\!\! 2} \bigg\rangle_{\!\! t,R} \,\frac{dR}{s_n}}\;. \label{lowerbound_change_variable}
\end{align}
By Proposition~\ref{prop:concentration_L_on_<L>} in Appendix \ref{app:concentration_overlap} and the inequality \eqref{lowerbound_change_variable}, we get (remember that $s_n \triangleq n^{-\eta}$):
\begin{multline}
\frac{\lambda \Vert \varphi \Vert_{\infty}^4}{6} \int_{s_n}^{2s_n} \sqrt{\E\,\bigg\langle \!\bigg( Q - \bigg\Vert \frac{\langle \bx \rangle_{t,\epsilon}}{\sqrt{n}}\bigg\Vert^2 \bigg)^{\!\! 2} \bigg\rangle_{\!\! t,\epsilon}}\,\frac{d\epsilon}{s_n}\\
\leq \frac{\sqrt{\lambda} \Vert \varphi \Vert_{\infty}^4}{3} \sqrt{\frac{\Vert \varphi \Vert_{\infty}^3}{s_n}\sqrt{\frac{\lambda (s_n + \rho_x^2)}{2n}}}
= \frac{\lambda^{\frac{3}{4}} \Vert \varphi \Vert_{\infty}^{\nicefrac{11}{2}}}{3}
\bigg(\frac{s_n + \rho_x^2}{2}\bigg)^{\frac{1}{4}}
\frac{1}{n^{\frac{1-2\eta}{4}}} \;.
\end{multline}
Therefore, we see that the remainder on the right-hand side of \eqref{sumrule_lowerbound_integration_epsilon} vanishes as $\cO(n^{-\nicefrac{1}{6}})$ if we pick $\eta = \nicefrac{1}{6}$.
Passing to the limit superior on both sides of the inequality \eqref{sumrule_lowerbound_integration_epsilon} then yields:
$$
\limsup_{n \to +\infty} \frac{I(\bX ; \bY \vert \bW)}{n} \leq \adjustlimits{\inf}_{q_s \in [0,\rho_s]} {\sup}_{r_s \geq 0}\;\psi_{\lambda,\alpha}(q, q_s, r_s) \;.
$$
This inequality is true for all $q \in [0, \rho_x]$ and Proposition~\ref{prop:upperbound_mutual_info} follows directly.
\end{IEEEproof}
\subsection{Matching lower bound on the asymptotic normalized mutual information}
We now prove a matching lower bound by considering a different choice for $R(\cdot,\epsilon)$ in the sum-rule \eqref{sum_rule}.
$R(\cdot,\epsilon)$ will be the solution to a first-order ordinary differential equations (ODE).
We first describe this ODE and give the derivation of the lower bound.

\subsubsection{An ordinary differential equation}

For $t \in [0,1]$ and $R \in [0, +\infty)$, consider the problem of estimating $\bS$ from the observations:
\begin{align}\label{interpolation_model_R}
\begin{cases}
\bY^{(t)} \;\; = \frac{\sqrt{\lambda (1-t)}}{n} \bX^{\otimes 3} + \bZ\\
\widetilde{\bY}^{(t,R)} = \:\sqrt{\frac{\lambda R}{2}}\, \bX \;\, + \widetilde{\bZ}
\end{cases};
\end{align}
where $\bX \triangleq \varphi(\nicefrac{\bW \bS}{\sqrt{p}})$, $S_1,\dots,S_p \iid P_S$.
The noise vector $\widetilde{\bZ} \in \R^n$ has entries $\widetilde{Z}_1,\dots,\widetilde{Z}_n \iid \cN(0,1)$, while the symmetric noise tensor $\bZ \in (\R^n)^{\otimes 3}$ has entries $\bZ_{\bi} \iid \cN(0,1)$ for $\bi \in \cI \triangleq \{(i_1, i_2, i_3) \in [n]: i_1 \leq i_2 \leq i_3\}$.
The posterior distribution of $\bS$ given $(\bY^{(t)},\widetilde{\bY}^{(t,R)}, \bW)$ is:
\begin{equation}\label{posterior_H_t_R}
dP(\bs ; \bY^{(t)},\widetilde{\bY}^{(t,R)}, \bW) =
\frac{1}{\cZ_{t,R}(\bY^{(t)},\widetilde{\bY}^{(t,R)}, \bW)}dP_{S}(\bs) \, e^{-\cH_{t,R}(\bs ; \bY^{(t)},\widetilde{\bY}^{(t,R)}, \bW)} \;.
\end{equation}
where $\cZ_{t,R}(\bY^{(t)},\widetilde{\bY}^{(t,R)}, \bW) = \int  dP_{S}(\bs) \, e^{-\cH_{t,R}(\bs ; \bY^{(t)},\widetilde{\bY}^{(t,R)}, \bW)}$ and 
\begin{multline}\label{hamiltonian_model_R}
\cH_{t,R}(\bs ; \bY^{(t)}, \widetilde{\bY}^{(t,R)}, \bW)
\triangleq
\sum_{\bi \in \cI}
\frac{\lambda(1-t)}{2n^2} x_{i_1}^2 x_{i_2}^2 x_{i_3}^2 - \frac{\sqrt{\lambda(1-t)}}{n}\, Y_{\bi}^{(t)} x_{i_1} x_{i_2} x_{i_3}\\
+
\sum_{j=1}^{n} \frac{\lambda R}{4} x_j^2 - \sqrt{\frac{\lambda R}{2}}\,\widetilde{Y}_j^{(t,R)} x_j \;.
\end{multline}
Again, \eqref{hamiltonian_model_R} has the interpretation of a Hamiltonian and \eqref{posterior_H_t_R} a Gibbs distribution.
The Gibbs bracket notation $\langle - \rangle_{t,R}$ denotes the expectation with respect to this last posterior.
Finally, we define the following function used to formulate the ODE satisfied by the interpolation path:
\begin{equation}
G(t, R) = (\E \langle Q \rangle_{t,R})^2\;.
\end{equation}
\begin{lemma}\label{lemma:ode_for_interpolation_path}
Assume $\varphi:\R \to \R$ is continuous and bounded.
For all $\epsilon \in [0,+\infty)$, there exists a unique global solution $R(\cdot,\epsilon): [0,1] \to [0,+\infty)$ to the first-order ODE:
\begin{equation}\label{eq:ODE}
\forall \,t \in [0,1]: \frac{d g(t)}{dt}=G(t,g(t)) \,, \quad g(0)=\epsilon\,.
\end{equation}
This solution is continuously differentiable with bounded derivative (w.r.t.\ $t$) $R'(\cdot, \epsilon)$ and, for any $\delta>0$, $R'([0,1],\epsilon) \subseteq [0, (\rho_x+\delta)^2]$ for $n$ large enough independent of $\epsilon$.
Besides, $\forall \,t \in [0,1]$, $R(t,\cdot)$ is a $\mathcal{C}^1$-diffeomorphism from $[0, +\infty)$ into its image whose derivative w.r.t.\ $\epsilon$ is greater than or equal to one, i.e.,
	\begin{equation}\label{sol_ode_derivative_gretaer_one}
	\forall \,\epsilon \in [0, +\infty): \frac{\partial R}{\partial \epsilon}\Big\vert_{t,\epsilon} \geq 1 \,.
	\end{equation}
\end{lemma}
\begin{remark}
This lemma guarantees a unique global solution $R_n(t,\epsilon)$ for each finite $n$. Slightly abusively we do not indicate the $n$-dependence and simply write $R(t, \epsilon)$ for the solution.
\end{remark}
\begin{IEEEproof}
The function $G: (t, R) \in [0,1] \times [0,+\infty) \mapsto G(t,R)$ is continuous in $t$ and uniformly Lipschitz continuous in $R$ (meaning the Lipschitz constant is independent of $t$).
The later is readily checked by computing the derivative of $G(t,\cdot)$ and showing it is uniformly bounded in $(t, R)$:
\begin{align}
\frac{\partial G}{\partial R}\bigg\vert_{t,R}
&= \frac{\lambda \E\langle Q \rangle_{t, R}}{n} \sum_{i,j=1}^{n} \E[(\langle x_i x_j \rangle_{t,R} - \langle x_i \rangle_{t,R} \langle x_j \rangle_{t,R})^2]
\in [0, 4 \lambda \Vert \varphi \Vert_{\infty}^6 n ] \;.\label{derivative_G_wrt_R}
\end{align}
Therefore, by the Cauchy-Lipschitz theorem, for all $\epsilon \geq 0$ there exists a unique solution $R(\cdot, \epsilon): [0,\gamma] \to [0,+\infty)$ to the initial value problem \eqref{eq:ODE}.
Here $\gamma \in [0,1]$ is such that $[0,\gamma]$ is the maximal interval of existence of the solution.
By the Cauchy-Schwarz inequality and Nishimory identity, we have:
$$
\E \langle Q \rangle_{t, R}
\leq \frac{\E \langle \Vert \bx \Vert \Vert \bX \Vert \rangle_{t,R}}{n}
\leq \frac{1}{n} \sqrt{\E \langle \Vert \bx \Vert^2 \rangle_{t,R} \, \E \Vert \bX \Vert^2 }
= \frac{\E \Vert \bX \Vert^2 }{n}
= \E\bigg[\varphi\bigg(\frac{\bW_{1,\cdot}\,\bS}{\sqrt{p}}\bigg)^{\! 2}\bigg]
\xrightarrow[n \to +\infty]{} \rho_x \;.
$$
See \cite[Lemma 3 of Supplementary material]{Gabrie_TwoLayerGLM_JSTAT_2019} for a proof of the later limit.
Besides, by Nishimori identity, $\E \langle Q \rangle_{t, R} = n^{-1}\E \Vert \langle \bx \rangle_{t,R}\Vert^2$ is nonnegative.
Hence, for any $\delta > 0$, $G$ has its image in $[0, (\rho_x + \delta)^2]$ and $R([0, \gamma],\epsilon) \subseteq [\epsilon, \epsilon + \gamma (\rho_x + \delta)^2]$ as long as $n$ is large enough. It implies that $\gamma = 1$ (the solution never leaves the domain of definition of $G$).
	
Each initial condition $\epsilon \in [0, +\infty)$ is tied to a unique solution $R(\cdot,\epsilon)$. This implies that the function $\epsilon \mapsto R(t,\epsilon)$ is injective. Its derivative is given by Liouville's formula \cite{hartman1982ordinary}
\begin{equation}\label{liouville_formula_ode_lowerbound}
\frac{\partial R}{\partial \epsilon}\bigg\vert_{t,\epsilon}
	= \exp \biggl\{\int_0^t ds \, \frac{\partial G}{\partial R}\bigg\vert_{s,R(s,\epsilon)}\biggr\}
\end{equation}
and is greater than, or equal to one, by positivity of $\frac{\partial G}{\partial R}$ -- see \eqref{derivative_G_wrt_R} above --.
The fact that this partial derivative is bounded away from $0$ uniformly in $\epsilon$ implies by the inverse function theorem that the injective function  $\epsilon \mapsto R(t,\epsilon)$ is a $\mathcal{C}^1$-diffeomorphism from $[0, +\infty)$ onto its image.
\end{IEEEproof}

\subsubsection{Derivation of the lower bound}
\begin{proposition}\label{prop:lowerbound_mutual_info}
Suppose that \ref{hyp:S_bounded_support} and \ref{hyp:varphi} hold. Then:
	\begin{equation}
	\limsup_{n \to +\infty} \frac{I(\bX ; \bY \vert \bW)}{n}
	\geq \mathop{\vphantom{p}\inf}_{q_x \in [0,\rho_x]} \adjustlimits{\inf}_{q_s \in [0,\rho_s]} {\sup}_{r_s \geq 0}\;\psi_{\lambda,\alpha}(q_x, q_s, r_s)\;.
	\end{equation}
\end{proposition}
\begin{IEEEproof}
For all $\epsilon \in [0, +\infty)$, choose for the interpolation path the unique solution $R(\cdot, \epsilon)$ to the first-order ODE \eqref{eq:ODE}.
Fix $\nu > 0$ and let $n$ be large enough so that $\forall \epsilon \in [0, +\infty):R'(\cdot, \epsilon) \subseteq [0,(\rho_x+\nu)^2]$.
The interpolation path satisfies $R'(t,\epsilon) = (\E \langle Q \rangle_{t,\epsilon})^2$ and the sum-rule of Proposition~\ref{prop:sum_rule} yields:
\begin{multline}\label{sum_rule_sol_ode}
\frac{I(\bX ; \bY \vert \bW)}{n}
= \cO(\epsilon) + \smallO_n(1) 
+ \frac{1}{\alpha} \adjustlimits{\inf}_{q_s \in [0,\rho_s]} {\sup}_{r_s \geq 0}\;
\widetilde{\psi}_\alpha\bigg(\frac{\lambda \epsilon}{2} + \int_0^1 \frac{\lambda R'(t,\epsilon)}{2}dt, r_s, q_s\bigg)\\
+ \int_0^1 \bigg(\frac{\lambda}{12} \rho_x^3 + \frac{\lambda}{6} (\E\,\langle Q \rangle_{t,\epsilon})^3
- \frac{\lambda}{4} (\E\,\langle Q \rangle_{t,\epsilon})^2 \rho_x \bigg)dt
-\frac{\lambda}{12} \int_0^1 \Big( \E\,\langle Q^3 \rangle_{t,\epsilon} -  (\E\,\langle Q \rangle_{t,\epsilon})^3  \Big) dt \;.
\end{multline}
By Lemma~\ref{lemma:properties_functions} in Appendix \ref{app:useful_lemmas}, the map $r \mapsto \inf_{q_s \in [0,\rho_s]} \sup_{r_s \geq 0}\; \widetilde{\psi}_\alpha(r , r_s, q_s)$ is nondecreasing and concave. Therefore:
\begin{equation}\label{lowerbound_psi_tilde_jensen}
\adjustlimits{\inf}_{q_s \in [0,\rho_s]} {\sup}_{r_s \geq 0} \;\widetilde{\psi}_\alpha\bigg(\frac{\lambda \epsilon}{2} + \int_0^1 \frac{\lambda R'(t,\epsilon)}{2}dt, r_s, q_s\bigg)
\geq \int_0^1 \adjustlimits{\inf}_{q_s \in [0,\rho_s]} {\sup}_{r_s \geq 0} \;\widetilde{\psi}_\alpha\bigg(\frac{\lambda R'(t,\epsilon)}{2}, r_s, q_s\bigg) dt \;.
\end{equation}
Combining the identity \eqref{sum_rule_sol_ode} with \eqref{lowerbound_psi_tilde_jensen} yields:
\begin{align}
&\frac{I(\bX ; \bY \vert \bW)}{n}\nonumber\\
&\qquad\geq
\int_0^1 \!\bigg\{\adjustlimits{\inf}_{q_s \in [0,\rho_s]} {\sup}_{r_s \geq 0} \frac{1}{\alpha}\widetilde{\psi}_\alpha\bigg(\frac{\lambda(\E \langle Q \rangle_{t,\epsilon})^2}{2}, r_s, q_s \!\bigg) + \frac{\lambda \rho_x^3}{12}  + \frac{\lambda(\E\,\langle Q \rangle_{t,\epsilon})^3}{6} 
- \frac{\lambda (\E\,\langle Q \rangle_{t,\epsilon})^2 \rho_x}{4}  \bigg\}dt
\nonumber\\&
\qquad\qquad\qquad\qquad\qquad\qquad\quad
-\frac{\lambda}{12} \int_0^1 \Big( \E\,\langle Q^3 \rangle_{t,\epsilon} -  (\E\,\langle Q \rangle_{t,\epsilon})^3  \Big) dt
+ \cO(\epsilon) + \smallO_n(1)  \nonumber\\
&\qquad\geq 
\mathop{\vphantom{p}\inf}_{q_x \in [0,\rho_x + \nu]} \adjustlimits{\inf}_{q_s \in [0,\rho_s]} {\sup}_{r_s \geq 0} \; \psi_{\lambda,\alpha}(q_x, q_s, r_s)\nonumber\\
&\qquad\qquad\qquad\qquad\qquad\qquad\quad
-\frac{\lambda}{12} \int_0^1 \Big( \E\,\langle Q^3 \rangle_{t,\epsilon} -  (\E\,\langle Q \rangle_{t,\epsilon})^3  \Big) dt
+ \cO(\epsilon) + \smallO_n(1) \;.\label{lowerbound_I(X,Y)_with_remainder}
\end{align}
The second inequality follows from identity \eqref{link_psi_tilde_and_psi} and $\E\langle Q \rangle_{t, \epsilon} \in [0, \rho_x + \nu]$. 

The result of the proposition will follow if we can get rid of the integral term on the right-hand side of \eqref{lowerbound_I(X,Y)_with_remainder}
This is achieved by proceeding exactly as in the proof of the upper bound in Section~\ref{subsec:upper_bound}, that is, we integrate \eqref{lowerbound_I(X,Y)_with_remainder} over $\epsilon \in [s_n, 2 s_n]$ where $s_n = n^{-\eta}$, $\eta > 0$.
Then:
\begin{align}
&\frac{I(\bX ; \bY \vert \bW)}{n} = \int_{s_n}^{2s_n} \frac{I(\bX ; \bY \vert \bW)}{n}\,\frac{d\epsilon}{s_n}\nonumber\\
&\geq \smallO_n(1) \!
+  \! \mathop{\vphantom{p}\inf}_{q_x \in [0,\rho_x + \nu]} \adjustlimits{\inf}_{q_s \in [0,\rho_s]} {\sup}_{r_s \geq 0} \; \psi_{\lambda,\alpha}(q_x, q_s, r_s)
-\frac{\lambda}{12} \int_0^1 \! dt \! \int_{s_n}^{2s_n} \! \frac{d\epsilon}{s_n} 
\Big( \E\,\langle Q^3 \rangle_{t,\epsilon} -  (\E\,\langle Q \rangle_{t,\epsilon})^3  \Big)
\nonumber \\ &
\geq \smallO_n(1) \!
+  \! \mathop{\vphantom{p}\inf}_{q_x \in [0,\rho_x + \nu]} \adjustlimits{\inf}_{q_s \in [0,\rho_s]} {\sup}_{r_s \geq 0} \; \psi_{\lambda,\alpha}(q_x, q_s, r_s)
-\frac{\lambda \Vert \varphi \Vert_{\infty}^4}{6} \int_0^1 \! dt \! \int_{s_n}^{2s_n} \! \frac{d\epsilon}{s_n}
\sqrt{\E\,\langle (Q -\E\,\langle Q \rangle_{t,\epsilon})^2 \rangle_{t,\epsilon}} .
\label{integral_perturbation_upperbound}
\end{align}
The last inequality is simply due to:
\begin{align*}
 \E\langle Q^3 \rangle_{t,\epsilon} -  (\E\langle Q \rangle_{t,\epsilon})^3
= \E\langle Q(Q + \E\langle Q \rangle_{t,\epsilon})(Q -\E\langle Q \rangle_{t,\epsilon} ) \rangle_{t,\epsilon}
\leq 2 \Vert \varphi \Vert_{\infty}^4 \sqrt{\E\langle (Q -\E\langle Q \rangle_{t,\epsilon})^2 \rangle_{t,\epsilon}} \;.
\end{align*}
After the change of variables $\epsilon \to R \equiv R(t,\epsilon)$, which is justified by $R(t,\cdot)$ being a $\mathcal{C}^1$-diffeomorphism from $[0,+\infty)$ to its image (see Lemma~\ref{lemma:ode_for_interpolation_path}), we can upper bound the remainder on the right-side of \eqref{integral_perturbation_upperbound} in a way similar to \eqref{lowerbound_change_variable}:
$$
\bigg\vert  \int_{s_n}^{2s_n} \sqrt{\E\,\langle (Q -\E\,\langle Q \rangle_{t,\epsilon})^2 \rangle_{t,\epsilon}}\, \frac{d\epsilon}{s_n} \bigg\vert
\leq \sqrt{\int_{s_n}^{2s_n + \rho_x^2}
	\E\,\big\langle \big( Q - \E\,\langle Q \rangle_{t,R}\big)^{2} \big\rangle_{t,R} \,\frac{dR}{s_n}}\;.
$$
Finally, applying Proposition~\ref{prop:concentration_overlap} in Appendix \ref{app:concentration_overlap} with $M=2 + \rho_x^2, a=s_n, b=2s_n + \rho_x^2$ and $\delta=s_n n^{\frac{2\eta-1}{3}}$, we can further bound the right-hand side of the last inequality to obtain:
$$
\bigg\vert  \frac{\lambda \Vert \varphi \Vert_{\infty}^4}{6} \int_{s_n}^{2s_n} \sqrt{\E\,\langle (Q -\E\,\langle Q \rangle_{t,\epsilon})^2 \rangle_{t,\epsilon}}\, \frac{d\epsilon}{s_n} \bigg\vert
\leq C n^{\frac{5\eta-1}{6}}
$$
for $n$ large enough and $C$ a positive constant which does not depend on $t$ and $n$.
Thus, the remaining term on the right-hand side of \eqref{integral_perturbation_upperbound} vanishes when $n$ goes to infinity as long as $\eta < \nicefrac{1}{5}$.
Passing to the limit inferior on both sides of the inequality \eqref{integral_perturbation_upperbound} yields:
\begin{equation*}
\liminf_{n \to +\infty} \frac{I(\bX ; \bY \vert \bW)}{n} \geq
 \mathop{\vphantom{p}\inf}_{q_x \in [0,\rho_x + \nu]} \adjustlimits{\inf}_{q_s \in [0,\rho_s]} {\sup}_{r_s \geq 0} \; \psi_{\lambda,\alpha}(q_x, q_s, r_s) \;.
\end{equation*}
This is true for all $\nu > 0$ and Proposition~\ref{prop:lowerbound_mutual_info} follows directly.
\end{IEEEproof}

\section{Derivation of the asymptotic Tensor-MMSE}\label{sec:proofthm2}
The derivation of the asymptotic Tensor-MMSE rests on the following preliminary proposition.
\begin{proposition}\label{proposition:properties_h}
Suppose that \ref{hyp:S_bounded_support} and \ref{hyp:varphi} hold.
Define for all $\lambda \in (0, +\infty)$:
\begin{align*}
h(\lambda)
&\triangleq \mathop{\vphantom{p}\inf}_{q_x \in [0,\rho_x]}\adjustlimits{\inf}_{q_s \in [0,\rho_s]} {\sup}_{r_s \geq 0}\; \psi_{\lambda,\alpha}(q_x , q_s, r_s) \;;\\
\mathcal{Q}_x^*(\lambda)
&\triangleq \bigg\{q_x^* \in [0,\rho_x] :
\adjustlimits{\inf}_{q_s \in [0,\rho_s]} {\sup}_{r_s \geq 0}\; \psi_{\lambda,\alpha}(q_x^* , q_s, r_s) 
= h(\lambda)
\bigg\}\;.
\end{align*}
For every $\lambda > 0$, $\mathcal{Q}_x^*(\lambda)$ is nonempty.
The function $h$ is differentiable at $\lambda$ if, and only if, the set $\mathcal{Q}_x^*(\lambda)$ is a singleton.
In this case, letting $\mathcal{Q}_x^*(\lambda) = \{q_x^*(\lambda)\}$, the derivative of $h$ at $\lambda$ satisfies:
\begin{equation}\label{derivative_h}
h'(\lambda)
= \frac{1}{12} \bigg(\rho_x^3 - \big(q_x^*(\lambda)\big)^3\bigg)\;.
\end{equation}
\end{proposition}
The proof of this result is given in Appendix~\ref{app:proof_mmse}.
We can now prove Theorem~\ref{theorem:tensor_MMSE}.
\begin{IEEEproof}[Proof of Theorem~\ref{theorem:tensor_MMSE}]
Let $n \in \mathbb{N}^*$.
The angular brackets $\langle - \rangle_{n,\lambda}$ denote the expectation with respect to the posterior distribution of $\bS$ given $(\bY, \bW)$.
Define $h_n: \lambda \in (0,+\infty) \mapsto \frac{I(\bX,\bY \vert \bW)}{n}$ (the mutual information depends on $\lambda$ through the observation $\bY$).
We have for all $\lambda \in (0, +\infty)$:
\begin{align}
h_n(\lambda)
&=  \frac{\lambda}{2 n^3 }\sum_{\bi \in \cI }\E[X_{i_1}^2 X_{i_2}^2 X_{i_3}^2]
- \frac{1}{n}\E \ln \! \int \! dP_S(\bs)
 e^{\sum\limits_{\bi \in \cI}x_{i_1} x_{i_2} x_{i_3}\big(-
	\frac{\lambda}{2n^2} x_{i_1} x_{i_2} x_{i_3}
	+\frac{\lambda}{n^2}\, X_{i_1} X_{i_2} X_{i_3} 
	+\frac{\sqrt{\lambda}}{n}\, Z_{\bi}\big)}\nonumber\\
h'_n(\lambda)
&=  \frac{1}{2 n^3 }\sum_{\bi \in \cI }\E[(X_{i_1}^2 X_{i_2}^2 X_{i_3} - \langle x_{i_1} x_{i_2} x_{i_3} \rangle_{n,\lambda})^2]
=  \frac{\MMSE_n(\bX^{\otimes 3}\vert \bY, \bW)}{12} + \cO(n^{-1})\label{I-MMSE_relation}\\
h''_n(\lambda)
&=  -\frac{1}{2 n^5}\sum_{\bi, \mathbf{j} \in \cI}\E[(\langle x_{i_1} x_{i_2} x_{i_3} x_{j_1} x_{j_2} x_{j_3} \rangle_{n,\lambda}
-\langle x_{i_1} x_{i_2} x_{i_3} \rangle_{n,\lambda}\langle x_{j_1} x_{j_2} x_{j_3} \rangle_{n,\lambda})^2]\label{2nd_derivative_hn}
\end{align}
Differentiations under the integral sign yielding \eqref{I-MMSE_relation} and \eqref{2nd_derivative_hn} are justified by the domination properties implied by \ref{hyp:S_bounded_support}, \ref{hyp:varphi}.
$h_n''$ is nonpositive so $h_n$ is concave on $(0,+\infty)$.
By Theorem~\ref{theorem:limit_mutual_information},
$h: \lambda \mapsto \mathop{\vphantom{p}\inf}\limits_{q_x \in [0,\rho_x]}\adjustlimits{\inf}_{q_s \in [0,\rho_s]} {\sup}_{r_s \geq 0}\; \psi_{\lambda,\alpha}(q_x , q_s, r_s)$
is the pointwise limit of the sequence of differentiable concave functions $(h_n)_{n \in \mathbb{N}^*}$.
Hence, $h$ is concave and thus differentiable on $(0,+\infty)$ minus a countable set, and at every $\lambda$ where $h$ is differentiable we have:
\begin{equation}\label{limit_derivative_hn}
\lim_{n \to +\infty} h_n^\prime(\lambda) = h'(\lambda)
= \frac{\rho_x^3 - (q_x^*(\lambda))^3}{12} \;.
\end{equation}
The last equality follows from Proposition~\ref{proposition:properties_h} and $q_x^*(\lambda)$ denotes the unique element of $\mathcal{Q}_x^*(\lambda)$.
Combining \eqref{limit_derivative_hn} with the I-MMSE relation \eqref{I-MMSE_relation} yields the theorem.
\end{IEEEproof}
\section{Limit at \texorpdfstring{$\alpha = 0$}{alpha = 0}}\label{section:alphatendstozero}
In this section we give a \textit{non-rigorous} derivation of the limit of the asymptotic normalized mutual information when $\alpha$ goes to $0$.
\noindent Fix $\lambda > 0$. We define the function
\begin{align*}
&\qquad\qquad\qquad\qquad\qquad\qquad\;\;
\Psi_*: \alpha \mapsto \inf\limits_{q_x \in [0,\rho_x]}\adjustlimits{\inf}_{q_s \in [0,\rho_s]} {\sup}_{r_s \geq 0}\; \Psi(q_x , q_s, r_s, \alpha)\\ 
\shortintertext{where}
&\Psi(q_x , q_s, r_s, \alpha)
\triangleq
\alpha \psi_{\lambda,\alpha}(q_x , q_s, r_s)
= I(S;\sqrt{r_s}\,S + Z) - \frac{r_s(\rho_s - q_s)}{2}
+ \alpha \frac{\lambda}{12} (\rho_x - q_x)^2(\rho_x + 2 q_x)\\
&\qquad\qquad\qquad\qquad\qquad\qquad\qquad\qquad\quad
+ \alpha I\big(U; \sqrt{\lambda q_x^2/2}\,\varphi(\sqrt{\rho_s - q_s} \, U + \sqrt{q_s} \, V) + \widetilde{Z} \,\big\vert\, V \big)\;.
\end{align*}
The function $\Psi_*$ is convex on $[0,+\infty)$ so it is continuous on $[0,+\infty)$ and differentiable almost everywhere on $(0,+\infty)$.
Note that:
\begin{equation*}
\frac{\partial \Psi}{\partial \alpha}\bigg\vert_{(q_x , q_s, r_s, \alpha)}
= I\big(U; \sqrt{\lambda q_x^2/2}\,\varphi(\sqrt{\rho_s - q_s} \, U + \sqrt{q_s} \, V) + \widetilde{Z} \,\big\vert\, V \big)
+ \frac{\lambda}{12} (\rho_x - q_x)^2(\rho_x + 2 q_x)\;.
\end{equation*}
Hence, assuming that we can apply some envelope theorem\cite{Milgrom_Envelope_Theorems} as in Appendix~\ref{app:proof_mmse}, it comes
\begin{equation*}
\Psi_*^\prime(\alpha)
=  I\big(U; \sqrt{\nicefrac{\lambda q_x^*(\alpha)^2}{2}}\,\varphi(\sqrt{\rho_s - q_s^*(\alpha)} \, U + \sqrt{q_s^*(\alpha)} \, V) + \widetilde{Z} \big\vert V \big)
+ \frac{\lambda}{12} (\rho_x - q_x^*(\alpha))^2(\rho_x + 2 q_x^*(\alpha))
\end{equation*}
whenever $(q_x^*(\alpha), q_s^*(\alpha), r_s^*(\alpha))$ is the unique triple satisfying $\Psi_*(\alpha) = \Psi(q_x^*(\alpha), q_s^*(\alpha), r_s^*(\alpha), \alpha)$.
At $\alpha = 0$,
$\Psi(q_x , q_s, r_s, \alpha) = I(S;\sqrt{r_s}\,S + Z) - r_s(\rho_s - q_s)/2$ so $\Psi_*(0) = \Psi(q_x , q_s^*(0), r_s^*(0), \alpha) = 0$
where $q_s^*(0) = (\E_{S \sim P_S}\,S)^2 \triangleq m_s^2$, $r_s^*(0) = 0$.
By Theorem~\ref{theorem:limit_mutual_information}, $\lim_{\substack{n \to +\infty\\ \nicefrac{n}{p}\to \alpha}}  \frac{I(\bX ; \bY \vert \bW)}{n} = \frac{\Psi_*(\alpha)}{\alpha}$.
Using L'H\^opital's rule, it follows that (provided that the limit on the right-hand side exists):
\begin{equation}\label{eq:hopital_rule}
\lim_{\alpha \to 0^+} \lim_{\substack{n \to +\infty\\ \nicefrac{n}{p}\to \alpha}}  \frac{I(\bX ; \bY \vert \bW)}{n}
= \lim_{\alpha \to 0^+} \Psi_*^\prime(\alpha)\;.
\end{equation}
Assuming that $\lim_{\alpha \to 0^+} (q_s^*(\alpha), r_s^*(\alpha))=  (q_s^*(0),r_s^*(0)) = (m_s^2,0)$, we have:
\begin{align*}
&\lim_{\alpha \to 0^+} \psi_{\lambda,\alpha}(q_x^*(\alpha), q_s^*(\alpha), r_s^*(\alpha))
= \lim_{\alpha \to 0^+} \psi_{\lambda,\alpha}(q_x^*(\alpha), m_s^2, 0)\\
&=\lim_{\alpha \to 0+}
I\big(U; (\nicefrac{\lambda q_x^*(\alpha)^2}{2})^{\frac12}\,\varphi(\sqrt{\rho_s - m_s^2} \, U + \vert m_s \vert\, V) + \widetilde{Z} \big\vert V\big)
+ \frac{\lambda}{12} (\rho_x - q_x^*(\alpha))^2(\rho_x + 2 q_x^*(\alpha))\\
&=\lim_{\alpha \to 0+} \Psi_*^\prime(\alpha)
\shortintertext{as well as}
&\lim_{\alpha \to 0^+} \psi_{\lambda,\alpha}(q_x^*(\alpha), q_s^*(\alpha), r_s^*(\alpha))
= \! \lim_{\alpha \to 0^+} \inf_{q_x \in [0,\rho_x]} \psi_{\lambda,\alpha}(q_x ,  q_s^*(\alpha), r_s^*(\alpha))
= \! \inf_{q_x \in [0,\rho_x]} \psi_{\lambda,\alpha}(q_x ,  m_s^2, 0)
\\&
=
\inf_{q_x \in [0,\rho_x]}\; 
I\big(U; (\nicefrac{\lambda q_x^2}{2})^{\frac{1}{2}} \,\varphi(\sqrt{\rho_s - m_s^2} \, U + \vert m_s \vert\, V) + \widetilde{Z} \big\vert V\big)
+ \frac{\lambda}{12} (\rho_x - q_x)^2(\rho_x + 2 q_x)\;.
\end{align*}
Both chains of equalities together with \eqref{eq:hopital_rule} give: 
\begin{multline}\label{limit_mutual_info_alpha=0}
\lim_{\alpha \to 0} \lim_{\substack{n \to +\infty\\ \nicefrac{n}{p}\to \alpha}}  \frac{I(\bX ; \bY \vert \bW)}{n}\\
= \inf_{q_x \in [0,\rho_x]}
I\big(U; \sqrt{\nicefrac{\lambda q_x^2}{2}}\,\varphi(\sqrt{\rho_s - m_s^2} \, U + \vert m_s \vert\, V) + \widetilde{Z} \big\vert V\big)
+ \frac{\lambda}{12} (\rho_x - q_x)^2(\rho_x + 2 q_x)\;.
\end{multline}
Thus, we conjecture that the asymptotic normalized multual information converges when $\alpha \to 0^+$ to the asymptotic normalized mutual information of the following channel:
\begin{align*}
\widetilde{Y}_{ijk} = \frac{\sqrt{\lambda}}{n} \widetilde{X}_i\widetilde{X}_j \widetilde{X}_k + \widetilde{Z}_{ijk} \;, 1 \leq i \leq j \leq k \leq n\;,
\end{align*}
with $\widetilde{X}_i = \varphi(\sqrt{\rho_s - m_s^2} \, U_i + \vert m_s \vert\, V_i)$
where $U_1,\dots, U_n, V_1,\dots, V_n\iid \cN(0,1)$ and $\bV$ is known.
The second moment of the i.i.d.\ random variables $\widetilde{X}_i$ is $\rho_x \triangleq \E[\varphi(\cN(0,\rho_s))^2]$.
Proofs in the literature can be easily adapted to show that $\lim\limits_{n \to +\infty}  \frac{I(\widetilde{\bX} ; \widetilde{\bY} \vert \bV )}{n}$ is given by the right-hand side of \eqref{limit_mutual_info_alpha=0}.

\section*{Acknowledgment}
\noindent C. L acknowledges funding from Swiss National Foundation for Science grant 200021E 175541.
%
\ifCLASSOPTIONcaptionsoff
  \newpage
\fi
%


\bibliographystyle{IEEEtran}
\bibliography{IEEEabrv,bibliography}
%
%
%
\newpage
\appendices
\section{Auxiliary lemmas}\label{app:useful_lemmas}
\begin{lemma}\label{lemma:properties_functions}
Assume $\varphi: \R \to \R$ is a bounded continuous function.
Let $U,V,\widetilde{Z} \iid \cN(0,1)$.
For all $\rho \geq 0$, the function
\begin{align*}
I_{\varphi}(\cdot\, ,\cdot \,;\rho): [0,+\infty) \times [0,\rho] \longrightarrow [0,+\infty), \, (r,q) \longmapsto I(U;\sqrt{r}\varphi(\sqrt{\rho - q}U + \sqrt{q}V) + \widetilde{Z}\vert V)
\end{align*}
is continuous, and $\forall q \in [0,\rho]: r \mapsto I_{\varphi}(\cdot\, , q\,;\rho)$ is twice-differentiable, nondecreasing, concave and $\nicefrac{\Vert \varphi \Vert_{\infty}^2}{2}$-Lipschitz on $\R_+$.
Let $S \sim P_S$ (a probability distribution on $\R$) and $Z \sim \cN(0,1)$. Fix $\alpha, \rho_s \geq 0$ and define
\begin{align*}
\widetilde{\psi}_\alpha: \!  [0,+\infty)^2 \times [0,\rho_s] \! \longrightarrow \! [0,+\infty), 
(r,r_s,q_s) \! \longmapsto \! I(S;\sqrt{r_s}S + Z) \! + \! \alpha I_{\varphi}(r, q_s;\rho_s) - \frac{r_s(\rho_s - q_s)}{2}.
\end{align*}
Both functions $r \mapsto \sup_{r_s \geq 0}  \widetilde{\psi}_\alpha(r,r_s,q_s)$ and $r \mapsto \inf_{q_s \in [0,\rho_s]} \sup_{r_s \geq 0}  \widetilde{\psi}_\alpha(r,r_s,q_s)$
are nondecreasing, concave and $\nicefrac{\alpha \Vert \varphi \Vert_{\infty}^2}{2}$-Lipschitz on $\R_+$.
\end{lemma}
\begin{IEEEproof}
Fix $\rho \geq 0$ and $q \in [0,\rho]$.
Define $Y^{(r)} \triangleq \sqrt{r}\varphi(\sqrt{\rho - q}U + \sqrt{q}V) + \widetilde{Z}$.
Then, $I_{\varphi}(r, q; \rho) = I(U;Y^{(r)}\vert V)$.
We have:
\begin{multline*}
I_{\varphi}(r, q;\rho) \\
 = 
-\E
\ln \int du \frac{e^{-\frac{u^2}{2}}}{\sqrt{2\pi}} \,e^{-\frac{r}{2}(\varphi(\sqrt{\rho - q} U + \sqrt{q}V ) - \varphi(\sqrt{\rho - q} u + \sqrt{q}V ))^2
- \sqrt{r}(\varphi(\sqrt{\rho - q} U + \sqrt{q}V) - \varphi(\sqrt{\rho - q} u + \sqrt{q}V ))\widetilde{Z}
} \:.
\end{multline*}
Let $\langle - \rangle_r$ denote the expectation with respect to the posterior distribution of $U$ given $(Y^{(r)}, V)$.
The assumptions on $\varphi$ imply domination assumptions justifying the continuity of $I_{\varphi}(\cdot\, ,\cdot \,;\rho)$ and the (twice) differentiability of $r \mapsto I_{\varphi}(r, q;\rho)$.
Differentiating w.r.t.\ $r$, it comes:
\begin{align}
\frac{\partial I_{\varphi}}{\partial r}(r,q;\rho) 
&= \frac{1}{2}\,\E\,\big\langle(\varphi(\sqrt{\rho - q}\,U + \sqrt{q}V ) - \varphi(\sqrt{\rho - q}\,u + \sqrt{q}V ))^2\big\rangle_{r}\nonumber\\
&\qquad\qquad\qquad\qquad\qquad\qquad\qquad\qquad
- \frac{1}{2\sqrt{r}}\E\big[\big\langle \varphi(\sqrt{\rho - q}\,u + \sqrt{q}V ) \big\rangle_{r}\widetilde{Z}\big]\nonumber\\
&= \frac{1}{2} \E\big[\varphi^2(\sqrt{\rho - q}\,U + \sqrt{q}V ) - \big\langle\varphi(\sqrt{\rho - q}\,u + \sqrt{q}V ))\big\rangle_{r}^2\,\big]
\geq 0 \;.\label{sign_derivative_I}
\end{align}
	The second equality is obtained using integration by parts w.r.t.\ $\widetilde{Z}$ and Nishimori identity
	$$
	\E\big[\varphi(\sqrt{\rho - q}\,U + \sqrt{q}V ) \big\langle \varphi(\sqrt{\rho - q}\,u + \sqrt{q}V ) \big\rangle_{r}\big]
	=\E\,\big\langle \varphi(\sqrt{\rho - q}\,u + \sqrt{q}V ) \big\rangle_{r}^2\;.
	$$
	The nonnegativity of the derivative follows from  Jensen's inequality and Nishimori identity:
	$$
	\E\,\big\langle \varphi(\sqrt{\rho - q}\,u + \sqrt{q}V ) \big\rangle_{r}^2
	\leq \E\,\big\langle \varphi^2(\sqrt{\rho - q}\,u + \sqrt{q}V ) \big\rangle_{r}
	= \E\,\varphi^2(\sqrt{\rho - q}\,U + \sqrt{q}V ) \;.
	$$
	Further differentiating and using integration by parts w.r.t.\ $\widetilde{Z}$ and Nishimori identity where necessary yields:
	\begin{equation}
	\frac{\partial^2 I_\varphi}{\partial r^2}(r,q;\rho) 
	= -\frac{1}{2}\E\,\big\langle \big(\varphi(\sqrt{\rho - q} u + \sqrt{q}V) - \langle \varphi(\sqrt{\rho - q} u + \sqrt{q}V) \rangle_r \big)^2\,\big\rangle_r^2
	\leq 0 \;. \label{sign_second_derivative_I}
	\end{equation}
	From \eqref{sign_derivative_I},\eqref{sign_second_derivative_I} $I_{\varphi}(\cdot, q;\rho)$ is concave nondecreasing.
	The Lipschitzianity follows simply from:
	\begin{equation*}
	0 \leq\frac{\partial I_{\varphi}}{\partial r}(r, q;\rho)
	\leq \frac{1}{2} \E\big[\varphi^2(\sqrt{\rho - q}\,U + \sqrt{q}V )\big]
	\leq \frac{\Vert \varphi \Vert_{\infty}^2}{2}  \;.
	\end{equation*}
The properties of $r \mapsto \sup_{r_s \geq 0} \widetilde{\psi}_\alpha(r,r_s,q_s)$ follow directly from the ones of $I_{\varphi}(\cdot, q_s; \rho_s)$ as
$$
\sup_{r_s \geq 0} \widetilde{\psi}_\alpha(r,r_s,q_s)
= \alpha I_{\varphi}(r, q_s; \rho_s) + \sup_{r_s \geq 0} I(S;\sqrt{r_s}S + Z) - \frac{r_s(\rho_s - q_s)}{2} \;.
$$
Finally, $r \mapsto \inf_{q_s \in [0,\rho_s]}\sup_{r_s \geq 0} \widetilde{\psi}_\alpha(r,r_s,q_s)$ is the infimum of nondecreasing, concave, $\nicefrac{\alpha \Vert \varphi \Vert_{\infty}^2}{2}$-Lipschitzian functions, hence its properties.
\end{IEEEproof}
\begin{lemma}\label{lemma:properties_I_PS}
	Assume that $P_S$ is a probability distribution on $\mathbb{R}$ with bounded support $\mathrm{supp}(P_S) \subseteq [-M_S,M_S]$, $M_S > 0$.
	Let $S \sim P_S, Z \sim \cN(0,1)$ be random variables independent of each other.
	Define the functions
	\begin{eqnarray*}
		& I_{P_S}:[0,+\infty) \longrightarrow [0,+\infty), \quad r_s \longmapsto I(S;\sqrt{r_s}S + Z)
		\\&
		I_{P_S}^*:\mathbb{R} \longrightarrow [0,+\infty], \quad  x  \longmapsto \sup_{r_s \geq 0}  I_{P_S}(r_s) + xr_s \;.
	\end{eqnarray*}
Then, $I_{P_S}$ is twice-differentiable, concave and nondecreasing, while $I_{P_S}^*$ is convex, nondecreasing, finite on $(-\infty,0)$, equal to $+\infty$ on $(0,+\infty)$ and $I_{P_S}^*(0) = \lim_{r_s \to +\infty} I_{P_S}(r_s) \in [0,+\infty]$.
\end{lemma}
\begin{IEEEproof}
Let $Y^{(r_s)} = \sqrt{r_s}\,S + Z$. We have:
\begin{equation*}
	I_{P_S}(r_s) = \frac{r_s \E S^2}{2}
		-\E\ln \int dP_S(s) \,e^{r_s S s +\sqrt{r_s}\,Z s -\frac{r_s s^2}{2}} \;.
\end{equation*}
Let $\langle - \rangle_{r_s}$ denote the expectation with respect to the posterior distribution of $S$ given $Y^{(r_s)}$.
The assumption on the support of $P_S$ ensures domination properties, thus allowing to show that $I_{P_S}$ is twice differentiable.
Differentiating w.r.t.\ $r_s$, it comes:
\begin{equation}\label{sign_derivative_I_PS}
I'_{P_S}(r_s) 
= \frac{\E S^2}{2}-\,\E\,S \langle s \rangle_{r_s} + \frac{\E\,\langle s^2 \rangle_{r_s}}{2} + \frac{\E\,Z \langle s \rangle_{r_s}}{2\sqrt{r_s}}
= \frac{1}{2} \E\big[(S - \langle s \rangle_{r_s})^2\big] \geq 0 \;.
\end{equation}
Further differentiating and using integration by parts w.r.t.\ $Z$ and Nishimori identity where necessary yields:
\begin{equation}\label{sign_second_derivative_I_PS}
I''_{P_S}(r_s) 
= -\frac{1}{2}\E\big[\big\langle (s - \langle s \rangle_{r_s})^2\,\big\rangle_{r_s}^2\,\big]
\leq 0 \;.
\end{equation}
From \eqref{sign_derivative_I_PS} and \eqref{sign_second_derivative_I_PS}, $I_{P_S}$ is nondecreasing and concave.

The function $I_{P_S}^*$ is the Legendre transform of the convex function $-I_{P_S}$, hence it is well-defined and convex. Besides, $I_{P_S}^*$ is defined as the supremum of nondecreasing affine functions of $x$ so it is nondecreasing.
The trivial lower bound $I_{P_S}^*(x) \geq \sup_{r_s \geq 0} xr_s$ shows that $I_{P_S}^*$ is nonnegative and is equal to $+\infty$ on $(0,+\infty)$.
Because the support of $P_S$ is included in $[-M_S,M_S]$, its differential entropy is upper bounded by $\ln(2M_s)$ (the differential entropy of the uniform distribution on the segment $[-M_S,M_S]$) and we have $\forall x \in (-\infty,0)$:
$$
I_{P_S}^*(x)
\; \leq \;  \sup_{r_s \geq 0} \ln\bigg(2M_s\sqrt{\frac{r_s}{2\pi e}}\bigg) +xr_s \; = \; \ln\bigg(\frac{M_S^2}{\pi e^2 \vert x \vert}\bigg) < +\infty \;.
$$
Finally, $I_{P_S}^*(0) = \sup_{r_s \geq 0}  I_{P_S}(r_s) = \lim_{r_s \to +\infty} I_{P_S}(r_s)$.
\end{IEEEproof}

\section{Establishing the sum-rule}\label{app:establishing_sum_rule}
\begin{lemma}[Link between average free entropy and normalized mutual information]\label{lemma:link_i_n_f_n}
Suppose that \ref{hyp:S_bounded_support} and \ref{hyp:varphi} hold,  and that $R'(t,\epsilon)$ is uniformly bounded in ${(t,\epsilon) \in [0,1] \times [0,+\infty)}$ where $R'(\cdot,\epsilon)$ denotes the derivative of $R(\cdot,\epsilon)$.
The normalized mutual information $\eqref{definition_i_n}$ and its partial derivative with respect to $t$, which we denote $i'_n(t,\epsilon)$, satisfy:
\begin{align}
i_n(t,\epsilon) &= - f_n(t,\epsilon) + \frac{\lambda R(t,\epsilon)}{4} \rho_x + \frac{\lambda (1-t)}{12} \rho_x^3 + (1-t)\,\smallO_n(1) \label{link_i_n_f_n}
\\
i'_n(t,\epsilon) &= - f'_n(t,\epsilon) + \frac{\lambda R'(t,\epsilon)}{4} \rho_x - \frac{\lambda}{12} \rho_x^3 \, + \,  \smallO_n(1)\;.\label{link_i'_n_f'_n}
\end{align}
The quantity $\smallO_n(1)$ does not depend on $(t,\epsilon)$ and vanishes when $n$ goes to infinity. Besides, at $t=0$, for all $\epsilon \in [0,+\infty)$:
\begin{equation}\label{i_n(t=0,epsilon)}
\bigg\vert i_n(0,\epsilon) - \frac{I(\bX ; \bY \vert \bW)}{n} \bigg\vert \leq \frac{\lambda \Vert \varphi \Vert_{\infty}^2}{2}\,\epsilon\;.
\end{equation}
\end{lemma}
\begin{IEEEproof}
By definition of the normalized mutual information \eqref{definition_i_n}, we have:
\begin{align}
i_n(t,\epsilon)
&= \frac{1}{n}H\big(\bY^{(t)},\widetilde{\bY}^{(t,\epsilon)}\big\vert \bW \big)
-\frac{1}{n}H\big(\bY^{(t)},\widetilde{\bY}^{(t,\epsilon)}\big\vert \bS, \bW \big)\nonumber\\
&= -\frac{1}{n}\E\,\ln\Big(\cZ_{t,\epsilon}(\bY^{(t)},\widetilde{\bY}^{(t,\epsilon)}, \bW)e^{-\frac{1}{2} (\sum_{\bi \in \cI } Y_{\bi}^2+ \Vert \widetilde{\bY} \Vert^2 )}\Big)
+ \frac{1}{n}\E\Big[\ln e^{-\frac{1}{2} (\sum_{\bi \in \cI } Z_{\bi}^2+ \Vert \widetilde{\bZ} \Vert^2 )}\Big]\nonumber\\
&= -f_n(t,\epsilon)
+ \frac{\lambda R(t,\epsilon)}{4n}\sum_{j=1}^n \E[ X_j^2 ] + \frac{\lambda (1-t)}{2 n^3 }\sum_{\bi \in \cI }\E[X_{i_1}^2 X_{i_2}^2 X_{i_3}^2]\nonumber\\
&= -f_n(t,\epsilon) + \frac{\lambda R(t,\epsilon)}{4} \E[ X_1^2 ]
\nonumber\\&
\qquad\qquad\quad 
+ \frac{\lambda (1-t)}{2 n^3 } \bigg(\binom{n}{3}\E[X_{1}^2 X_{2}^2 X_{3}^2] + n(n-1)\E[X_{1}^2 X_{2}^4] + n\E[X_{1}^6]\bigg)\nonumber\\
&= -f_n(t,\epsilon) + \frac{\lambda R(t,\epsilon)}{4} \E[ X_1^2 ] + \frac{\lambda (1-t)}{12} \E[X_{1}^2 X_{2}^2 X_{3}^2] + \lambda (1-t)\cO\big(n^{-1}\big)\;.\label{link_i_n_f_n_proof}
\end{align}
The quantity $\cO(n^{-1})$ appearing in the last equality does not depend on $(t,\lambda)$ and is such that $\big\vert \cO(n^{-1}) \big\vert \leq \nicefrac{C}{n}$ with $C \triangleq \Vert \varphi \Vert_{\infty}^6/2$.
It directly follows that
\begin{equation}\label{link_i'_n_f'_n_proof}
i'_n(t,\epsilon) = -f'_n(t,\epsilon) + \frac{\lambda R'(t,\epsilon)}{4} \E[ X_1^2 ] - \frac{\lambda }{12n^3 } \E[X_{1}^2 X_{2}^2 X_{3}^2] - \lambda \cO\big(n^{-1}\big)\;;
\end{equation}
where the quantity $\cO(n^{-1})$ on the right-hand side of \eqref{link_i'_n_f'_n_proof} is the same as the one appearing on the right-hand side of \eqref{link_i_n_f_n_proof}.

Note that $\E[X_{1}^2 X_{2}^2 X_{3}^2] = \E[\E[X_{1}^2 \vert \bS]^3]$ converges to $\rho_x^3$ as $n$ goes to infinity (the proof of this limit is similar to \cite[Lemma 3 of Supplementary material]{Gabrie_TwoLayerGLM_JSTAT_2019}).
This limit together with \eqref{link_i_n_f_n_proof} and \eqref{link_i'_n_f'_n_proof} ends the proofs of \eqref{link_i_n_f_n} and \eqref{link_i'_n_f'_n}.

At $t=0$, we can use \eqref{link_i_n_f_n_proof} to obtain (remember that $R(0,\epsilon) = \epsilon$):
\begin{equation}\label{upperbound_i_n(0,epsilon)-i_n(0,0)}
\vert i_n(0,\epsilon) - i_n(0,0)\vert \leq \vert f_n(0,\epsilon) - f_n(0,0)\vert + \frac{\lambda \epsilon}{4}\E[X_1^2]\;.
\end{equation}
It is clear that $i_n(0,0) = n^{-1}I(\bX ; \bY \vert \bW)$ where $\bY, \bX$ are defined in \eqref{eq:entries_Y}, \eqref{eq:definition_X}. At $t=0$, the free entropy \eqref{interpolating_free_entropy} reads:
\begin{equation}\label{f_n(0,epsilon)}
f_n(0,\epsilon) = \frac1n \E \ln \int dP_s(\bs) \, e^{-\cH_{0,\epsilon}(\bs \,;\, \bZ, \widetilde{\bZ}, \bX, \bW)}\,
\end{equation}
with (remember that $x_1,\dots,x_n$ are the entries of $\bx \triangleq \varphi(\nicefrac{\bW \bs}{\sqrt{p}})$):
\begin{multline*}
\cH_{0,\epsilon}(\bs \,;\, \bZ, \widetilde{\bZ}, \bX, \bW)
\triangleq
\sum_{\bi \in \cI}
\frac{\lambda}{2n^2} x_{i_1}^2 x_{i_2}^2 x_{i_3}^2
- \frac{\lambda}{n^2}\, X_{i_1} X_{i_2} X_{i_3} x_{i_1} x_{i_2} x_{i_3}
- \frac{\sqrt{\lambda}}{n}\, Z_{\bi} x_{i_1} x_{i_2} x_{i_3}\\
+
\sum_{j=1}^{n} \frac{\lambda \epsilon}{4} x_j^2
- \frac{\lambda \epsilon}{2}\,X_j x_j 
 - \sqrt{\frac{\lambda \epsilon}{2}}\,\widetilde{Z}_j x_j \;.
\end{multline*}
Differentiating \eqref{f_n(0,epsilon)} under the integral sign yields 
$
\nicefrac{\partial f_n}{\partial \epsilon}\vert_{0,\epsilon} =  - \E\,\langle \nicefrac{\partial \cH_{0,\epsilon}}{\partial \epsilon}\rangle_{0,\epsilon} = - \E\,\langle \cL \rangle_{0,\epsilon}
$
where 
$$
\cL \triangleq \frac{1}{n}\sum_{j=1}^{n} \frac{\lambda}{4} x_j^2
- \frac{\lambda}{2}\,X_j x_j 
- \frac{1}{2}\sqrt{\frac{\lambda}{2\epsilon}}\,\widetilde{Z}_j x_j\,.
$$
We show in Lemma~\ref{lemma:computation_E<L>_and_others} in Appendix \ref{app:concentration_overlap}, that  
$\E\,\langle \cL \rangle_{0,\epsilon} = -\frac{\lambda}{4} \E\,\langle Q \rangle_{0,\epsilon}$
with $Q \triangleq n^{-1}\bx^T \bX$ the  overlap. Hence $\big\vert \nicefrac{\partial f_n}{\partial \epsilon}\vert_{0,\epsilon} \big\vert \leq \nicefrac{\lambda \Vert \varphi \Vert_{\infty}^2}{4}$.
This upper bound together with the mean value theorem and the inequality \eqref{upperbound_i_n(0,epsilon)-i_n(0,0)} yields \eqref{i_n(t=0,epsilon)}.
\end{IEEEproof}
\begin{lemma}[Derivative of the normalized mutual information]\label{lemma:formula_derivative_free_entropy}
Suppose that \ref{hyp:S_bounded_support} and \ref{hyp:varphi} hold, and that $R'(t,\epsilon)$ is uniformly bounded in $(t,\epsilon) \in [0,1] \times [0,+\infty)$ where $R'(\cdot,\epsilon)$ denotes the derivative of $R(\cdot,\epsilon)$.
Recall
$$
Q \triangleq \frac{1}{n} \sum_{i=1}^{n} \varphi\bigg(\bigg[\frac{\bW \bs}{\sqrt{p}}\bigg]_i \,\bigg) \varphi\bigg(\bigg[\frac{\bW \bS}{\sqrt{p}}\bigg]_i \,\bigg)
= \frac{1}{n} \sum_{i=1}^{n} x_i X_i
$$ 
denotes the overlap.
Then, the derivative with respect to $t$ of the normalized mutual information \eqref{definition_i_n} satisfies $\forall (t, \epsilon) \in [0,1] \times (0,+\infty)^2$:
\begin{equation}\label{formula_derivative_free_entropy}
i'_{n}(t,\epsilon)
= \frac{\lambda}{12} \big(\E\,\langle Q^3 \rangle_{t,\epsilon} - \rho_x^3\big)
+ \frac{\lambda R'(t,\epsilon)}{4} \big(\rho_x - \E\,\langle Q \rangle_{t,\epsilon}\big)
+ \smallO_n(1) \;,
\end{equation}
where $\smallO_{n}(1)$ vanishes uniformly in $(t, \epsilon)$ as $n$ goes to infinity.
\end{lemma}
\begin{IEEEproof}
The average interpolating free entropy satisfies:
\begin{equation}\label{eq:precise_formula_f}
f_n(t,\epsilon)
= \frac{1}{n} \E_{\bS, \bW}\bigg[\int d\by d\widetilde{\by} \:
\frac{e^{-\frac{1}{2} (\sum_{\bi \in \cI } y_{\bi}^2+ \Vert \widetilde{\by} \Vert^2 )}}{\sqrt{2\pi}^{n + \vert \cI \vert}} e^{-\cH_{t,\epsilon}(\bs \,;\, \bY^{(t)},\widetilde{\bY}^{(t,\epsilon)}, \bW)}
\ln \cZ_{t,\epsilon}\big(\by,\widetilde{\by}, \bW\big) \bigg] \;.
\end{equation}
Taking the derivative with respect to $t$ of \eqref{eq:precise_formula_f}, we get:
\begin{align}
f'_n(t,\epsilon)
&= -\frac{1}{n} \E\big[\cH'_{t,\epsilon}\big(\bS; \bY^{(t)},\widetilde{\bY}^{(t,\epsilon)}, \bW\big) \ln \cZ_{t,\epsilon}\big(\bY^{(t)},\widetilde{\bY}^{(t,\epsilon)} , \overline{\bY}^{(t,\epsilon)}\big)\big]\nonumber\\
&\qquad\qquad\qquad\qquad\qquad\quad
-\frac{1}{n} \E\big[\big\langle \cH'_{t,\epsilon}\big(\bs; \bY^{(t)},\widetilde{\bY}^{(t,\epsilon)}, \bW\big)\big\rangle_{\!t,\epsilon} \,\big]\,,
\label{f'_sum_2_expectations}\\
\shortintertext{with}
\cH'_{t,\epsilon}(\bs ; \by, \widetilde{\by}, \bW)
&\triangleq
\sum_{\bi \in \cI}
-\frac{\lambda}{2n^2} x_{i_1}^2 x_{i_2}^2 x_{i_3}^2 + \frac{1}{2n}\sqrt{\frac{\lambda}{1-t}}\, y_{\bi} x_{i_1} x_{i_2} x_{i_3}\nonumber\\
&\qquad\qquad\qquad\qquad\qquad\quad
+\sum_{j=1}^{n} \frac{\lambda R'(t,\epsilon)}{4} x_i^2 - \frac{R'(t,\epsilon)}{2}\sqrt{\frac{\lambda}{2 R(t,\epsilon)}}\,\widetilde{y}_j x_j \;.
\label{derivative_hamiltonian}
\end{align}
Equation \eqref{derivative_hamiltonian} comes from differentiating the interpolating Hamiltonian \eqref{interpolating_hamiltonian}.
Evaluating \eqref{derivative_hamiltonian} at
$(\bs, \by, \widetilde{\by}) = (\bS, \bY^{(t)},\widetilde{\bY}^{(t,\epsilon)})$ yields:
\begin{equation}\label{eq:H_timederivative}
\cH'_{t,\epsilon}\big(\bS; \bY^{(t)},\widetilde{\bY}^{(t,\epsilon)} , \bW \big) =
\sum_{\bi \in \cI} \frac{1}{2n}\sqrt{\frac{\lambda}{1-t}}\, Z_{\bi} X_{i_1} X_{i_2} X_{i_3}
-
\sum_{j=1}^{n} \frac{R'(t,\epsilon)}{2}\sqrt{\frac{\lambda}{2 R(t,\epsilon)}}\,\widetilde{Z}_j X_j \;.
\end{equation}
The second expectation on the right-hand side of \eqref{f'_sum_2_expectations} is now easily shown to be zero thanks to the Nishimori identity:
$
\E\,\big\langle \cH'_{t,\epsilon}\big(\bs; \bY^{(t)},\widetilde{\bY}^{(t,\epsilon)}, \bW\big)\big\rangle_{\!t,\epsilon}
= \E\,\cH'_{t,\epsilon}\big(\bS; \bY^{(t)},\widetilde{\bY}^{(t,\epsilon)}, \bW\big)
=0$.
Therefore, the identity \eqref{f'_sum_2_expectations} simplifies to:
\begin{multline}\label{eq:non_final_formula_derivative}
f'_{n}(t,\epsilon)
= -\frac{1}{2n^2}\sqrt{\frac{\lambda}{1-t}}\sum_{\bi \in \cI} \E[Z_{\bi} X_{i_1} X_{i_2} X_{i_3}\ln \cZ_{t,\epsilon}\big(\bY^{(t)},\widetilde{\bY}^{(t,\epsilon)} , \bW\big)]\\
+\frac{R'(t,\epsilon)}{2n}\sqrt{\frac{\lambda}{2 R(t,\epsilon)}}\,
\sum_{j=1}^{n} \E[\widetilde{Z}_j X_j  \ln \cZ_{t,\epsilon}\big(\bY^{(t)},\widetilde{\bY}^{(t,\epsilon)} , \bW\big)]\;.
\end{multline}
The two expectations appearing on the right-hand side of~\eqref{eq:non_final_formula_derivative} are simplified thanks to Stein's lemma, i.e., by integrating by parts w.r.t.\ the Gaussian noises:
\begin{align*}
\E[Z_{\bi} X_{i_1} X_{i_2} X_{i_3}\ln \cZ_{t,\epsilon}\big(\bY^{(t)},\widetilde{\bY}^{(t,\epsilon)} , \bW\big)]
&= \frac{\sqrt{\lambda (1-t)}}{n}\E\,\langle x_{i_1} X_{i_1} x_{i_2}X_{i_2} x_{i_3} X_{i_3} \rangle_{t,\epsilon} \;;\\
 \E[\widetilde{Z}_j X_j  \ln \cZ_{t,\epsilon}\big(\bY^{(t)},\widetilde{\bY}^{(t,\epsilon)} , \bW \big)]
&= \sqrt{\frac{ \lambda R(t,\epsilon)}{2}}\E\,\langle x_j X_j \rangle_{t,\epsilon} \;.
\end{align*}
Hence, we have
\begin{align*}
f'_{n}(t,\epsilon)
&= -\frac{\lambda}{2n^3} \sum_{\bi \in \cI} \E\,\langle x_{i_1} X_{i_1} x_{i_2}X_{i_2} x_{i_3} X_{i_3} \rangle_{t,\epsilon}
+\frac{\lambda R'(t,\epsilon)}{4 n}\,
\sum_{j=1}^{n} \E\,\langle x_j X_j \rangle_{t,\epsilon}\\
&= -\frac{\lambda}{12} \E\,\langle Q^3 \rangle_{t,\epsilon}
+ \frac{\lambda R'(t,\epsilon)}{4} \E\,\langle Q \rangle_{t,\epsilon} + \frac{\lambda}{2}\cO(n^{-1}) \;,
\end{align*}
with $\vert \cO(n^{-1}) \vert = \frac{1}{n^3} \Big\vert \sum\limits_{\bi \in \cI} \E\,\langle x_{i_1} X_{i_1} x_{i_2}X_{i_2} x_{i_3} X_{i_3} \rangle_{t,\epsilon}
- \frac{1}{6}\sum\limits_{i_1,i_2,i_3=1}^n \E\,\langle x_{i_1} X_{i_1} x_{i_2}X_{i_2} x_{i_3} X_{i_3} \rangle_{t,\epsilon} \Big\vert \leq \frac{\Vert \varphi \Vert_{\infty}^6}{n}$.
\end{IEEEproof}
\section{Concentration of the overlap}\label{app:concentration_overlap}
One important result in order to prove Propositions~\ref{prop:upperbound_mutual_info} and \ref{prop:lowerbound_mutual_info} is the concentration of the scalar overlap $Q$ around its expectation $\E \langle Q \rangle_{t,R}$ as long as we integrate over $R$ in a bounded subset of $(0,+\infty)$.
Remember that the angular brackets $\langle - \rangle_{t,R}$ denote the expectation with respect to the posterior distribution \eqref{posterior_H_t_R}.
\begin{proposition}[Concentration of the overlap around its expectation]\label{prop:concentration_overlap}
Suppose that \ref{hyp:S_bounded_support} and \ref{hyp:varphi} hold.
Let $M >0$.
For $n$ large enough, there exists a constant $C$, which depends only on $\Vert \varphi \Vert_{\infty}$, $\Vert \varphi' \Vert_{\infty}$, $\Vert \varphi'' \Vert_{\infty}$, $M_S$, $\lambda$ and $M$, such that $\forall (a,b) \in (0,M)^2: a < \min\{1,b\}$, $\forall \delta \in (0,a)$, $\forall t \in [0,1]$:
\begin{equation}\label{eq:concentration_overlap}
\int_{a}^{b} \E\,\big\langle \big(Q -\E\,\langle Q \rangle_{t,R}\,\big)^2\,\big\rangle_{t,R}\, dR
\leq C\bigg(\frac{1}{\delta^2 n} - \frac{\ln(a)}{n} + \frac{\delta}{a-\delta}\bigg)\;.
\end{equation}
\end{proposition}
The concentration of the scalar overlap around its expectation will follow from the concentration of the quantity:
\begin{equation}\label{def_L}
\cL
= \frac{1}{n}\sum_{j=1}^{n} \frac{\lambda}{4} x_j^2 - \frac{\lambda}{2}\, x_j X_j - \frac{1}{2}\sqrt{\frac{\lambda}{2 R}}\, x_j\widetilde{Z}_j \:.
\end{equation}
%
\begin{lemma}[Link between the fluctuations of \texorpdfstring{$\cL$}{L} and \texorpdfstring{$Q$}{Q}]\label{lemma:computation_E<L>_and_others}\\
Assume $\varphi: \R \to \R$ is continuous and bounded. For all $(t,R) \in [0,1] \times (0, +\infty)$:
\begin{align}
\E\,\langle \cL \rangle_{t,R} &=  -\frac{\lambda}{4} \E\,\langle Q \rangle_{t,R} \;;\label{formula_E<L>}\\
\frac{\lambda}{4}\E\,\langle (Q - \langle Q \rangle_{t,R})^2 \rangle_{t,R}
&\leq \frac{\Vert \varphi \Vert_{\infty}^2}{\sqrt{2}}\sqrt{\E\,\big\langle \big(\cL - \langle \cL \rangle_{t,R}\big)^{2}\,\big\rangle_{t,R}
	-\frac{1}{n}\E\,\bigg\langle \frac{\partial \cL }{\partial R} \bigg\rangle_{\!\! t,R}} \;\:;\label{upperbound_fluctuation_Q_<Q>}\\
\frac{\lambda^2}{16}\,\E\langle (Q - \E\,\langle Q\rangle_{t,R})^2 \rangle_{t,R}
&\leq\E\langle (\cL - \E\,\langle \cL \rangle_{t,R})^2 \rangle_{t,R}\;.\label{upperbound_fluctuation_Q_E<Q>}
\end{align}
\end{lemma}
\begin{IEEEproof}
Fix $(t,R) \in [0,1] \times (0,+\infty)$.
By the definition \eqref{def_L} of $\cL$, we have:
\begin{align}
\E\,\langle \cL \rangle_{t,R}
	&= \frac{1}{n}\sum_{j=1}^{n} \frac{\lambda}{4} \E \langle x_j^2 \rangle_{t,R} - \frac{\lambda}{2} \, \E\big[ \langle x_j \rangle_{t,R} X_j \big]
	- \frac{1}{2}\sqrt{\frac{\lambda}{2 R}}\, \E\big[ \langle x_j \rangle_{t,R} \widetilde{Z}_j \big] \;;\label{def_E<L>}\\
\E\,\langle Q \cL \rangle_{t,R}
	&= \frac{1}{n}\sum_{j=1}^{n}  \frac{\lambda}{4}\E \langle Q x_j^2 \rangle_{t,R} -  \frac{\lambda}{2} \E\big[ \langle Q x_j \rangle_{t,R} X_j \big] 	- \frac{1}{2}\sqrt{\frac{\lambda}{2 R}}\, \E\big[ \langle Q x_j \rangle_{t,R} \widetilde{Z}_j \big]\,.\label{def_E<Q L>}
\end{align}
Integrating by parts with respect to the Gaussian random variable $\widetilde{Z}_j$, the last expectation on the right-hand side of each of \eqref{def_E<L>} and \eqref{def_E<Q L>} reads:
\begin{align}
\E\big[ \langle x_j \rangle_{t,R} \widetilde{Z}_j \big]
	&= \sqrt{\frac{\lambda R}{2}}\E\big[ \langle x_j^2 \rangle_{t,R} \big]
	-\sqrt{\frac{\lambda R}{2}} \E\big[ \langle x_j \rangle_{t,R}^2\big] \;;\label{stein_lemma_last_term_E<L>}\\
\E\big[ \langle Q x_j \rangle_{t,R} \widetilde{Z}_j \big]
	&= \sqrt{\frac{\lambda R}{2}} \E\big[ \langle Q x_j^2 \rangle_{t,R} \big]
	-\sqrt{\frac{\lambda R}{2}} \E\big[ \langle Q x_j \rangle_{t,R} \langle x_j \rangle_{t,R} \big] \;.\label{stein_lemma_last_term_E<Q L>}
\end{align}
Plugging \eqref{stein_lemma_last_term_E<L>} in \eqref{def_E<L>} yields:
\begin{equation*}
	\E\,\langle \cL \rangle_{t,R}
= \frac{\lambda}{2n}\sum_{j=1}^{n} \frac{1}{2} \E\big[ \langle x_j \rangle_{t,R}^2\big] - \, \E\big[ \langle x_j \rangle_{t,R} X_j \big]
= -\frac{\lambda}{4n} \sum_{j=1}^{n} \frac{\E\big[ \langle x_j \rangle_{t,R} X_j \big]}{n}
= -\frac{\lambda}{4} \E\,\langle Q \rangle_{t,R} \;,
\end{equation*}
where the second equality follows from Nishimori identity: $\E[ \langle x_j \rangle_{t,R}^2] = \E[ \langle x_j \rangle_{t,R} X_j]$.
This ends the proof of \eqref{formula_E<L>}.
Plugging \eqref{stein_lemma_last_term_E<Q L>} in \eqref{def_E<Q L>}, it comes:
\begin{multline}
\E\,\langle Q \cL \rangle_{t,R}
= \frac{\lambda}{2n}\sum_{j=1}^{n_v} \frac{1}{2}\E\big[ \langle Q x_j \rangle_{t,R} \langle x_j \rangle_{t,R} \big] -  \E\big[ \langle Q x_j \rangle_{t,R} X_j \big]\\
= \frac{\lambda}{2n}\sum_{j=1}^{n_v} \frac{1}{2}\E\big[ \langle Q \rangle_{t,R} \langle x_j X_j\rangle_{t,R} \big] -  \E\big[ \langle Q x_j \rangle_{t,R} X_j \big]
= \frac{\lambda}{2}\bigg(\frac{1}{2}\E\big[ \langle Q \rangle_{t,R}^2\big] -  \E\big[ \langle Q^2 \rangle_{t,R}\big]\bigg) \;.\label{formula_E<QL>}
\end{multline}
The second equality follows once again from Nishimori identity.
Combining \eqref{formula_E<QL>} and \eqref{formula_E<L>} yields:
\begin{align*}
\E\,\langle (Q - \E\,\langle Q \rangle_{t,R}) (\cL - \E\,\langle \cL \rangle_{t,R}) \rangle_{t,R}
&= \E\,\langle Q \cL \rangle_{t,R} - \E\,\langle Q \rangle_{t,R} \E\,\langle \cL \rangle_{t,R}\\
&= \frac{\lambda}{4} \Big(\E\big[ \langle Q \rangle_{t,R}^2\big] - 2\E\big[ \langle Q^2 \rangle_{t,R}\big] + (\E\,\langle Q \rangle_{t,R})^2\Big)\\
&= -\frac{\lambda}{4} \Big(\E\,\big\langle (Q - \langle Q \rangle_{t,R})^2 \big\rangle_{t,R} + \E\,\langle (Q - \E\,\langle Q\rangle_{t,R})^2 \rangle_{t,R}\Big) \;.
\end{align*}
The upper bound \eqref{upperbound_fluctuation_Q_E<Q>} on the fluctuation of $Q$ simply follows from this last identity:
\begin{align*}
\frac{\lambda}{4} \E\,\big\langle (Q - \E\,\langle Q\rangle_{t,R})^2 \big\rangle_{t,R}
&\leq
-\E\,\langle (Q - \E\,\langle Q \rangle_{t,R}) (\cL - \E\,\langle \cL \rangle_{t,R}) \rangle_{t,R}\\
&\leq \sqrt{\E\,\langle (Q - \E\,\langle Q \rangle_{t,R})^2 \rangle_{t, R} \,\E\,\langle (\cL - \E\,\langle \cL \rangle_{t,R})^2 \rangle_{t,R}}\;.
\end{align*}
The second inequality is a simple application of Cauchy-Schwarz inequality.

The proof of the inequality \eqref{upperbound_fluctuation_Q_<Q>} is more involved.
These two identities will be useful (just replace $Q$ by its definition):
\begin{align}
\E\,\langle (Q - \langle Q \rangle_{t,R})^2 \rangle_{t,R}
&= \frac{1}{n^2}\sum_{i,j=1}^n \E\big[X_iX_j(\langle x_i x_j \rangle_{t, R} - \langle x_i \rangle_{t, R} \langle x_j \rangle_{t, R})\big]
\label{expression_E(Q-<Q>)^2}\\
\E\bigg[\!\bigg(\langle Q \rangle_{t,R} - \bigg\Vert \frac{\langle \bx \rangle_{t,R}}{\sqrt{n}} \bigg\Vert^2\,\bigg)^{\!\! 2}\,\bigg]
&= \frac{1}{n^2}\sum_{i,j=1}^n \E\big[X_iX_j \langle x_i \rangle_{t, R}\langle x_j \rangle_{t, R}\big] - 2 \E\big[X_i \langle x_i \rangle_{t, R} \langle x_j \rangle_{t, R}^2\big]\nonumber\\
&\qquad\qquad\qquad\qquad\qquad\qquad\quad\;\;\,
+ \E\big[\langle x_i \rangle_{t, R}^2 \langle x_j \rangle_{t, R}^2\big]\nonumber\\
&= \frac{1}{n^2}\sum_{i,j=1}^n \E\big[X_iX_j \langle x_i \rangle_{t, R}\langle x_j \rangle_{t, R}\big] - \E\big[\langle x_i \rangle_{t, R}^2 \langle x_j \rangle_{t, R}^2\big]\;.\label{expression_E(Q-<x><x>/n)^2}
\end{align}
Differentiating with respect to $R$ on both side of the identity \eqref{formula_E<L>} yields:
\begin{align}
-n\E\,\big\langle \big(\cL - \langle \cL \rangle_{t,R}\big)^{2}\,\big\rangle_{t,R}
+\E\,\bigg\langle \frac{\partial \cL }{\partial R} \bigg\rangle_{\!\! t,R}
&= \frac{n\lambda}{4}(\E\,\big\langle Q \cL \rangle_{t,R}
- \E\,\langle Q \rangle_{t,R}\langle \cL \rangle_{t,R})\nonumber\\
\Leftrightarrow\qquad
-\frac{\lambda}{4}\Big(\E\,\big\langle Q \cL \rangle_{t,R}
- \E\,\langle Q \rangle_{t,R}\langle \cL \rangle_{t,R}\Big)
&= \E\,\big\langle \big(\cL - \langle \cL \rangle_{t,R}\big)^{2}\,\big\rangle_{t,R}
-\frac{1}{n}\E\,\bigg\langle \frac{\partial \cL }{\partial R} \bigg\rangle_{\!\! t,R} \;.\label{identity_E<QL>-E<Q><L>}
\end{align}
Next, we simplify the left-hand side of \eqref{identity_E<QL>-E<Q><L>}. By definition, we have:
\begin{multline}
\E\,\langle Q \rangle_{t,R}\langle \cL \rangle_{t,R}\\
= \frac{1}{n}\sum_{j=1}^{n}  \frac{\lambda}{4}\E\big[\langle Q \rangle_{t,R} \langle x_j^2 \rangle_{t,R}\big] -  \frac{\lambda}{2} \E\big[ \langle Q \rangle_{t,R} \langle x_j \rangle_{t,R} X_j \big] 	- \frac{1}{2}\sqrt{\frac{\lambda}{2 R}}\, \E\big[ \langle Q \rangle_{t,R} \langle x_j \rangle_{t,R} \widetilde{Z}_j \big]\;.\label{def_E<Q><L>}
\end{multline}
After an integration by parts with respect to $\widetilde{Z}_j$ the third expectation in the summand of \eqref{def_E<Q><L>} reads:
\begin{align*}
&\E\big[ \langle Q \rangle_{t,R} \langle x_j \rangle_{t,R} \widetilde{Z}_j \big]\\
&\qquad\qquad=
\sqrt{\frac{\lambda R}{2}}\E\big[ \langle Q x_j \rangle_{t,R} \langle x_j \rangle_{t,R} \big]
+ \sqrt{\frac{\lambda R}{2}}\E\big[ \langle Q \rangle_{t,R} \langle x_j^2 \rangle_{t,R} \big]
-2 \sqrt{\frac{\lambda R}{2}}\E\big[ \langle Q \rangle_{t,R} \langle x_j \rangle_{t,R}^2 \big]\\
&\qquad\qquad=
\sqrt{\frac{\lambda R}{2}}\E\big[ \langle Q \rangle_{t,R} \langle x_j X_j \rangle_{t,R} \big]
+ \sqrt{\frac{\lambda R}{2}}\E\big[ \langle Q \rangle_{t,R} \langle x_j^2 \rangle_{t,R} \big]
-2 \sqrt{\frac{\lambda R}{2}}\E\big[ \langle Q \rangle_{t,R} \langle x_j \rangle_{t,R}^2 \big]\;.
\end{align*}
Plugging this result back in \eqref{def_E<Q><L>} gives:
\begin{align}
\E\langle Q \rangle_{t,R}\langle \cL \rangle_{t,R}
&= \frac{\lambda}{2n}\sum_{j=1}^{n} \E\big[ \langle Q \rangle_{t,R} \langle x_j \rangle_{t,R}^2 \big]
- \frac{3}{2}\E\big[ \langle Q \rangle_{t,R} \langle x_j X_j \rangle_{t,R} \big]\nonumber\\
&= \frac{\lambda}{2}\E\bigg[ \langle Q \rangle_{t,R} \bigg\Vert \frac{\langle \bx \rangle_{t,R}}{\sqrt{n}} \bigg\Vert^2 \bigg]
- \frac{3\lambda}{4}\E\big[ \langle Q \rangle_{t,R}^2 \big]\;.\label{formula_E<Q><L>}
\end{align}
Finally, combining \eqref{formula_E<QL>} and \eqref{formula_E<Q><L>} yields the following expression for the left-hand side of \eqref{identity_E<QL>-E<Q><L>}:
\begin{align}
&-\frac{\lambda}{4}\Big(\E\,\big\langle Q \cL \rangle_{t,R}
- \E\,\langle Q \rangle_{t,R}\langle \cL \rangle_{t,R}\Big)\nonumber\\
&\qquad\qquad\qquad= \frac{\lambda^2}{8} \bigg(\E\big[ \langle Q^2 \rangle_{t,R}\big]
-\E\big[ \langle Q \rangle_{t,R}^2\big]
+ \E\bigg[ \langle Q \rangle_{t,R} \bigg\Vert \frac{\langle \bx \rangle_{t,R}}{\sqrt{n}} \bigg\Vert^2 \,\bigg]
-\E\big[ \langle Q \rangle_{t,R}^2\big]\bigg)\nonumber\\
&\qquad\qquad\qquad= \frac{\lambda^2}{8} \Bigg(\E\big[ \langle (Q-\langle Q \rangle_{t,R})^2 \rangle_{t,R}\big]
-\E\bigg[ \bigg(\langle Q \rangle_{t,R}- \bigg\Vert \frac{\langle \bx \rangle_{t,R}}{\sqrt{n}} \bigg\Vert^2\,\bigg)^{\!\! 2}\, \bigg]\Bigg)\nonumber\\
&\qquad\qquad\qquad= \frac{\lambda^2}{8n^2} \sum_{i,j=1}^n
 \E\big[X_iX_j\langle x_i x_j \rangle_{t, R}\big]
-2\E\big[X_iX_j \langle x_i \rangle_{t, R}\langle x_j \rangle_{t, R}\big] +\E\big[\langle x_i \rangle_{t, R}^2 \langle x_j \rangle_{t, R}^2\big]\nonumber\\
&\qquad\qquad\qquad= \frac{\lambda^2}{8n^2} \sum_{i,j=1}^n
\E\big[\big(\langle x_i x_j \rangle_{t, R} - \langle x_i \rangle_{t, R}\langle x_j \rangle_{t, R}\big)^2\big]\label{formula_E<QL>-E<Q><L>}\;.
\end{align}
The second-to-last equality follows from \eqref{expression_E(Q-<Q>)^2} and \eqref{expression_E(Q-<x><x>/n)^2}, while the factorization in the last equality is easily obtained after applying Nishimori identity: $\E[X_iX_j \langle x_i x_j \rangle_{t, R}] = \E\langle x_i x_j \rangle_{t, R}^2$ and
$\E[X_iX_j \langle x_i \rangle_{t, R}\langle x_j \rangle_{t, R}] = \E[\langle x_i x_j \rangle_{t, R}\langle x_i \rangle_{t, R}\langle x_j \rangle_{t, R}]$.
We now come back to the identity \eqref{expression_E(Q-<Q>)^2} and apply Jensen's inequality to its right-hand side to get:
\begin{multline}\label{upperbound_fluctuations_Q_<Q>_proof}
\E\,\langle (Q - \langle Q \rangle_{t,R})^2 \rangle_{t,R}
\leq 
\frac{\Vert \varphi \Vert_{\infty}^2}{n^2}\sum_{i,j=1}^n\E\big[\big\vert \langle x_i x_j \rangle_{t, R} - \langle x_i \rangle_{t, R} \langle x_j \rangle_{t, R}\big\vert\big]\\
\leq \Vert \varphi \Vert_{\infty}^2\sqrt{
\frac{1}{n^2}\sum_{i,j=1}^n\E\big[\big( \langle x_i x_j \rangle_{t, R} - \langle x_i \rangle_{t, R} \langle x_j \rangle_{t, R}\big)^2\big]}\;.
\end{multline}
Combining \eqref{identity_E<QL>-E<Q><L>}, \eqref{formula_E<QL>-E<Q><L>} and \eqref{upperbound_fluctuations_Q_<Q>_proof} yields the inequality \eqref{upperbound_fluctuation_Q_<Q>}.
\end{IEEEproof}
\subsection{Concentration of \texorpdfstring{$\cL$}{L} around its expectation}
To prove concentration results on $\cL$, it will be useful to work with the free entropy $\frac{1}{n} \ln \cZ_{t,R}(\bY^{(t)},\widetilde{\bY}^{(t,R)}, \bW)$ where $\cZ_{t,R}(\bY^{(t)},\widetilde{\bY}^{(t,R)}, \bW)$ is the normalization factor of the Gibbs posterior distribution \eqref{posterior_H_t_R}.
In Appendix \ref{app:concentration_free_entropy}, we prove that this free entropy concentrates around its expectation when $n \to +\infty$. 
In order to shorten notations, we define:
\begin{equation*}
F_n(t,R) \triangleq \frac{1}{n} \ln \cZ_{t,R}\big(\bY^{(t)},\widetilde{\bY}^{(t,R)}, \bW \big)\,;\:
f_n(t,R) \triangleq \frac{1}{n} \E\big[\ln \cZ_{t,R}\big(\bY^{(t)},\widetilde{\bY}^{(t,R)}, \bW \big)\big] = \E\,F_n(t,R) \,.
\end{equation*}
\begin{proposition}[Thermal fluctuations of $\cL$ and $Q$]\label{prop:concentration_L_on_<L>}
Assume $\varphi: \R \to \R$ is continuous and bounded.
For all positive real numbers $a < b$ and $t \in [0,1]$, we have:
\begin{align*}
	\int_a^b \E\,\big\langle \big(\cL - \langle \cL \rangle_{t,R}\big)^{2}\,\big\rangle_{t,R}\,dR
&\;\leq\; \frac{\lambda \Vert \varphi \Vert_{\infty}^2}{4n} \bigg(\frac{\ln(b/a)}{2} + 1 \bigg)\;;\\
\frac{\lambda}{4}\,\int_a^b \E\,\bigg\langle \!\bigg( Q - \bigg\Vert \frac{\langle \bx \rangle_{t,R}}{\sqrt{n}} \bigg\Vert^2\,\bigg)^{\!\! 2} \,\bigg\rangle_{\!\! t,R}\,dR
&\;\leq\; \Vert \varphi \Vert_{\infty}^3\sqrt{\frac{\lambda (b-a)}{2n}}\;.
\end{align*}
\end{proposition}
\begin{IEEEproof}
Fix $(n,t) \in \mathbb{N}^* \times [0,1]$.
Note that $\forall R \in (0, +\infty)$:
\begin{equation}\label{first_derivative_fn}
\frac{\partial f_n}{\partial R}\bigg\vert_{t,R}
= -\frac{1}{n}\E\Bigg[\Bigg\langle \frac{\partial \cH_{t,R}(\bx;\bY^{(t)},\widetilde{\bY}^{(t,R)}, \bW)}{\partial R} \Bigg\rangle_{\!\! t,R}\,\Bigg]
=-\E\,\langle \cL \rangle_{t,R} \;.
\end{equation}
Further differentiating, we obtain:
\begin{align}
\frac{\partial^2 f_n}{\partial R^2}\bigg\vert_{t,R}
	&= \E\bigg[\bigg\langle \cL \,\frac{\partial \cH_{t,R}}{\partial R} \bigg\rangle_{\!\! t,R}\,\bigg]
	-\E\bigg[\langle \cL \rangle_{t,R}\,
	\bigg\langle \frac{\partial \cH_{t,R}}{\partial R} \bigg\rangle_{\!\! t,R}\,\bigg]
	-\E\,\bigg\langle \frac{\partial \cL }{\partial R} \bigg\rangle_{\!\! t,R}\nonumber\\
	&=n\E\,\big\langle \big(\cL - \langle \cL \rangle_{t,R}\big)^{2}\,\big\rangle_{t,R}
	-\E\,\bigg\langle \frac{\partial \cL }{\partial R} \bigg\rangle_{\!\! t,R}\;.\label{second_derivative_f_n_R}
\end{align}
It follows directly from \eqref{second_derivative_f_n_R} and the definition \eqref{def_L} of $\cL$ that:
\begin{equation}\label{thermal_fluctuation_L}
\E\,\big\langle \big(\cL - \langle \cL \rangle_{t,R}\big)^{2}\,\big\rangle_{t,R}
	= \frac{1}{n}\frac{\partial^2 f_n}{\partial R^2}\bigg\vert_{t,R}
	+ \frac{1}{4 R}\sqrt{\frac{\lambda}{2R}}\frac{\E\big[\langle \bx \rangle_{t,R}^T \widetilde{\bZ}\,\big]}{n^2}
\end{equation}
We start with upper bounding the integral over the second summand on the right-hand side of \eqref{thermal_fluctuation_L}.
Thanks to an integration by parts with respect to $\widetilde{Z}_j$, $j \in \{1,\dots,n_u\}$, it comes:
\begin{equation}
\frac{1}{4 R}\sqrt{\frac{\lambda}{2R}}\frac{\E\big[\langle \bx \rangle_{t,R}^T \widetilde{\bZ}\,\big]}{n^2}
= \frac{\lambda}{8 R}\frac{\E\big[\langle \Vert \bx \Vert^2 \rangle_{t,R} - \Vert \langle \bx \rangle_{t,R} \Vert^2\,\big]}{n^2}
\leq \frac{\lambda \Vert \varphi \Vert_{\infty}^2}{8 R n}\:.
\end{equation}
Therefore:
\begin{equation}\label{upperbound_int_dL/dR}
\int_a^b \frac{dR}{4 R}\sqrt{\frac{\lambda}{2R}}\frac{\E\big[\langle \bx \rangle_{t,R}^T \widetilde{\bZ}\,\big]}{n^2}
\leq \frac{\lambda \Vert \varphi \Vert_{\infty}^2}{8 n} \ln(b/a) \:.
\end{equation}
It remains to upper bound $\int_a^b \frac{dR}{n}\frac{\partial^2 f_n}{\partial R^2}\big\vert_{t,R}
		= \frac{1}{n}\frac{\partial f_n}{\partial R}\big\vert_{t,R=b} - \frac{1}{n}\frac{\partial f_n}{\partial R}\big\vert_{t,R=a}$.
Note that $\forall R \in (0,+\infty)$:
\begin{equation}\label{upperbound_derivative_fn_R}
0 \leq \frac{\partial f_n}{\partial R}\bigg\vert_{t,R}
= - \E\,\langle \cL \rangle_{t,R}
= \frac{\lambda}{4} \E\,\langle Q \rangle_{t,R}
= \frac{\lambda}{4n} \E\,\Vert \langle \bx \rangle_{t,R} \Vert^2
\leq \frac{\lambda}{4}\Vert \varphi \Vert_{\infty}^2 \:,
\end{equation}
where the first equality follows from \eqref{first_derivative_fn}, the second one from \eqref{formula_E<L>} in Lemma~\ref{lemma:computation_E<L>_and_others}, and the third one from Nishimori identity.
Combining both \eqref{upperbound_int_dL/dR} and \eqref{upperbound_derivative_fn_R}, we finally get the first inequality:
\begin{equation*}
\int_a^b \E\,\big\langle \big(\cL - \langle \cL \rangle_{t,R}\big)^{2}\,\big\rangle_{t,R}\,dR
\leq \frac{\lambda \Vert \varphi \Vert_{\infty}^2}{4n} \bigg(\frac{\ln(b/a)}{2} + 1 \bigg)\;.
\end{equation*}
To prove the second inequality, we first integrate both sides of the inequality \eqref{upperbound_fluctuation_Q_<Q>} with respect to $R$ and then use Cauchy-Schwarz inequality. We obtain:
\begin{multline}
\frac{\lambda}{4}\int_a^b \E\,\langle (Q - \langle Q \rangle_{t,R})^2 \rangle_{t,R}\,dR\\
\leq \Vert \varphi \Vert_{\infty}^2\sqrt{\frac{b-a}{2}\int_a^b \bigg(\E\,\big\langle \big(\cL - \langle \cL \rangle_{t,R}\big)^{2}\,\big\rangle_{t,R}
	-\frac{1}{n}\E\,\bigg\langle \frac{\partial \cL }{\partial R} \bigg\rangle_{\!\! t,R}\,\bigg)dR}\\
= \Vert \varphi \Vert_{\infty}^2\sqrt{\frac{b-a}{2}\int_a^b \frac{dR}{n}\frac{\partial^2 f_n}{\partial R^2}\bigg\vert_{t,R}}
\leq \Vert \varphi \Vert_{\infty}^3\sqrt{\frac{\lambda (b-a)}{8n}}\;.\label{upperbound_integral_E<(Q_<Q>)^2>}
\end{multline}
Finally, note that:
\begin{align}
\E\,\bigg\langle \bigg( Q - \bigg\Vert \frac{\langle \bx \rangle_{t,R}}{\sqrt{n}} \bigg\Vert^2\bigg)^{\!\! 2} \bigg\rangle_{\!\! t,R}
&= \E\,\big\langle \big( Q - \langle Q \rangle_{t,R}\big)^2\big\rangle_{t,R}
+ \E \bigg[\bigg( \langle Q \rangle_{t,R} - \bigg\Vert \frac{\langle \bx \rangle_{t,R}}{\sqrt{n}} \bigg\Vert^2\bigg)^{\!\! 2} \bigg]\nonumber\\
&= \E\,\big\langle \big( Q - \langle Q \rangle_{t,R}\big)^2\big\rangle_{t,R}
+ \E \bigg[\bigg\langle Q  - \frac{\langle \bx \rangle_{t,R}\bx}{n} \bigg\rangle_{\!\! t,R}^{\!\! 2}\,\bigg]\nonumber\\
&\leq \E\,\big\langle \big( Q - \langle Q \rangle_{t,R}\big)^2\big\rangle_{t,R}
+ \E \bigg[\bigg\langle \bigg(Q  - \frac{\langle \bx \rangle_{t,R}\bx}{n}\bigg)^{\!\! 2} \bigg\rangle_{\!\! t,R}\,\bigg]\label{proof1_upperbound_fluctuation_Q_<x><x>/n}\\
&= \E\,\big\langle \big( Q - \langle Q \rangle_{t,R}\big)^2\big\rangle_{t,R}
+ \E \bigg[\bigg\langle \bigg(Q  - \frac{\langle \bx \rangle_{t,R}\bX}{n}\bigg)^{\!\! 2} \bigg\rangle_{\!\! t,R}\,\bigg]\label{proof2_upperbound_fluctuation_Q_<x><x>/n}\\
&= 2 \E\,\big\langle \big( Q - \langle Q \rangle_{t,R}\big)^2\big\rangle_{t,R} \;.\label{upperbound_fluctuation_Q_<x><x>/n}
\end{align}
The inequality \eqref{proof1_upperbound_fluctuation_Q_<x><x>/n} follows from Jensen's inequality, while the equality \eqref{proof2_upperbound_fluctuation_Q_<x><x>/n} is a simple application of Nishimori identity.
The inequalities \eqref{upperbound_integral_E<(Q_<Q>)^2>} and \eqref{upperbound_fluctuation_Q_<x><x>/n} together give the second inequality of the proposition.
\end{IEEEproof}

\begin{proposition}[Quenched fluctuations of $\cL$]\label{prop:concentration_<L>_on_E<L>}
Suppose that \ref{hyp:S_bounded_support} and \ref{hyp:varphi} hold.
Let $M >0$.
For $n$ large enough, there exists a constant $C$, which depends only on $\Vert \varphi \Vert_{\infty}$, $\Vert \varphi' \Vert_{\infty}$, $\Vert \varphi'' \Vert_{\infty}$, $M_S$, $\lambda$ and $M$, such that $\forall (a,b) \in (0,M)^2: a < \min\{1,b\}$, $\forall \delta \in (0,a)$, $\forall t \in [0,1]$:
\begin{equation}
	\int_{a}^{b} \E\,\big[\big(\langle \cL \rangle_{t,R}-\E\,\langle \cL \rangle_{t,R}\,\big)^2\,\big]\,dR
	\leq C\bigg(\frac{1}{\delta^2 n} - \frac{\ln(a)}{n} + \frac{\delta}{a-\delta} \bigg)\,.
\end{equation}
\end{proposition}
\begin{IEEEproof}
	Fix $(n,t) \in \mathbb{N}^* \times [0,1]$. For all $R \in (0,+\infty)$, we have:
	\begin{align}
		\frac{\partial F_n}{\partial R}\bigg\vert_{t,R}
		&=-\langle \cL \rangle_{t,R} \;;\\
		\frac{\partial^2 F_n}{\partial R^2}\bigg\vert_{t,R}
		&=n \big\langle \big(\cL - \langle \cL\rangle_{t,R}\big)^{2}\,\big\rangle_{t,R}
	- \frac{1}{4 R}\sqrt{\frac{\lambda}{2R}}\frac{\langle \bx \rangle_{t,R}^T \widetilde{\bZ}}{n}\;;\label{quenched:2ndDeriv_Fn}\\
		\frac{\partial f_n}{\partial R}\bigg\vert_{t,R}
		&=-\E\,\langle \cL \rangle_{t,R}\;;\\
		\frac{\partial^2 f_n}{\partial R^2}\bigg\vert_{t,R}
		&=n \E\,\big\langle \big(\cL - \langle \cL\rangle_{t,R}\big)^{2}\,\big\rangle_{t,R}
	- \frac{1}{4 R}\sqrt{\frac{\lambda}{2R}}\frac{\E\big[\langle \bx \rangle_{t,R}^T \widetilde{\bZ}\,\big]}{n}\,.
	\end{align}
	The second term on the right-hand side of \eqref{quenched:2ndDeriv_Fn} can be upper bounded with Cauchy-Schwarz inequality:
	\begin{equation}\label{upperbound_2ndterm_2ndDeriv_Fn}
	\Bigg\vert\frac{1}{4 R}\sqrt{\frac{\lambda}{2R}}\frac{\langle \bx \rangle_{t,R}^T \widetilde{\bZ}}{n}\Bigg\vert
	\leq \frac{1}{4 R}\sqrt{\frac{\lambda}{2R}} \frac{\Vert \langle \bx \rangle_{t,R} \Vert\, \Vert \widetilde{\bZ} \Vert}{n}
	\leq \frac{\Vert \varphi \Vert_{\infty}}{4 R}\sqrt{\frac{\lambda}{2R}} \frac{\Vert \widetilde{\bZ} \Vert}{\sqrt{n}}\;.
	\end{equation}
	We now define for all $R \in (0, +\infty)$:
	\begin{align}
		F(R) &\triangleq F_n(t,R) - \Vert \varphi \Vert_{\infty}\sqrt{\frac{\lambda R}{2}} \frac{\Vert \widetilde{\bZ} \Vert}{\sqrt{n}}\:;\\
		f(R) &\triangleq f_n(t,R) - \Vert \varphi \Vert_{\infty}\sqrt{\frac{\lambda R}{2}}\frac{\E\,\Vert \widetilde{\bZ} \Vert}{\sqrt{n}}\,.
	\end{align}
	$F$ is convex on $(0, +\infty)$ as it is twice differentiable with a nonnegative second derivative by \eqref{quenched:2ndDeriv_Fn} and \eqref{upperbound_2ndterm_2ndDeriv_Fn}.
	The same holds for $f$.
	We will apply the following standard result to these two convex functions (we refer to \cite{barbier_adaptive_2019} for a proof):
	\begin{lemma}[An upper bound for differentiable convex functions]\label{lemma:diff_convex_functions}
		Let $g$ and $G$ be two differentiable convex functions defined on an interval $I\subseteq \mathbb{R}$.
		Let $r \in I$ and $\delta > 0$ such that $r \pm \delta \in I$. Then
		\begin{equation}
		\vert G'(r) - g'(r) \vert
		\leq C_{\delta}(r) + \frac{1}{\delta}\sum_{u \in \{-\delta,0,\delta\}} \vert G(r+u) - g(r+u) \vert \,,
		\end{equation}
		where $C_{\delta}(r) = g'(r+\delta) - g'(r-\delta) \geq 0$.
	\end{lemma}
	\noindent For all $R \in (0, +\infty)$, we have:
	\begin{align}
		F(R) - f(R)
		= F_n(t,R) - f_n(t,R)
		 - \Vert \varphi \Vert_{\infty}\sqrt{\frac{\lambda R}{2}} \frac{\Vert \widetilde{\bZ} \Vert - \E\,\Vert \widetilde{\bZ} \Vert}{\sqrt{n}}\:;\label{F_minus_f}\\
		F'(R) - f'(R)
		= -\Big(\langle \cL \rangle_{t,R} - \E\,\langle \cL\rangle_{t,R}\Big)
		 - \frac{\Vert \varphi \Vert_{\infty}}{2}\sqrt{\frac{\lambda}{2R}} \frac{\Vert \widetilde{\bZ} \Vert - \E\,\Vert \widetilde{\bZ} \Vert}{\sqrt{n}} \;. \label{F'_minus_f'}
	\end{align}
	Let $C_{\delta}(r) = f'(r+\delta) - f'(r-\delta)$, which is nonnegative by convexity of $f$.
	It follows from Lemma \ref{lemma:diff_convex_functions} and the two identities \eqref{F_minus_f} and \eqref{F'_minus_f'} that $\forall R \in (0, +\infty)$, $\forall \delta \in (0,R)$:
	\begin{align*}
		\big\vert \langle \cL \rangle_{t,R} - \E\,\langle\cL \rangle_{t,R}\big\vert
		&\leq
		\frac{\Vert \varphi \Vert_{\infty}}{2}\sqrt{\frac{\lambda}{2R}} \frac{\big\vert \Vert \widetilde{\bZ} \Vert - \E\,\Vert \widetilde{\bZ} \Vert \big\vert}{\sqrt{n}}
		+ C_{\delta}(R)\\
		&\qquad\qquad\qquad\qquad\qquad\qquad
		+ \frac{1}{\delta}\sum_{x \in \{-\delta,0,\delta\}} \vert F(R+x) - f(R+x)\vert\\
		&\leq
		\Vert \varphi \Vert_{\infty}\sqrt{\frac{\lambda}{2}} \bigg(\frac{1}{2\sqrt{R}} + 3\sqrt{R}\bigg)
		\frac{\big\vert \Vert \widetilde{\bZ} \Vert - \E\,\Vert \widetilde{\bZ} \Vert \big\vert}{\sqrt{n}}
		+ C_{\delta}(R)\\
		&\qquad\qquad\qquad\qquad\qquad\qquad
		+ \frac{1}{\delta}\sum_{x \in \{-\delta,0,\delta\}} \vert F_n(t,R + x) - f_n(t,(R + x)\vert \;.
	\end{align*}
	Thanks to the inequality $(\sum_{i=1}^{m} v_i)^2 \leq m \sum_{i=1}^{m} v_i^2$, this directly implies $\forall R \in (0, +\infty)$, $\forall \delta \in (0,R)$:
	\begin{multline}\label{upperbound_variance_thermal_L}
		\E\big[\big( \langle \cL \rangle_{t,R} - \E\,\langle\cL \rangle_{t,R}\big)^{2}\,\big]
		\leq 		5 \Vert \varphi \Vert_{\infty}^2\frac{\lambda}{2}
		\bigg(\frac{1}{4 R} + 3 + 9 R\bigg) \, \frac{\Var \Vert \widetilde{\bZ} \Vert}{n}
		+ 5C_{\delta}(R)^2\\
		+ \frac{5}{\delta^2}\sum_{x \in \{-\delta,0,\delta\}} \E\big[\big(F_n(t,R + x) - f_n(t,R + x)\big)^2\big]\,.
	\end{multline}
	The next step is to bound the integral of the three summands on the right-hand side of \eqref{upperbound_variance_thermal_L}.
	By \cite[Theorem 3.1.1]{vershynin_2018}, there exists $C_1$ such that $\Var\,\Vert\widetilde{\bZ}\Vert \leq C_1$ independently of the dimension $n$. Then:
	\begin{equation}\label{upperbound_1st_summand}
	\int_{a}^{b} dR \, 5 \Vert \varphi \Vert_{\infty}^2\frac{\lambda}{2}
	\bigg(\frac{1}{4 R} + 3 + 9 R\bigg) \, \frac{\Var \Vert \widetilde{\bZ} \Vert}{n}
	\leq
	5 \Vert \varphi \Vert_{\infty}^2\frac{\lambda}{2}
	\bigg(\frac{\ln(b/a)}{4} + 3b + \frac{9}{2} b^2 \bigg) \, \frac{C_1}{n}
	\,.
	\end{equation}
	Note that $C_{\delta}(R) = \vert C_{\delta}(R)\vert \leq \vert f'(R+\delta) \vert + \vert f'(R-\delta) \vert$.
	For all $R \in (0, +\infty)$, we have:
	\begin{equation}\label{upperbound_f'}
		\vert f'(R) \vert
		\leq
		\big\vert \E\,\langle \cL \rangle_{t,R} \big\vert
		+ \frac{\Vert \varphi \Vert_{\infty}}{2}\sqrt{\frac{\lambda}{2R}} \frac{\E\,\Vert \widetilde{\bZ} \Vert}{\sqrt{n}}
		\leq \frac{\Vert \varphi \Vert_{\infty}}{2}\sqrt{\frac{\lambda}{2}} 
		\bigg( \sqrt{\frac{\lambda}{2}}\,\Vert \varphi \Vert_{\infty}  + \frac{1}{\sqrt{R}}\bigg)\;,
	\end{equation}
	The second inequality in \eqref{upperbound_f'} follows from
	the upper bounds
	$\vert \E\,\langle \cL \rangle_{t,R} \vert \leq \nicefrac{\lambda \Vert \varphi \Vert_{\infty}^2}{4}$ (see \eqref{upperbound_derivative_fn_R}) and
	$\E \Vert\widetilde{\bZ}\Vert \leq \sqrt{n}$.
	Thus, for the second summand, we obtain $\forall \delta \in (0,a)$:
	\begin{align}\label{upperbound_2nd_summand}
		&\int_a^b dR \, C_{\delta}(R)^2\nonumber\\
		&\qquad\quad\leq
		\Vert \varphi \Vert_{\infty}\sqrt{\frac{\lambda}{2}} 
		\bigg( \sqrt{\frac{\lambda}{2}}\,\Vert \varphi \Vert_{\infty}  + \frac{1}{\sqrt{a - \delta}}\bigg)
		\int_a^b dR \, C_{\delta}(R)\nonumber\\
		&\qquad\quad= \Vert \varphi \Vert_{\infty}\sqrt{\frac{\lambda}{2}} 
		\bigg( \sqrt{\frac{\lambda}{2}}\,\Vert \varphi \Vert_{\infty}  + \frac{1}{\sqrt{a-\delta}}\bigg)
		\big[\big(f(b+\delta) - f(b-\delta)\big) - \big(f(a + \delta) - f(a -\delta)\big)\big]\nonumber\\
		&\qquad\quad= \lambda \Vert \varphi \Vert_{\infty}^2\delta
\bigg( \sqrt{\frac{\lambda}{2}}\,\Vert \varphi \Vert_{\infty}  + \frac{1}{\sqrt{a-\delta}}\bigg)^{\! 2}\;.
	\end{align}
	The last inequality is a simple application of the mean value theorem.
	We finally turn to the third summand.
	By Proposition~\ref{prop:concentration_free_entropy} of Appendix \ref{app:concentration_free_entropy}, there exists a positive constant $C_2$ depending only on $a$, $b$, $\Vert \varphi \Vert_{\infty}$, $\Vert \varphi' \Vert_{\infty}$, $\Vert \varphi'' \Vert_{\infty}$, $M_S$ and $\lambda$ such that $\forall(t,R) \in [0,1] \times (0,b+a)$:
	\begin{equation}\label{upperbound_variance_free_entropy}
	\E\big[\big(F_n(t,R) - f_n(t,R)\big)^2 \,\big] \leq \frac{C_2}{n}\,.
	\end{equation}
	Using \eqref{upperbound_variance_free_entropy}, we see that the third summand satisfies $\forall \delta \in (0,a)$:
	\begin{equation}\label{upperbound_3rd_summand}
		\int_{a}^{b} \! dR \,
		\frac{5}{\delta^2}\sum_{x \in \{-\delta,0,\delta\}} \E\big[\big(F_n(t,R+x) - f_n(t,R+x)\big)^2 \,\big]
		\leq \frac{15C_2}{\delta^2 n} b \;.
	\end{equation}
	To end the proof it remains to integrate \eqref{upperbound_variance_thermal_L} over $R \in [a,b]$ and use the three upper bounds \eqref{upperbound_1st_summand}, \eqref{upperbound_2nd_summand} and \eqref{upperbound_3rd_summand}.
\end{IEEEproof}

\subsection{Concentration of \texorpdfstring{$Q$}{Q} around its expectation: proof of Proposition~\ref{prop:concentration_overlap}}
Using the upper bound \eqref{upperbound_fluctuation_Q_E<Q>}, it directly comes:
\begin{equation}
\frac{\lambda^2}{16}\int_{a}^{b} \E\,\big\langle \big(Q-\E\,\langle Q \rangle_{t,R}\,\big)^2\,\big\rangle_{t,R}\,dR
\leq \int_a^b \E\,\langle (\cL - \E\,\langle \cL \rangle_{t,R})^2 \rangle_{t,R} \,dR \;.
\end{equation}
We then use the concentration results for $\cL$, that is, Propositions~\ref{prop:concentration_L_on_<L>} and~\ref{prop:concentration_<L>_on_E<L>}, to upper bound
$$
\int_a^b \E\,\langle (\cL - \E\,\langle \cL \rangle_{t,R})^2 \rangle_{t,R} \,dR
= \int_a^b \E\,\langle (\cL - \langle \cL \rangle_{t,R})^2 \rangle_{t,R} \, dR + \int_a^b \E[(\langle \cL \rangle_{t,R} - \E\,\langle \cL \rangle_{t,R})^2\,]\,dR
$$
and prove Proposition~\ref{prop:concentration_overlap}.\hfill\IEEEQED

\section{Concentration of the free entropy}\label{app:concentration_free_entropy}
Consider the inference problem~\eqref{interpolation_model_R}.
Once both observations $\bY^{(t)}$ and $\widetilde{\bY}^{(t,R)}$ have been replaced by their definitions, the associated Hamiltonian reads:
\begin{multline}
\cH_{t,R}(\bs ; \bS, \bZ, \widetilde{\bZ}, \bW)
\triangleq
\sum_{j=1}^{n} \frac{\lambda R}{4} x_j^2 - \frac{\lambda R}{2}\,X_j x_j - \sqrt{\frac{\lambda R}{2}}\,\widetilde{Z}_j x_j \\
+\sum_{\bi \in \cI}
\frac{\lambda(1-t)}{2n^2} x_{i_1}^2 x_{i_2}^2 x_{i_3}^2
- \frac{\lambda(1-t)}{n^2}\, X_{i_1} X_{i_2} X_{i_3} x_{i_1} x_{i_2} x_{i_3}
- \frac{\sqrt{\lambda(1-t)}}{n}\, Z_{\bi} x_{i_1} x_{i_2} x_{i_3} \;.
\end{multline}
In this section, we show that the free entropy
\begin{equation}
\frac{1}{n} \ln \cZ_{t,R}\big(\bY^{(t)},\widetilde{\bY}^{(t,R)}, \bW\big)
= \frac{1}{n} \ln\int dP_s(\bs) \, e^{-\cH_{t,R}(\bs ; \bS, \bZ, \widetilde{\bZ}, \bW)} \;.
\end{equation}
concentrates around its expectation.
We will sometimes write $\frac{1}{n} \ln \cZ_{t,R}$, omitting the arguments, to shorten notations.
\begin{proposition}[Concentration of the free entropy]\label{prop:concentration_free_entropy}
Suppose that \ref{hyp:S_bounded_support}, \ref{hyp:varphi} hold.
There exists a polynomial $C(\Vert \varphi \Vert_{\infty}, \Vert \varphi' \Vert_{\infty}, \Vert \varphi'' \Vert_{\infty}, M_S, \lambda, R)$ with positive coefficients such that $\forall t \in [0,1]$:
\begin{equation}\label{bound_variance_free_entropy}
\E \Bigg[\Bigg(\frac{ \ln \cZ_{t,R}}{n}
	- \E\bigg[\frac{\ln \cZ_{t,R}}{n} \bigg]
	\Bigg)^{\!\! 2}\:\Bigg]
	\leq \frac{C(\Vert \varphi \Vert_{\infty}, \Vert \varphi' \Vert_{\infty}, \Vert \varphi'' \Vert_{\infty}, M_S, \lambda, R)}{n} \;.
\end{equation}
\end{proposition}
\begin{IEEEproof}
First, we show that the free entropy concentrates on its conditional expectation given $(\bW, \bS)$.
Thus, $\nicefrac{\ln  \cZ_{t,R}}{n}$ is seen as a function of the Gaussian random variables $\bZ$, $\widetilde{\bZ}$ and we work conditionally to $(\bW, \bS)$:
$g(\bZ, \widetilde{\bZ}) \equiv \nicefrac{\ln  \cZ_{t,R}}{n}$.
By the Gaussian-Poincar\'{e} inequality (see \cite[Theorem 3.20]{boucheron_concentration}), we have:
\begin{equation*}
\E \Bigg[\Bigg(\frac{ \ln \cZ_{t,R}}{n} - \E\bigg[\frac{\ln  \cZ_{t,R}}{n} \bigg\vert \bS, \bW \bigg]\Bigg)^{\!\! 2}\,\Bigg]
\leq \E\big[\big\Vert \nabla g(\bZ,\widetilde{\bZ}) \big\Vert^2 \,\big] \;.
\end{equation*}
The squared norm of the gradient of $g$ reads
$\Vert \nabla g\Vert^2 =
\sum_{\bi \in \cI} \vert\nicefrac{\partial g}{\partial Z_{\bi}}\vert^2
+ \sum_{j} \vert\nicefrac{\partial g}{\partial \widetilde{Z}_{j}}\vert^2$.
Each of these partial derivatives takes the form 
$\nicefrac{\partial g}{\partial x} = -n^{-1} \big\langle \nicefrac{\partial \mathcal{H}_{t,R}}{\partial x} \big\rangle$. More precisely:
\begin{equation*}
\frac{\partial g}{\partial Z_{\bi}} 
= n^{-1}  \frac{\sqrt{\lambda (1-t)}}{n} \, \langle  x_{i_1} x_{i_2} x_{i_3} \rangle_{t,R} \quad ; \quad
\frac{\partial g}{\partial \widetilde{Z}_{j}} 
= n^{-1} \sqrt{\frac{\lambda R}{2}} \, \langle x_j \rangle_{t,R} \;.
\end{equation*}
We see that $\vert \nicefrac{\partial g}{\partial Z_{\bi}} \vert \leq \frac{\sqrt{\lambda}}{n^2} \Vert \varphi \Vert_{\infty}^3$
and $\vert \frac{\partial g}{\partial \widetilde{Z}_{j}} \vert \leq n^{-1}  \sqrt{\frac{\lambda R}{2}} \Vert \varphi \Vert_{\infty}$.
Therefore:
$$
\Vert \nabla g(\bZ,\widetilde{\bZ}) \Vert^2 \leq \frac{\lambda^{\nicefrac{3}{2}}}{6n} \Vert \varphi \Vert_{\infty}^6 + \frac{\lambda R}{2n}\Vert \varphi \Vert_{\infty}^2 + \cO(n^{-2})\;.
$$
Making use of the Gaussian-Poincar\'{e} inequality, we obtain (the term $\cO(n^{-2})$ is neglected):
\begin{equation}\label{bound_variance_GP_1}
\E \Bigg[\Bigg(\frac{ \ln \cZ_{t,R}}{n} - \E\bigg[\frac{\ln  \cZ_{t,R}}{n} \bigg\vert \bS, \bW \bigg]\Bigg)^{\!\! 2}\,\Bigg]
\leq \frac{\lambda \Vert \varphi \Vert_{\infty}^4}{6n} (\sqrt{\lambda} \Vert \varphi \Vert_{\infty}^2 +  3 R) \;.
\end{equation}
Next we show that $\E[\nicefrac{\ln  \cZ_{t,R}}{n} \vert \bS, \bW]$ concentrates on its conditional expectation given $\bS$.
Thus, $\nicefrac{\ln  \cZ_{t,R}}{n}$ is seen as a function of the Gaussian random variables $\bW$ and we work conditionally to $\bS$:
$g(\bW) \equiv \E[\nicefrac{\ln  \cZ_{t,R}}{n} \vert \bW, \bS]$.
We will again invoke the Gaussian-Poincar\'{e} inequality (see \cite[Theorem 3.20]{boucheron_concentration}).
To lighten the equations we drop the subscripts in the Gibbs bracket $\langle - \rangle_{t,R}$, we introduce the notation $\widetilde{\E} \triangleq \E[\cdot \vert \bS,\bW]$ and we define the following quantities:
\begin{align*}
\bX' = \varphi'\bigg(\frac{\bW \bS}{\sqrt{p}}\bigg) \quad ; \quad \bx' = \varphi'\bigg(\frac{\bW \bS}{\sqrt{p}}\bigg) \;.
\end{align*}
The squared norm of the gradient of $g$ reads
$\Vert \nabla g\Vert^2 = \sum_{i, j} \vert\nicefrac{\partial g}{\partial W_{ij}}\vert^2$ where $\forall (i,j) \in \{1,\dots,n\} \times \{1,\dots, p\}$:
\begin{align*}
\frac{\partial g}{\partial W_{ij}}
&= \cO(n^{-\nicefrac{5}{2}})+
\frac{1}{2n}
\sum_{\substack{k = 1\\ k \neq i}}^n \sum_{\substack{\ell = 1\\ \ell \neq k,i}}^{n}
\bigg(-\frac{\lambda(1-t)}{n^2\sqrt{p}}  \widetilde{\E} \langle x_{i} x'_i s_j x_{k}^2 x_{\ell}^2 \rangle
+\frac{\lambda(1-t)}{n^2\sqrt{p}}\, S_j X'_{i} X_{k} X_{\ell} \widetilde{\E}\langle x_{i} x_{k} x_{\ell}\rangle\\
&\qquad\qquad\qquad\qquad\quad\;\:
+\frac{\lambda(1-t)}{n^2\sqrt{p}}\, X_{i} X_{k} X_{\ell}\,\widetilde{\E} \langle s_j x'_{i} x_{k} x_{\ell} \rangle
+\frac{\sqrt{\lambda(1-t)}}{n\sqrt{p}}\, \widetilde{\E} Z_{i k \ell} \langle s_j x'_{i} x_{k} x_{\ell} \rangle\bigg)\\
&\quad
+\frac{1}{n} \sqrt{\frac{\lambda R}{2p}}
\bigg(\!-\sqrt{\frac{\lambda R}{2}} \widetilde{\E}\langle s_j x'_i x_i\rangle +  \sqrt{\frac{\lambda R}{2}}\,S_j X'_i \,\widetilde{\E}\langle x_i \rangle
+  \sqrt{\frac{\lambda R}{2}}\, X_i \,\widetilde{\E}\langle s_j x'_i\rangle
+ \widetilde{\E}\widetilde{Z}_i \langle s_j x'_i \rangle\bigg).
\end{align*}
The term $\cO(n^{-\nicefrac{5}{2}})$ comes from those triplets $\bi$ in $\cI \triangleq \{(i,j,k) \in [n]^3: i \leq j \leq k \}$ whose elements are non unique.
In order to further simplify these partial derivatives, we do an integration by parts with respect to the Gaussian noises $\bZ$ and $\widetilde{\bZ}$. It yields:
\begin{align*}
\widetilde{\E} Z_{i k \ell} \langle s_j x'_{i} x_{k} x_{\ell} \rangle
&= \frac{\sqrt{\lambda (1-t)}}{n}\widetilde{\E}\langle s_j x'_{i} x_i x_{k}^2 x_{\ell}^2 \rangle
-  \frac{\sqrt{\lambda (1-t)}}{n}\widetilde{\E} \langle s_j x'_{i} x_{k} x_{\ell} \rangle  \langle x_{i} x_{k} x_{\ell} \rangle \;;\\
\widetilde{\E}\widetilde{Z}_i \langle s_j x'_i \rangle
&= \sqrt{\frac{\lambda R}{2}}\widetilde{\E}\langle s_j x'_i x_i\rangle - \sqrt{\frac{\lambda R}{2}}\widetilde{\E}\langle s_j x'_i\rangle \langle x_i\rangle\;.
\end{align*}
Therefore, $\forall (i,j) \in \{1,\dots,n\} \times \{1,\dots, p\}$:
\begin{multline*}
\frac{\partial g}{\partial W_{ij}}
= \cO(n^{-\nicefrac{5}{2}})
+\frac{\lambda R}{2 n\sqrt{p}}
\bigg(S_j X'_i \,\widetilde{\E}\langle x_i \rangle
+  X_i \,\widetilde{\E}\langle s_j x'_i\rangle
- \widetilde{\E}\langle s_j x'_i\rangle \langle x_i\rangle\bigg)\\
+\frac{\lambda(1-t)}{2n^3\sqrt{p}}  
\sum_{\substack{k = 1\\ k \neq i}}^n \sum_{\substack{\ell = 1\\ \ell \neq k,i}}^{n}
\bigg(\!
S_j X'_{i} X_{k} X_{\ell} \widetilde{\E}\langle x_{i} x_{k} x_{\ell}\rangle
+ X_{i} X_{k} X_{\ell}\,\widetilde{\E} \langle s_j x'_{i} x_{k} x_{\ell} \rangle
-\widetilde{\E} \langle s_j x'_{i} x_{k} x_{\ell} \rangle  \langle x_{i} x_{k} x_{\ell} \rangle \!\bigg)\,.
\end{multline*}
Making use of the boundedness assumptions, we obtain the following uniform bound on the partial derivatives:
\begin{equation*}
\bigg\vert \frac{\partial g}{\partial W_{ij}} \bigg\vert
\leq \cO(n^{-\nicefrac{5}{2}}) +
\frac{3 \lambda M_S}{2n\sqrt{p}}\Vert \varphi \Vert_{\infty} \Vert \varphi' \Vert_{\infty}\big(\Vert \varphi \Vert_{\infty}^4 + R \big)\;.
\end{equation*}
Therefore, $\Vert \nabla g(\bW) \Vert^2 \leq \frac{9 \lambda^2 M_S^2}{4n}\Vert \varphi \Vert_{\infty}^2 \Vert \varphi' \Vert_{\infty}^2\big(\Vert \varphi \Vert_{\infty}^4 + R \big)^2 + \cO(n^{-3})$ and the Gaussian-Poincar\'{e} inequality yields (the term $\cO(n^{-3})$ is neglected):
\begin{equation}\label{bound_variance_GP_2}
\E \Bigg[\Bigg(\E\bigg[\frac{\ln  \cZ_{t,R}}{n} \bigg\vert \bS, \bW \bigg] - \E\bigg[\frac{\ln  \cZ_{t,R}}{n} \bigg\vert \bS\bigg]\Bigg)^{\!\! 2}\,\Bigg]
\leq
\frac{9 \lambda^2 M_S^2}{4n}\Vert \varphi \Vert_{\infty}^2 \Vert \varphi' \Vert_{\infty}^2\big(\Vert \varphi \Vert_{\infty}^4 + R \big)^2 \;.
\end{equation}
Finally, it remains to show that $\E[\nicefrac{\ln  \cZ_{t,R}}{n} \vert \bS]$ concentrates on its expectation.
We will show that the function
$$
g: \mathtt{S} \in [-M_S,M_S]^p \mapsto \E[\nicefrac{\ln  \cZ_{t,R}}{n} \vert \bS = \mathtt{S} ]
$$
has bounded differences. To do so, we will show that the partial derivatives of $g$ are uniformly bounded.
Then we will apply the bounded differences inequality, also called McDiarmid's inequality, to get the concentration result (see \cite{McDiarmid}, \cite[Corollary 3.2]{boucheron_concentration}).
Similarly to what has be done with the random vector $\bS$, we define $\mathtt{X} = \varphi\big(\frac{\bW \ttS}{\sqrt{p}}\big)$, $\mathtt{X}' = \varphi'\big(\frac{\bW \ttS}{\sqrt{p}}\big)$ and $\mathtt{X}'' = \varphi''\big(\frac{\bW \ttS}{\sqrt{p}}\big)$.
For $\ell \in \{1,\dots,p\}$, the partial derivative of $g$ with respect to its $\ell^{\mathrm{th}}$ coordinate:
\begin{align}
\frac{\partial g}{\partial \ttS_\ell}
&= \frac{\lambda(1-t)}{n^3\sqrt{p}}
\sum_{\bi \in \cI}
\E\big[(W_{i_1 \ell} \ttX'_{i_1} \ttX_{i_2} \ttX_{i_3} + W_{i_2 \ell} \ttX_{i_1} \ttX'_{i_2} \ttX_{i_3} + W_{i_3 \ell} \ttX_{i_1} \ttX_{i_2} \ttX'_{i_3} )
\langle x_{i_1} x_{i_2} x_{i_3}\rangle\big\vert \bS = \ttS \big]\nonumber\\
&\qquad\qquad\qquad\qquad\qquad\qquad\qquad\qquad\qquad\quad
+\frac{\lambda R}{2n\sqrt{p}}
\sum_{i=1}^{n} \E\big[W_{i\ell}\ttX'_i \langle x_i \rangle\big\vert \bS = \ttS \big]\nonumber\\
&= \cO(n^{-\nicefrac{3}{2}}) + \frac{\lambda(1-t)}{2n^3\sqrt{p}}
\sum_{i=1}^{n}\sum_{\substack{j = 1\\ j \neq i}}^n \sum_{\substack{k = 1\\ k \neq i,j}}^{n}
\E\big[W_{i \ell} \ttX'_{i} \ttX_{j} \ttX_{k}
\langle x_{i} x_{j} x_{k}\rangle\big\vert \bS = \ttS \big]\nonumber\\
&\qquad\qquad\qquad\qquad\qquad\qquad\qquad\qquad\qquad\quad
+\frac{\lambda R}{2n\sqrt{p}}
\sum_{i=1}^{n} \E\big[W_{i\ell}\ttX'_i \langle x_i \rangle\big\vert \bS = \ttS \big]\label{formula_dg/dS_l}\;.
\end{align}
Once again the triplets $\bi \in \cI$ whose elements are non unique are accounted for with the term $\cO(n^{-\nicefrac{3}{2}})$, which is negligible compared to the others.
An integration by parts with respect to $\bW$ gives for all $(i,j,k,\ell) \in \{1,\dots,n\}^3 \times \{1,\dots,p\}$ such that $j \neq i$ and $k \neq i,j$:
\begin{align}
&\E\big[W_{i \ell} \ttX'_{i} \ttX_{j} \ttX_{k}
\langle x_{i} x_{j} x_{k}\rangle\big\vert \bS = \ttS \big]\nonumber\\
&\qquad= \frac{1}{\sqrt{p}} \E\big[ \ttS_\ell \ttX''_{i} \ttX_{j} \ttX_{k}
\langle x_{i} x_{j} x_{k}\rangle\big\vert \bS = \ttS \big]
+\frac{1}{\sqrt{p}} \E\big[ \ttX'_{i} \ttX_{j} \ttX_{k}
\langle s_\ell x'_{i} x_{j} x_{k}\rangle\big\vert \bS = \ttS \big]\nonumber\\
&\qquad\qquad
-\E\bigg[\ttX'_{i} \ttX_{j} \ttX_{k} \bigg\langle \!\! x_{i} x_{j} x_{k} \frac{\partial \cH_{t,R}}{\partial W_{i\ell}} \bigg\rangle\bigg\vert \bS = \ttS \bigg]
+\E\bigg[\ttX'_{i} \ttX_{j} \ttX_{k} \langle x_{i} x_{j} x_{k} \rangle \bigg\langle\frac{\partial \cH_{t,R}}{\partial W_{i\ell}} \bigg\rangle\bigg\vert \bS = \ttS \bigg]\;;
\label{1st_expectation_dg/dS_l}
\end{align}
\begin{align}
\E\big[W_{i\ell}\ttX'_i \langle x_i \rangle\big\vert \bS = \ttS \big]
&= \frac{1}{\sqrt{p}} \E\big[\ttS_\ell \ttX''_i \langle x_i \rangle\big\vert \bS = \ttS \big]
+ \frac{1}{\sqrt{p}} \E\big[\ttX'_i \langle s_\ell x'_i \rangle\big\vert \bS = \ttS \big]\nonumber\\
&\qquad\quad
-\E\bigg[\ttX'_i \bigg\langle\! x_i \frac{\partial \cH_{t,R}}{\partial W_{i\ell}} \bigg\rangle\bigg\vert \bS = \ttS \bigg]
+\E\bigg[\ttX'_i \langle x_i \rangle \bigg\langle\frac{\partial \cH_{t,R}}{\partial W_{i\ell}} \bigg\rangle\bigg\vert \bS = \ttS \bigg]\;.
\label{2nd_expectation_dg/dS_l}
\end{align}
Here $\cH_{t,R} \equiv \cH_{t,R}(\bs; \ttS, \bZ, \widetilde{\bZ}, \bW)$.
In order to prove the concentration result that we aim for, we need to make sure that both expectations\eqref{1st_expectation_dg/dS_l} and \eqref{2nd_expectation_dg/dS_l} are $\cO(n^{-\nicefrac{1}{2}})$.
The main difficulty resides in managing the terms where partial derivatives $\nicefrac{\partial \cH_{t,R}}{\partial W_{i \ell}}$ appear.
We have already dealt with these partial derivatives when proving the concentration with respect to $\bW$, and we found:
\begin{align*}
\frac{\partial \cH_{t,R}}{\partial W_{i \ell}}
&= \cO(n^{-\nicefrac{3}{2}})+
\frac{1}{2}
\sum_{\substack{j = 1\\ j \neq i}}^n \sum_{\substack{k = 1\\ k \neq i,j}}^{n}
\bigg(-\frac{\lambda(1-t)}{n^2\sqrt{p}} x_{i} x'_i s_\ell x_{j}^2 x_{k}^2
+\frac{\lambda(1-t)}{n^2\sqrt{p}}\, \ttS_\ell \ttX'_{i} \ttX_{j} \ttX_{k} x_{i} x_{j} x_{k}\\
&\qquad\qquad\qquad\qquad\qquad\quad
+\frac{\lambda(1-t)}{n^2\sqrt{p}}\, \ttX_{i} \ttX_{j} \ttX_{k} s_\ell x'_{i} x_{j} x_{k} 
+\frac{\sqrt{\lambda(1-t)}}{n\sqrt{p}}\, Z_{i j k}  s_\ell x'_{i} x_{j} x_{k}\bigg)\\
&\qquad\qquad\quad
+ \sqrt{\frac{\lambda R}{2p}}
\bigg(-  \sqrt{\frac{\lambda R}{2}} s_\ell x'_i x_i +  \sqrt{\frac{\lambda R}{2}}\,\ttS_\ell \ttX'_i x_i
+  \sqrt{\frac{\lambda R}{2}}\, \ttX_i \, s_\ell x'_i
+ \widetilde{Z}_i s_\ell x'_i \bigg)\;.
\end{align*}
For $(i,\ell) \in \{1,\dots,n\} \times \{1,\dots,p\}$ define
\begin{multline}
\mathcal{A}_{i\ell}
\triangleq \frac{\lambda(1-t)}{2n^2\sqrt{p}} 
\sum_{\substack{j = 1\\ j \neq i}}^n \sum_{\substack{k = 1\\ k \neq i,j}}^{n}
\big(-x_{i} x'_i s_\ell x_{j}^2 x_{k}^2
+\ttS_\ell \ttX'_{i} \ttX_{j} \ttX_{k} x_{i} x_{j} x_{k}
+ \ttX_{i} \ttX_{j} \ttX_{k} s_\ell x'_{i} x_{j} x_{k} \big)\\
+ \frac{\lambda R}{2\sqrt{p}}
\big(- s_\ell x'_i x_i + \ttS_\ell \ttX'_i x_i
+ \ttX_i \, s_\ell x'_i \big)\;.
\end{multline}
Note these two simple facts about $\mathcal{A}_{i\ell}$:
\begin{align}
\frac{\partial \cH_{t,R}}{\partial W_{i \ell}}
&= \cO(n^{-\nicefrac{3}{2}}) + \mathcal{A}_{i\ell}
+
\frac{\sqrt{\lambda(1-t)}}{2n\sqrt{p}}
\sum_{\substack{j = 1\\ j \neq i}}^n \sum_{\substack{k = 1\\ k \neq i,j}}^{n} Z_{i j k}  s_\ell x'_{i} x_{j} x_{k}
+ \sqrt{\frac{\lambda R}{2p}} \widetilde{Z}_i s_\ell x'_i \;;\label{link_A_dG/dW}\\
\vert \mathcal{A}_{i\ell} \vert
&\leq \frac{3\lambda}{2\sqrt{p}} \, M_S \Vert \varphi \Vert_{\infty} \Vert \varphi' \Vert_{\infty}\big(\Vert \varphi \Vert_{\infty}^4 + R \big)\;.\label{upperbound_A}
\end{align}
Plugging the identity \eqref{link_A_dG/dW} in both \eqref{1st_expectation_dg/dS_l} and \eqref{2nd_expectation_dg/dS_l} and making use of the upper bound \eqref{upperbound_A}, we obtain:
\begin{align*}
&\big\vert \E\big[W_{i \ell} \ttX'_{i} \ttX_{j} \ttX_{k}
\langle x_{i} x_{j} x_{k}\rangle\big\vert \bS = \ttS \big] \big\vert\\
&\;\leq \cO(n^{-\nicefrac{3}{2}}) + \frac{M_S \Vert \varphi \Vert_{\infty}^4 }{\sqrt{p}}
\Big(\Vert \varphi \Vert_{\infty} \Vert \varphi'' \Vert_{\infty}
+ \Vert \varphi \Vert_{\infty} \Vert \varphi' \Vert_{\infty}^2
+ 3\lambda\,\Vert \varphi \Vert_{\infty}^6\Vert \varphi' \Vert_{\infty}^2 
+  3\lambda\, \Vert \varphi \Vert_{\infty}^2\Vert \varphi' \Vert_{\infty}^2 R\Big)\\
&\quad\;
+\frac{\sqrt{\lambda(1-t)}}{2n\sqrt{p}}
\sum_{\substack{j' = 1\\ j' \neq i}}^n \sum_{\substack{k' = 1\\ k \neq i',j'}}^{n}\!
\big\vert \E\big[\ttX'_{i} \ttX_{j} \ttX_{k} Z_{i j' k'} \big(\langle x_{i} x_{j} x_{k} s_\ell x'_{i} x_{j'} x_{k'} \rangle
-  \langle x_{i} x_{j} x_{k} \rangle \langle  s_\ell x'_{i} x_{j'} x_{k'} \rangle\big)
\big\vert \bS = \ttS \big]\big\vert\\
&\quad\;
+\sqrt{\frac{\lambda R}{2p}} \big\vert\E\big[\ttX'_{i} \ttX_{j} \ttX_{k} \widetilde{Z}_i \big(\langle x_{i} x_{j} x_{k}  s_\ell x'_i \rangle
-\langle x_{i} x_{j} x_{k} \rangle \langle  s_\ell x'_i \rangle \big)\big\vert \bS = \ttS \big]\big\vert\;;
\end{align*}
\begin{align*}
&\big\vert \E\big[W_{i\ell}\ttX'_i \langle x_i \rangle\big\vert \bS = \ttS \big] \big\vert\\
&\qquad\leq \cO(n^{-\nicefrac{3}{2}}) + \frac{M_S}{\sqrt{p}}
\Big(\Vert \varphi \Vert_{\infty}\Vert \varphi'' \Vert_{\infty}
+ \Vert \varphi' \Vert_{\infty}^2
+ 3\lambda\,\Vert \varphi \Vert_{\infty}^6\Vert \varphi' \Vert_{\infty}^2 
+  3\lambda\, \Vert \varphi \Vert_{\infty}^2\Vert \varphi' \Vert_{\infty}^2 R\Big)\\
&\qquad\qquad\qquad
+\frac{\sqrt{\lambda(1-t)}}{2n\sqrt{p}}
\sum_{\substack{j' = 1\\ j' \neq i}}^n \sum_{\substack{k' = 1\\ k \neq i',j'}}^{n}
\big\vert \E\big[\ttX'_{i} Z_{i j' k'} \big(\langle x_{i} s_\ell x'_{i} x_{j'} x_{k'} \rangle
-  \langle x_{i} \rangle \langle  s_\ell x'_{i} x_{j'} x_{k'} \rangle\big)
\big\vert \bS = \ttS \big]\big\vert\\
&\qquad\qquad\qquad
+\sqrt{\frac{\lambda R}{2p}} \big\vert\E\big[\ttX'_i \widetilde{Z}_i \big(\langle x_i  s_\ell x'_i \rangle
-\langle x_i \rangle \langle  s_\ell x'_i \rangle \big)\big\vert \bS = \ttS \big]\big\vert\;.
\end{align*}
By integrating by parts with respect to $\bZ$ or $\widetilde{\bZ}$, we can now show that both upper bounds are $\cO(p^{-\nicefrac{1}{2}})$. This is because $Z_{ij'k'}$ and $\widetilde{Z}_i$ appear in the Hamiltonian $\cH_{t,R}$ via the terms $\frac{\sqrt{\lambda (1-t)}}{n}x_i x_{j'} x_{k'} Z_{ij'k'}$ and $\sqrt{\frac{\lambda R}{2}} x_i \widetilde{Z}_i$, respectively.
In the end, for all $(i,j,k,\ell) \in \{1,\dots,n\}^3 \times \{1,\dots,p\}$ such that $j \neq i$ and $k \neq i,j$:
\begin{align*}
\big\vert \E\big[W_{i \ell} \ttX'_{i} \ttX_{j} \ttX_{k}
\langle x_{i} x_{j} x_{k}\rangle\big\vert \bS = \ttS \big] \big\vert
&\leq \frac{M_S \Vert \varphi \Vert_{\infty}^4 }{\sqrt{p}}C_1\;;\\
\big\vert \E\big[W_{i\ell}\ttX'_i \langle x_i \rangle\big\vert \bS = \ttS \big] \big\vert
&\leq \frac{M_S}{\sqrt{p}} C_1\;;
\end{align*}
where
$C_1 \triangleq \Vert \varphi \Vert_{\infty}\Vert \varphi'' \Vert_{\infty}
+ \Vert \varphi' \Vert_{\infty}^2
+ 6\lambda\Vert \varphi \Vert_{\infty}^6\Vert \varphi' \Vert_{\infty}^2 
+  6\lambda \Vert \varphi \Vert_{\infty}^2\Vert \varphi' \Vert_{\infty}^2 R$.
Going back to the identity \eqref{formula_dg/dS_l}, these upper bounds yield
$\big\vert \frac{\partial g}{\partial \ttS_\ell} \big\vert \leq \frac{\lambda M_S}{2 p} (\Vert \varphi \Vert_{\infty}^4 + R)\,
C_1$
uniformly in $\ttS \in [-M_S,M_S]^p$ and $\ell \in \{1\dots,p\}$ (we neglect the term $\cO(n^{-\nicefrac{3}{2}})$ that should appear in the upper bound). Hence $g$ has bounded differences (this is a simple application of the mean value theorem):
\begin{equation*}
\forall \ell \in \{1,\dots,p\}:
\sup_{-M_S \leq \ttS_1,\dots,\ttS_p, \ttS'_\ell \leq M_S}
\big\vert g(\ttS) - g(\ttS_1,\dots,\ttS_{\ell-1},\ttS'_\ell,\ttS_{\ell+1},\dots,\ttS_p)\big\vert
\leq \frac{C_2}{p}\,;
\end{equation*}
where
$C_2 \triangleq \lambda M_S^2 (\Vert \varphi \Vert_{\infty}^4 + R)\,C_1$.
By McDiarmid's inequality:
\begin{equation}\label{bound_variance_McDiarmid}
\E \Bigg[\Bigg(\E\bigg[\frac{\ln  \cZ_{t,R}}{n} \bigg\vert \bS \bigg] - \E\bigg[\frac{\ln  \cZ_{t,R}}{n}\bigg]\Bigg)^{\!\! 2}\,\Bigg]
\leq
\frac{C_2^2}{4p}\;.
\end{equation}
Combining the inequalities \eqref{bound_variance_GP_1},  \eqref{bound_variance_GP_2} and \eqref{bound_variance_McDiarmid} yields the final result.
\end{IEEEproof}
\section{Proof of Proposition~\ref{proposition:properties_h}}\label{app:proof_mmse}
The proof is based on the envelope theorem \cite[Corollary 4]{Milgrom_Envelope_Theorems} to obtain the derivative of $h$.
We proceed as follows:
\begin{enumerate}
	\item We show that $h$ is equal to the minimization on a compact 
	subset of a function having sufficient regularity properties to apply \cite[Corollary 4]{Milgrom_Envelope_Theorems}.
	\item The later gives a formula for the derivative of $h$ at any point where it is differentiable.
	\item We use an optimality condition on $q_x^* \in \mathcal{Q}_x^*(\lambda)$ leading to simplified formula \eqref{derivative_h} for $h'(\lambda)$.
\end{enumerate}

\begin{IEEEproof}[Proof of Proposition \ref{proposition:properties_h}]
We proceed according to the plan outlined above.\\
\noindent\textbf{1)}
We define $f(q_x,q_s,\lambda) \triangleq {\sup}_{r_s \geq 0}\; \psi_{\lambda,\alpha}(q_x , q_s, r_s)$.
By the definition \eqref{def_potential_psi} of $\psi_{\lambda, \alpha}$, we have for all $(q_x, q_s,\lambda) \in [0,\rho_x] \times [0,\rho_s] \times (0,+\infty)$:
\begin{equation}\label{def_potential_f}
f(q_x,q_s,\lambda) = \frac{1}{\alpha}I_{P_S}^*\bigg(\frac{q_s - \rho_s}{2}\bigg) + I_\varphi\bigg(\frac{\lambda q_x^2}{2}, q_s;\rho_s\bigg) +\frac{\lambda}{12} (\rho_x - q_x)^2(\rho_x + 2 q_x)\;,
\end{equation}
where the functions $I_{P_S}^*$ and $I_{\varphi}(\cdot\, , \cdot\,; \rho_s)$ are defined in Lemma~\ref{lemma:properties_I_PS} and Lemma~\ref{lemma:properties_functions}, respectively.
By Lemma~\ref{lemma:properties_functions}, $I_{\varphi}(\cdot\, , \cdot\,; \rho_s)$ is continuous on $[0,+\infty) \times [0,\rho_s]$.
By Lemma~\ref{lemma:properties_I_PS}, $I_{P_S}^*$ is convex and finite on $(-\infty,0)$, hence continuous on $(-\infty,0)$.
Besides, $I_{P_S}^*$ is nondecreasing on $(-\infty,0)$ and we distinguish between two cases:
(i) $\lim_{\substack{x \to 0\\x<0}} I_{P_S}^*(x)$ exists and is finite, and (ii) $\lim_{\substack{x \to 0\\x<0}} I_{P_S}^*(x)$ diverges to $+\infty$.
If (i) then, by monotonicity of $I_{P_S}^*$, $\lim_{\substack{x \to 0\\x<0}} I_{P_S}^*(x) \leq I_{P_S}^*(0)$.
We can redefine $I_{P_S}^*$ at $x=0$ to make it left continuous while leaving $h$ unchanged: $I_{P_S}^*(0) \triangleq \lim_{\substack{x \to 0\\x<0}} I_{P_S}^*(x)$.
Hence, $f$ is continuous on $[0,\rho_x] \times [0,\rho_s] \times (0,+\infty)$ and
$h(\lambda) = \min_{(q_x,q_s) \in [0,\rho_x] \times [0,\rho_s]} \;f(q_x,q_s,\lambda)$.
If (ii), first note that
\begin{align*}
f(0,0,\lambda) = \frac{1}{\alpha}I_{P_S}^*\big(-\rho_s/2\big) +\frac{\lambda}{12} \rho_x^3
\quad \text{and} \quad
f(q_x,q_s,\lambda) \geq \frac{1}{\alpha}I_{P_S}^*\bigg(\frac{q_s - \rho_s}{2}\bigg) \xrightarrow[q_s \to \rho_s]{q_s < 0} +\infty \;.
\end{align*}
Then, for every positive $\bar{\lambda}$, there exists $\rho_s(\bar{\lambda}) \in (0,\rho_s)$ such that:
\begin{itemize}
	\item $\forall (q_x,q_s,\lambda) \in [0,\rho_x] \times [\rho_s(\bar{\lambda}),\rho_s] \times (0,\bar{\lambda}]$:
	$f(q_x,q_s,\lambda) > f(0,0,\lambda)$;
	\item $f$ is continuous on  $[0,\rho_x] \times [0,\rho_s(\bar{\lambda})]\times (0,+\infty)$.
\end{itemize}
Thus, $\forall \lambda \in [0,\bar{\lambda}]$:
$h(\lambda) = \min_{(q_x,q_s) \in [0,\rho_x] \times [0,\rho_s(\bar{\lambda})]} \;f(q_x,q_s,\lambda)$.

\noindent \textbf{2)}
Fix $\bar{\lambda} > 0$. The conclusion of step \textbf{1)} is that there exists $\rho_s(\bar{\lambda}) \in (0,\rho_s]$ such that
\begin{equation*}
\forall \lambda \in (0,\bar{\lambda}]:
h(\lambda) = \min_{(q_x,q_s) \in [0,\rho_x] \times [0,\rho_s(\bar{\lambda})]} \;f(q_x,q_s,\lambda) \;,
\end{equation*}
where $f$ is defined in \eqref{def_potential_f} with $I_{P_S}^*(0) \triangleq \lim_{\substack{x \to 0\\x<0}} I_{P_S}^*(x) \in [0,+\infty]$ and is continuous on $[0,\rho_x] \times [0,\rho_s(\bar{\lambda})] \times (0,+\infty)$.
By Lemma~\ref{lemma:properties_functions}, $f$ admits a derivative with respect to $\lambda$ and for all $(q_x, q_s, \lambda) \in [0,\rho_x] \times [0,\rho_s(\bar{\lambda})] \times (0,+\infty)$:
\begin{equation}\label{derivative_f_wrt_lambda}
\frac{\partial f}{\partial \lambda}\bigg\vert_{q_x,q_s,\lambda}
= \frac{q_x^2}{2}\frac{\partial I_\varphi}{\partial r} \bigg(\frac{\lambda q_x^2}{2},q_s;\rho_s\bigg)
+\frac{1}{12} (\rho_x - q_x)^2(\rho_x + 2 q_x) \;.
\end{equation}
This partial derivative is continuous on $[0,\rho_x] \times [0,\rho_s(\bar{\lambda})] \times (0,+\infty)$ ($\nicefrac{\partial I_\varphi}{\partial r}$ is given by \eqref{sign_derivative_I} and its continuity is justified by domination assumptions).
For all $\lambda \in (0,\bar{\lambda})$, define the following nonempty subset of $[0,\rho_x] \times [0,\rho_s(\bar{\lambda})]$:
$$
\mathcal{Q}_{x,s}^*(\lambda)
\triangleq \big\{(q_x^*,q_s^*) \in [0,\rho_x] \times [0,\rho_s]: f(q_x^* , q_s^*, \lambda) 
= h(\lambda)\big\}\;.
$$
By \cite[Corollary 4]{Milgrom_Envelope_Theorems}, $h$ is differentiable at $\lambda \in (0,\bar{\lambda})$ if, and only if, the set
\begin{equation*}
\mathcal{F}(\lambda)
\triangleq \bigg\{ \frac{\partial f}{\partial \lambda}\bigg\vert_{q_x^*,q_s^*,\lambda}: (q_x^* , q_s^*) \in \mathcal{Q}_{x,s}^*(\lambda)\bigg\}
\end{equation*}
is a singleton.
In this case, $\forall (q_x^* , q_s^*) \in \mathcal{Q}_{x,s}^*(\lambda): h'(\lambda) = \frac{\partial f}{\partial \lambda}\big\vert_{q_x^*,q_s^*,\lambda}$.
Note that $\mathcal{F}(\lambda)$ could be a singleton without $\mathcal{Q}_{x,s}^*(\lambda)$ being one.
However, in the next and final step, we derive a simple expression for $\frac{\partial f}{\partial \lambda}\big\vert_{q_x^*,q_s^*,\lambda}$ where $(q_x^* , q_s^*) \in \mathcal{Q}_{x,s}^*(\lambda)$ that shows that $\mathcal{F}(\lambda)$ is a singleton if, and only if, $\mathcal{Q}_{x,s}^*(\lambda)$ is one too.

\noindent \textbf{3)}
Let $\lambda \in (0,\bar{\lambda})$ and $(q_x^*, q_s^*) \in \mathcal{Q}_{x,s}^*(\lambda)$.
The function $q_x \mapsto f(q_x,q_s^*,\lambda)$ is differentiable on $[0,\rho_x]$ and $f(q_x^*,q_s^*,\lambda) = \min_{q_x^* \in [0,\rho_x]} f(q_x,q_s^*,\lambda)$.
If $q_x^* \in (0,\rho_x)$ then it satisfies the optimality condition $\frac{\partial f}{\partial q_x}\Big\vert_{q_x^*,q_s^*,\lambda} = 0$, i.e.,
\begin{equation}\label{optimality_condition_q_x*}
q_x^* \,\frac{\partial I_\varphi}{\partial r} \bigg(\frac{\lambda (q_x^*)^2}{2},q_s^*;\rho_s\bigg)
= \frac{q_x^*}{2} (\rho_x - q_x^*)\;.
\end{equation}
The identity \eqref{optimality_condition_q_x*} is trivially satisfied if $q_x^* = 0$.
If $q_x^* = \rho_x$ then the necessary optimality condition reads $\frac{\partial f}{\partial q_x}(\rho_x,q_s^*,\lambda) = \lambda \rho_x \frac{\partial I_\varphi}{\partial r} \big(\frac{\lambda \rho_x^2}{2},q_s^*;\rho_s\big) \leq 0$.
We also show in the proof of Lemma~\ref{lemma:properties_functions} that $\frac{\partial I_\varphi}{\partial r} \geq 0$. Hence, if $q_x^* = \rho_x$, the identity \eqref{optimality_condition_q_x*} is still satisfied.
Making use of the identity \eqref{optimality_condition_q_x*} in \eqref{derivative_f_wrt_lambda}, we have $\forall (q_x^*, q_s^*) \in \mathcal{Q}_{x,s}^*(\lambda)$:
\begin{align*}
\frac{\partial f}{\partial \lambda}\bigg\vert_{q_x^*,q_s^*,\lambda}
&= \frac{(q_x^*)^2}{2}\frac{\partial I_\varphi}{\partial r} \bigg(\frac{\lambda (q_x^*)^2}{2},q_s^*;\rho_s\bigg)
+\frac{1}{12} (\rho_x - q_x^*)^2(\rho_x + 2 q_x^*)\\
&= \frac{(q_x^*)^2}{4} (\rho_x - q_x^*)
+\frac{1}{12} (\rho_x - q_x^*)^2(\rho_x + 2 q_x^*)
= \frac{\rho_x^3 - (q_x^*)^3}{12}\;.
\end{align*}
It follows that $\mathcal{F}(\lambda)$ is a singleton if, and only if, $\mathcal{Q}_x^*(\lambda)$ is a singleton.
We conclude that $h$ is differentiable if, and only if, $\mathcal{Q}_x^*(\lambda)$ is a singleton in which case, letting $\mathcal{Q}_x^*(\lambda) = \{q_x^*(\lambda)\}$, $h'(\lambda) = \frac{\rho_x^3 - (q_x^*(\lambda))^3}{12}$.
\end{IEEEproof}

\end{document}